%% file: arXiv_version3.tex
\documentclass[12pt]{article}
\usepackage{rotating} 
\usepackage{adjustbox,lipsum}
\usepackage{float}
\usepackage{amssymb}
\usepackage{amsthm}
\usepackage{amsmath}
\usepackage{latexsym}
\usepackage[noend]{algpseudocode}
\usepackage{epsfig}
\usepackage{graphicx}
\usepackage{wasysym}
\usepackage{threeparttable}
\usepackage{natbib}
\usepackage{color}
\usepackage[table]{xcolor}
\usepackage{epstopdf}
\usepackage{caption}
\usepackage{subcaption}
\usepackage{booktabs}

\newcommand{\ind}{\stackrel{\mathrm{ind}}{\sim}}

\usepackage{multirow}
\usepackage[breaklinks]{hyperref}

\usepackage{enumitem}
\usepackage{caption}
\usepackage{algorithm}

\def\boxit#1{\vbox{\hrule\hbox{\vrule\kern6pt
          \vbox{\kern6pt#1\kern6pt}\kern6pt\vrule}\hrule}}

\def\bse{\begin{eqnarray*}}
\def\ese{\end{eqnarray*}}
\def\be{\begin{eqnarray}}
\def\ee{\end{eqnarray}}
\def\bq{\begin{equation}}
\def\eq{\end{equation}}
\def\bse{\begin{eqnarray*}}
\def\ese{\end{eqnarray*}}

\usepackage[linewidth=1pt]{mdframed}
\usepackage{mathptmx}      
\usepackage{bm}

 {\begin{list}{}%
         {\setlength{\leftmargin}{#1}}%
         \item[]%
 }
 {\end{list}}
\usepackage{setspace}

\renewcommand{\arraystretch}{1.3}

\include{notations4}

\begin{document}
\thispagestyle{empty} \baselineskip=28pt

\begin{center}
{\LARGE{\bf Bayesian Hierarchical Models with Conjugate Full-Conditional Distributions for Dependent Data from the Natural Exponential Family}}
\end{center}

\baselineskip=12pt

\vskip 2mm
\begin{center}
Jonathan R. Bradley\footnote{(\baselineskip=10pt to whom correspondence should be addressed) Department of Statistics, Florida State University, 117 N. Woodward Ave, Tallahassee, Fl 32306, bradley@stat.fsu.edu},
Scott H. Holan\footnote{\baselineskip=10pt  Department of Statistics, University of Missouri, 146 Middlebush Hall, Columbia, MO 65211-6100}\footnote{\baselineskip=10pt  U.S. Census Bureau, 4600 Silver Hill Road, Washington, D.C., 20233-9100},
Christopher K. Wikle$^2$
\end{center}
%
%
%
%
\vskip 4mm

\begin{center}
\large{{\bf Abstract}}
\end{center}
We introduce a Bayesian approach for analyzing (possibly) high-dimensional dependent data that are distributed according to a member from the natural exponential family of distributions. This problem requires extensive methodological advancements, as jointly modeling high-dimensional dependent data leads to the so-called ``big $n$ problem.'' The computational complexity of the ``big $n$ problem'' is further exacerbated when allowing for non-Gaussian data models, as is the case here. Thus, we develop new computationally efficient distribution theory for this setting. In particular, we introduce the ``conjugate multivariate distribution,'' which is motivated by the univariate distribution introduced in \citet{diaconis}. Furthermore, we provide substantial theoretical and methodological development including: results regarding conditional distributions, an asymptotic relationship with the multivariate normal distribution, conjugate prior distributions, and full-conditional distributions for a Gibbs sampler. To demonstrate the wide-applicability of the proposed methodology, we provide two simulation studies and three applications based on an epidemiology dataset, a federal statistics dataset, and an environmental dataset, respectively.
\baselineskip=12pt

%
%
%

\baselineskip=12pt
\par\vfill\noindent
\par\medskip\noindent
\chapter{\hfill}
\clearpage\pagebreak\newpage \pagenumbering{arabic}
\baselineskip=24pt
\singlespacing
\section{Introduction}
The multivariate normal distribution has become a fundamental tool for statisticians, as it provides a way to incorporate dependence for Gaussian and non-Gaussian data alike. Notice that many statistical models are defined hierarchically, where the joint distribution of the data, latent processes, and unknown parameters are written as the product of a data model, a latent Gaussian process model, and a parameter model \citep[e.g., see][among others]{cressie-wikle-book,banerjee-etal-2004}. Jointly modeling a member from the exponential family may be seen as straightforward to some. That is, one can simply define the data model to be the appropriate member of the exponential family and define latent Gaussian processes using the hierarchical modeling framework. Models of this form are often referred to as \textit{latent Gaussian process} (LGP) models; see \citet{diggle}, \citet{rue}, \citet[][Sections~4.1.2 and 7.1.5]{cressie-wikle-book}, and \citet{holan_glm}, among others. 

In the Bayesian context, LGPs can be nontrivial to implement using standard Markov chain Monte Carlo (MCMC) procedures when the dataset is high-dimensional. This is primarily because big data can lead to big parameter spaces, which allows parameters to be highly correlated. This in turn, creates a challenge for defining useful proposal distributions, tuning these proposal distributions, and assessing convergence of the Markov chain (e.g., see \citet{rue} and \citet{bradleyPMSTM} for a discussion on convergence issues of MCMC algorithms for LGPs). In this article, our primary goal is to introduce new distribution theory that facilitates Bayesian inference of dependent non-Gaussian data. In particular, we introduce a multivariate distribution that leads to conjugate forms of the full conditional distributions within a Gibbs sampler. 

We provide a multivariate extension of the class of distributions introduced the seminal paper by \citet{diaconis}, who developed the conjugate prior for distributions from the natural exponential family (EF), which leads to the well-known Poisson/gamma, binomial/beta, negative binomial/beta, and gamma/inverse-gamma hierarchical models. In this article, we develop a multivariate version of this distribution, which we call the \textit{conjugate multivariate} (CM) distribution. Similar to the special cases that emerged from \citet{diaconis} we obtain Poisson/multivariate log-gamma (MLG), binomial/multivariate logit beta, negative binomial/multivariate logit beta, and gamma/multivariate negative-inverse-gamma hierarchical models \citep{chen2003conjugate}. The hierarchical model that specifies the data model to be from the natural exponential family, and the latent process to be a CM distribution is referred to as a latent CM process (LCM) model. The LCM model constitutes a more general paradigm for modeling dependent data than LGPs, since the LGP is a special case of the LCM model. An important motivating feature of this more general framework is that the LCM model incorporates dependency and results in full-conditional distributions (within a Gibbs sampler) that are easy to simulate from. This allows one to avoid computationally inefficient and subjective tuning methods. 

An immediate issue that arises with the introduction of the LCM model is the need to define flexible prior distributions. One goal of this article is to describe the \textit{fully conjugate} Bayesian hierarchical model that has a data model that belongs to the natural exponential family. By ``fully conjugate'' we mean that each full conditional distribution, within a Gibbs  sampler, falls in the same class of distributions of the associated process or parameter models. To derive a fully conjugate statistical model, we introduce the LCM analogue to the prior distributions used in \citet{danielscov}, \citet{dunson}, and \citet{Pourahmadi} for covariance parameters. Additionally, extensions of the standard inverse-gamma priors for variances of a normal random variable \citep{gelmanprior} are discussed in context of the LCM.

There is an added benefit of the CM distribution besides providing conjugacy in the non-Gaussian dependent data setting. Namely, LGPs are not necessarily realistic for \textit{every} dataset. For example, \citet{Oliveira} shows that there are parametric limitations to the LGP paradigm for count-valued data (e.g., when spatial overdispersion is small). We support this claim by showing that if certain hyperparameters (defined in Section~2) of the CM distribution are ``large'' then the corresponding CM distribution gives a very good approximation to a Gaussian distribution. This indicates that if the data suggests small values of these hyperparameters, then the CM distribution should be used in place of the multivariate normal distribution.

 Reduced rank methods are extremely prevalent in the more general ``dependent data'' setting. For example, reduced rank assumptions are crucial for principle component analysis, which has become an established technique in multivariate data analysis \citep[e.g., see][among others]{Jolliffe,Cox,everitt}. Additionally, reduced rank models have been used to great effect within spatial and spatio-temporal settings to obtain precise predictions in a computationally efficient manner \citep[see, e.g.,][]{wiklecress_spt,johan-2006, cressie-tao,banerjee, johan,finley,katzfuss_1, cressie-shi-kang-2010,kang-cressie-shi-2010,kang-cressie-2011,katzfuss2012,bradleyMSTM}. Thus, an additional motivating feature of the LCM model is that it can easily be cast within the reduced rank modeling framework to obtain further computational gains. {The ability to specify a reduced rank LCM does not imply that the LCM can handle \textit{all} types of ``big data'' problems. One type of big data problem that we do not consider is the ``big $p$'' problem \citep{htf,handbookbigdata}. Here, our focus is on difficulties with incorporating dependence when $n$ is large. Specifically, inverses of $n\times n$ matrices often manifest in dependent data settings \citep[e.g., see][among others]{reviewmethods}. Our incorporation of reduced rank modeling allows one to avoid order $n^{3}$ computations needed for matrix inversion, and allows one to avoid storage of large $n\times n$ matrices.}

This computationally efficient fully conjugate distribution theory could have an important impact on a number of different communities within and outside statistics. High-dimensional non-Gaussian data are pervasive in official statistics \citep[e.g., see][]{bradleyPMSTM}, ecology \citep[e.g., see][among others]{hooten03,wuJABES}, climatology \citep[e.g., see][]{wikleanderson}, atmospheric sciences \citep[e.g., see][]{aritrajsm}, statistical genetics \citep[e.g., see][and the references therein]{biggenetics}, neuroscience \citep[e.g., see][]{neuroscience,neuroscience2}, and many other domains. {The size of modern datasets is becoming more and more high-dimensional, and the aforementioned computational difficulties with LGPs suggest that there is a growing need to develop methods that are straightforward to implement \citep[e.g., see][for a discussion]{bradley2014_comp}.} Hence, the methodology presented here offers an exciting avenue that makes new applied research for modeling dependent non-Gaussian data practical for modern big datasets.

The LCM model is a type of hierarchical generalized linear model (HGLM) from \citet{leenelder96}. However, the current HGLM literature specifies an LGP for the dependent data setting \citep{leenelderinbook,leenelder01}. Additionally, there are other alternatives to a Gibbs sampler with Metropolis-Hastings updates; in particular, integrated nested Laplace approximations (INLA) \citep{rue} and Hamiltonian MCMC have proven to be useful tools in the literature. These approaches can easily be applied to our new proposed distribution theory; however, the need to adapt INLA and Hamiltonian MCMC \citep{ne2} to the LCM is not immediately necessary since the full conditional distributions are straightforward to simulate from in this setting.

For Poisson counts there are a number of choices besides the LGP strategy available to incorporate dependence \citep[e.g., see][among others]{Lee,multgamma,loggama}. For example, \citet{wolpert}, introduced a spatial convolution of gamma random variables, and provide a data augmentation scheme for Gibbs sampling that produces spatial predictions. Similarly, \citet{poissonapproximate} have an approximate Bayesian method for Poisson counts with latent Gaussian random variables. The recently proposed multivariate log-gamma distribution of \citet{bradleyPMSTM} results in a special case of our modeling approach when the data model is Poisson, and the latent processes are distributed according to a type of CM distribution. {Additionally, in more specific settings (e.g., Pareto data spatio-temporal data), conjugate distribution theory has been developed \citep{pareto,hu2018bayesian}}. 

The remainder of this article is organized as follows. In Section~2, we introduce the conjugate multivariate distribution and provide the necessary  technical development for fully Bayesian inference of dependent data from the natural exponential family. Specifically, we define the CM distribution, give the specification of the LCM model, discuss important methodological properties, introduce additional hyperpriors, and derive the full conditional distributions for a Gibbs sampler. Then, in Section~3 we provide a simulated example and an in-depth simulation study to show the performance of the LCM model compared to LGPs. {Several illustrations from different subject matter areas are also presented in Section 3, which is done in an effort to demonstrate the wide-applicability of the LCM. Specifically, we provide an example analyzing an epidemiology dataset, a federal statistics dataset, and an environmental dataset.} Finally, Section~4 contains discussion. For convenience of exposition, proofs of the technical results, Matlab and R code, and instructions on implementation are given in the Supplemental Appendix.

\section{Distribution Theory for Dependent Data from the Natural Exponential Family} In this section, we propose methodology for Bayesian analysis of non-Gaussian dependent data from the natural exponential family. In Section~2.1, we review and develop the univariate distribution introduced in \citet{diaconis}. Then, in Section~2.2, this univariate distribution is used as the rudimentary quantity to develop the CM distribution. This new multivariate distribution theory is incorporated within a Bayesian hierarchical model (i.e., the aforementioned LCM model) in Section~2.3, and the corresponding methodological properties are discussed in Section 2.4. A collapsed Gibbs sampler is derived in Section~2.5, and additional properties associated with the Gibbs sampler are discussed in Section 2.6. Finally prior distributions on remaining parameters are discussed in Sections 2.7 and 2.8.

\subsection{The Diaconis and Ylvisaker Conjugate Distribution}Suppose $Z$ is distributed according to the natural exponential family \citep{diaconis,casellehman}, then
\begin{equation}
\label{EF}
{f(Z\vert Y)} =\mathrm{exp}\left\lbrace ZY - b\psi(Y) + c(Z)\right\rbrace; \hspace{4pt} Z \in \mathcal{Z}, Y \in \mathcal{Y},
\end{equation}
where $f$ will be used to denote a generic probability density function/probability mass function (pdf/pmf), $Z \in \mathcal{Z}$, $\mathcal{Z}$ is the support of $Z$, $\mathcal{Y}$ is the support of $Y$, $b$ is {possibly un}known, and both $\psi(\cdot)$ and $c(\cdot)$ are known real-valued functions. The function $b\psi(Y)$ is often called the log partition function \citep{casellehman}. It will be useful for us to discuss $\psi(Y)$ and not $b\psi(Y)$; hence, we refer to $\psi(Y)$ as the ``unit log partition function'' because it's coefficient is one and not $b$. Let $\mathrm{EF}(Y;\hspace{2pt}\psi)$ denote a shorthand for the pdf/pmf in (\ref{EF}). It follows from \citet{diaconis} that the conjugate prior distribution for $Y$ is given by,
\begin{equation}\label{univ_LG}
f(Y\vert \alpha, \kappa) = K(\alpha, \kappa)\mathrm{exp}\left\lbrace \alpha Y - \kappa \psi(Y)\right\rbrace; \hspace{4pt} Y \in \mathcal{Y}, \frac{\alpha}{\kappa} \in \mathcal{Z}, \kappa > 0,
\end{equation}
\noindent
where $K(\alpha, \kappa)$ is a normalizing constant. Let $\mathrm{DY}(\alpha,\kappa;\hspace{2pt}\psi)$ denote a shorthand for the pdf in (\ref{univ_LG}). Here ``DY'' stands for ``Diaconis-Ylvisaker,'' and we will refer to $Y$ as either a Diaconis-Ylvisaker random variable or a DY random variable. \citet{diaconis} proved that the pdf in (\ref{univ_LG}) is proper (i.e., yields a probability measure). We also call $\alpha$ and $\kappa$ ``DY parameters.''

Multiplying both sides of (\ref{univ_LG}) by $\mathrm{exp}(tY)$ and integrating, gives the moment generating function 
\begin{equation}\label{mgf}
E[\mathrm{exp}(tY)\vert \alpha, \kappa] = \frac{K(\alpha, \kappa)}{K(\alpha+t,\kappa)},
\end{equation}
\noindent
which exists provided that $(\alpha + t)/\kappa \in \mathcal{Z}$, $\kappa>0$, and the corresponding values of $K(\alpha+t,\kappa)$ and $K(\alpha,\kappa)$ are strictly positive and finite. This gives us that the mean and variance of $Y$ is
\begin{align}
\label{mean}
E(Y\vert \alpha, \kappa) &= K(\alpha,\kappa)K^{(1)}(\alpha, \kappa)\\
\label{varDY}
\mathrm{var}(Y\vert \alpha, \kappa) &=K(\alpha,\kappa) K^{(2)}(\alpha,\kappa) - K(\alpha,\kappa)^{2}K^{(1)}(\alpha, \kappa)^{2},
\end{align}
\noindent
assuming that the moment generating function exists at $t = 0$, where $K^{(1)}(\alpha, \kappa)\equiv \left[\frac{d}{dt}\frac{1}{K(\alpha+t,\kappa)}\right]_{t = 0}$ and $K^{(2)}(\alpha, \kappa)\equiv \left[\frac{d^{2}}{dt^{2}}\frac{1}{K(\alpha+t,\kappa)}\right]_{t = 0}$.

Finally, it is immediate from (\ref{EF}) and (\ref{univ_LG}) that 
\begin{align}
Y\vert Z, \alpha, \kappa  &\sim \mathrm{DY}\left( \alpha + Z, \kappa + b;\hspace{2pt} \psi\right).
\end{align}
\noindent
This conjugacy motivates the development of a multivariate version of the DY random variable to model dependent non-Gaussian data from the natural exponential family. Thus, in this section, we define a conjugate multivariate distribution and develop a distribution theory that we find useful for fully Bayesian analysis in the dependent non-Gaussian (natural exponential family) data setting.

\subsection{The Conjugate Multivariate (CM) Distribution} 
\begin{table}[htp]
\centering
{\renewcommand{\arraystretch}{3}%
\noindent\adjustbox{max width=\textwidth}{%
\begin{tabular}{|c|c|c|  }
\hline
	 \textbf{Unit Log Partition Function} (i.e., ${\psi}$) & \textbf{CM Distribution (i.e., $f(\by\vert \bm{\mu}, \textbf{V}, \bm{\alpha}, \bm{\kappa})$)}\\ \hline
	$\psi_{1}(Y) = \mathrm{log}\left(-\frac{1}{Y}\right)$ & 
$ \mathrm{det}(\textbf{V}^{-1})\left\lbrace\prod_{i = 1}^{n}\frac{\alpha_{i}^{\kappa_{i}+1}}{\Gamma(\kappa_{i}+1)}\right\rbrace\mathrm{exp}\left[\bm{\alpha}^{\prime}\textbf{s} - \bm{\kappa}^{\prime}\mathrm{log}\left(-\textbf{s}^{(-1)}\right)\right]I(-\textbf{s} \in \mathbb{R}_{n}^{+})$ \\ \hline
 $\psi_{2}(Y) = \mathrm{log}\left(1 + \mathrm{exp}(Y)\right)$ & 
$ \mathrm{det}(\textbf{V}^{-1})\left\lbrace\prod_{i = 1}^{n}\frac{\Gamma(\kappa_{i})}{\Gamma(\alpha_{i})\Gamma(\kappa_{i} - \alpha_{i})}\right\rbrace\mathrm{exp}\left[\bm{\alpha}^{\prime}\textbf{V}^{-1}(\textbf{Y} - \bm{\mu}) - \bm{\kappa}^{\prime}\mathrm{log}\left[\textbf{J}_{n,1} + \mathrm{exp}\left\lbrace\textbf{V}^{-1}(\textbf{Y} - \bm{\mu})\right]\right\rbrace\right]I(\by\in \mathbb{R}^{n})$
 	\\ \hline
 $\psi_{3}(Y) = \mathrm{exp}\left(Y\right)$ & 
$\mathrm{det}(\textbf{V}^{-1})\left\lbrace\prod_{i = 1}^{n}\frac{\kappa_{i}^{\alpha_{i}}}{\Gamma(\alpha_{i})}\right\rbrace\mathrm{exp}\left[\bm{\alpha}^{\prime}\textbf{V}^{-1}(\textbf{Y} - \bm{\mu}) - \bm{\kappa}^{\prime}\mathrm{exp}\left\lbrace\textbf{V}^{-1}(\textbf{Y} - \bm{\mu})\right\rbrace\right]I(\by\in \mathbb{R}^{n})$ \\ \hline
$\psi_{4}(Y) = Y^{2}$ & $ \mathrm{det}(\textbf{V}^{-1})\left\lbrace\prod_{i = 1}^{n}\left(\frac{\kappa_{i}}{\pi}\right)^{1/2}\right\rbrace\mathrm{exp}\left\lbrace -(\by - \bm{\mu} - \bm{\gamma})^{\prime}\textbf{V}^{-1}\bm{\Sigma}^{-1}\textbf{V}^{-1\prime}(\by - \bm{\mu} - \bm{\gamma})/2\right\rbrace I(\by\in \mathbb{R}^{n})$ \\ \hline
\end{tabular}}}
\caption{Special Cases: We list the form of the CM distribution by $\psi_{j}$ for $j = 1,\ldots, 4$. The first column has the unit log partition function $\psi_{j}$, and the second column has the form of the CM distribution with generic $\textbf{V}^{-1} \in \mathbb{R}^{n}\times \mathbb{R}^{n}$. Let $\textbf{J}_{m,g}$ denote a $m\times g$ matrix of ones, $\textbf{s} = (s_{1},\ldots, s_{n})^{\prime}\equiv \textbf{V}^{-1}(\by - \bm{\mu})$, $\bm{\gamma} = \left(\frac{\alpha_{1}}{2\kappa_{1}},\ldots, \frac{\alpha_{n}}{2\kappa_{n}}\right)^{\prime}$, $\textbf{s}^{(-1)} = (1/s_{1},\ldots, 1/s_{n})^{\prime}$, and $\bm{\Sigma} \equiv \mathrm{diag}\left(\frac{1}{2\kappa_{i}}: i = 1,\ldots, n\right)$.}
\end{table}

\citet{bradleyPMSTM} use a linear combination of independent log-gamma random variables to build their multivariate log-gamma distribution. In a similar manner we take linear combinations of DY random variables to generate a conjugate version of the DY distribution. Specifically, let the $n$-dimensional random vector $\textbf{w} = (w_{1},\ldots.,w_{n})^{\prime}$ consist of $n$ mutually independent DY random variables such that $w_{i}\sim\mathrm{DY}(\alpha_{i},\kappa_{i};\hspace{2pt}\psi)$ for $i = 1,\ldots,n$. Then, define $\by \equiv (Y_{1},\ldots, Y_{n})^{\prime}$ such that
\begin{equation}\label{linear_comb}
\textbf{Y} = \bm{\mu} + \textbf{V}\textbf{w},
\end{equation}
\noindent
where $\by \in \mathcal{M}^{n}$, the matrix $\textbf{V} \in \mathbb{R}^{n}\times \mathbb{R}^{n}$, and $\bm{\mu} \in\mathbb{R}^{n}$. The space $\mathcal{M}^{n}$ is not necessarily equal to $\mathcal{Y}^{n}\equiv\{\by =(Y_{1},\ldots, Y_{n})^{\prime}: Y_{i}\in \mathcal{Y}, i = 1,\ldots, n\}$; for example, if $\mathcal{Y}$ is strictly positive, we obtain a $\by$ that can have negative components since $\textbf{V} \in \mathbb{R}^{n}\times \mathbb{R}^{n}$. Call $\textbf{Y}$ in (\ref{linear_comb}) a \textit{conjugate multivariate} (CM) random vector. A special case of the CM random vector is the multivariate normal random vector. To see this, let $\alpha_{i} \equiv 0$, $\kappa_{i} \equiv 1/2$, and $\psi(Y) = Y^{2}$ for $Y \in \mathbb{R}$. Then, it follows that (\ref{linear_comb}) is a multivariate normal distribution with mean $\bm{\mu}$ and covariance matrix $\textbf{V}\textbf{V}^{\prime}$, since the elements of $\textbf{w}$ consist of i.i.d. standard normal random variables. Additionally, the aforementioned MLG distribution can be written as a CM distribution when $\alpha>0$, $\kappa>0$, and $\psi(Y) = \mathrm{exp}(Y)$.

To use the CM distribution in a Bayesian context, we require its pdf, which is formally stated below.\\

\noindent
\textit{Theorem 1: Let $\textbf{Y}=\bm{\mu} + \textbf{V}\textbf{w}$, where $\by = (Y_{1},\ldots, Y_{n})^{\prime}$, $\bm{\mu}\in \mathbb{R}^{n}$, the $n\times n$ real valued matrix $\textbf{V}$ is invertible, and the $n$-dimensional random vector $\textbf{w} = (w_{1},\ldots,w_{n})^{\prime}$ consists of $n$ mutually independent DY random variables such that $w_{i}\sim\mathrm{DY}(\alpha_{i},\kappa_{i};\hspace{2pt}\psi)$ for $i = 1,\ldots,n$. \begin{enumerate}[label=(\roman*)]
 \item Then $\textbf{Y}$ has the following pdf:
\begin{align}
\label{mlg_pdf}
\nonumber
& f(\textbf{Y}\vert \bm{\mu},\textbf{V},\bm{\alpha},\bm{\kappa}) =\\ &\mathrm{det}(\textbf{V}^{-1})\left\lbrace\prod_{i = 1}^{n}K(\kappa_{i},{\alpha_{i}})\right\rbrace\mathrm{exp}\left[\bm{\alpha}^{\prime}\textbf{V}^{-1}(\textbf{Y} - \bm{\mu}) - \bm{\kappa}^{\prime}\psi\left\lbrace\textbf{V}^{-1}(\textbf{Y} - \bm{\mu})\right\rbrace\right]I(\by\in \mathcal{M}^{n}),
\end{align}
\noindent
where $I(\cdot)$ is the indicator function, the $j$-th element of $\psi\left\lbrace\textbf{V}^{-1}(\textbf{Y} - \bm{\mu})\right\rbrace$ contains $\psi$ evaluated at the $j$-th element of the $n$-dimensional vector $\textbf{V}^{-1}(\textbf{Y} - \bm{\mu})$, ``det'' denotes the determinant function, $\bm{\alpha}\equiv (\alpha_{1},\ldots,\alpha_{n})^{\prime}$, and $\bm{\kappa} \equiv (\kappa_{1},\ldots,\kappa_{n})^{\prime}$.
\item The mean and variance of $\by$ is given by,
\begin{align}
\nonumber
& E(\by\vert \bm{\alpha},\bm{\kappa}) = \bm{\mu}+ \textbf{V}\textbf{k}(\bm{\alpha},\bm{\kappa})\\
\label{meancovar}
& \mathrm{cov}(\by\vert \bm{\alpha},\bm{\kappa}) = \textbf{V}\textbf{K}(\bm{\alpha},\bm{\kappa})\textbf{V}^{\prime},
\end{align}
\noindent
where, the $n$-dimensional real-valued vector 
\begin{equation*}
\textbf{k}(\bm{\alpha},\bm{\kappa}) = \left(K(\alpha_{1},\kappa_{1})K^{(1)}(\alpha_{1}, \kappa_{1}),\ldots, K(\alpha_{n},\kappa_{n})K^{(1)}(\alpha_{n}, \kappa_{n})\right)^{\prime}, 
\end{equation*}
\noindent
and the $n\times n$ diagonal matrix $\textbf{K}(\bm{\alpha},\bm{\kappa})\equiv \mathrm{diag}\left\lbrace K(\alpha_{i},\kappa_{i}) K^{(2)}(\alpha_{i},\kappa_{i}) - K(\alpha_{i},\kappa_{i})^{2}K^{(1)}(\alpha_{i}, \kappa_{i})^{2}\right\rbrace$.\\
\end{enumerate}
}

\noindent
The proof of Theorem~1($i$) can be found in the Supplemental Appendix. In general, we let $ \mathrm{CM}(\bm{\mu},\textbf{V},\bm{\alpha},\bm{\kappa};\hspace{2pt}\psi)$ denote the pdf in (\ref{mlg_pdf}). Theorem~1($ii$) follows immediately from Equations (\ref{mean}) and (\ref{varDY}), and thus, Theorem~1($ii$) is stated without proof. 

When comparing (\ref{EF}), (\ref{univ_LG}), and (\ref{mlg_pdf}) we see that the univariate natural exponential family, the DY pdf, and the CM pdf share a basic structure. Specifically, all three distributions have an exponential term and an ``exponential of $-\psi$ term.'' This pattern is the main reason why conjugacy exists between the distributions from the natural exponential family and the DY distribution, which we take advantage of in subsequent sections. Also, Proposition $1(ii)$, shows that if we restrict $\textbf{V}$ (or equivalently $\textbf{V}^{-1}$) to be a lower unit triangle matrix, then the expression of the covariance matrix of $\textbf{Y}$ in (\ref{meancovar}) is a type of LDL decomposition \citep{ravishank}. Hence, in subsequent sections we assume that $\textbf{V}$ is lower unit triangular.

Bayesian inference not only requires the pdf of $\by$, but also requires simulating from conditional distributions of $\textbf{Y}$.\\

\noindent
\textit{Theorem 2: Let $\textbf{Y} \sim \mathrm{CM}(\bm{\mu},\textbf{V},\bm{\alpha},\bm{\kappa};\hspace{2pt}\psi)$, and let $\textbf{Y} = (Y_{1},\ldots,Y_{n})^{\prime} = (\textbf{Y}_{1}^{\prime}, \textbf{Y}_{2}^{\prime})^{\prime}$, so that $\textbf{Y}_{1}$ is $r$-dimensional and $\textbf{Y}_{2}$ is $(n-r)$-dimensional. In a similar manner, partition $\textbf{V}^{-1} = [\textbf{H}\hspace{5pt}\textbf{B}]$ into an $n\times r$ matrix $\textbf{H}$ and an $n\times (n-r)$ matrix $\textbf{B}$. Also let $\bm{\mu}^{*} = \textbf{V}^{-1}\bm{\mu} - \textbf{B}\textbf{d}$ for $\textbf{d} \in \mathbb{R}^{n-r}$. Then, the conditional distribution $\textbf{Y}_{1}\vert \textbf{Y}_{2} = \textbf{d},\bm{\mu}^{*},\textbf{H},\bm{\alpha},\bm{\kappa}$ is given by 
	\begin{align}
	\label{cond_pdf2}
	f(\textbf{Y}_{1}\vert \textbf{Y}_{2} = \textbf{d},\bm{\mu}^{*},\textbf{H},\bm{\alpha},\bm{\kappa})&= M \hspace{5pt}\mathrm{exp}\left\lbrace\bm{\alpha}^{\prime}\textbf{H}\textbf{Y}_{1}- \bm{\alpha}^{\prime}\bm{\mu}^{*}- \bm{\kappa}^{\prime}\psi(\textbf{H}\textbf{Y}_{1} - \bm{\mu}^{*})\right\rbrace I\{(\textbf{Y}_{1}^{\prime},\textbf{d}^{\prime})^{\prime} \in \mathcal{M}^{n}\},
	\end{align}
	\noindent
	where $M$ is a strictly positive and finite normalizing constant. Let $\mathrm{CM_{c}}(\bm{\mu}^{*},\textbf{H},\bm{\alpha},\bm{\kappa};\hspace{2pt}\psi)$ be a shorthand for the pdf in (\ref{cond_pdf2}), where the subscript ``c'' represents the word ``conditional.''\\
}

\noindent
In Supplemental Appendix A we describe technical results on simulating from the conditional CM distribution.

In this article, we consider CM distributions that are implied by the unit log partition function of the data model including: the gamma data model, binomial data model, negative binomial data model, the Poisson data model, and the normal data model (see Tables 1 and 2). In the univariate case, each of these special cases lead to well-known hierarchical models (i.e., gamma/inverse-gamma, (negative) binomial/beta, Poisson/log-gamma, and normal/normal models) \citep{diaconis}. To delineate from the univariate setting, we shall refer to $\mathrm{CM}(\bm{\mu},\textbf{V},\bm{\alpha},\bm{\kappa}; \psi_{j})$ for $j = 1,\ldots, 4$ (see Table 1 for the definitions of $\psi_{1},\psi_{2},$ $\psi_{3},$ and $\psi_{4}$) as the multivariate negative-inverse-gamma distribution, multivariate logit-beta distribution, the multivariate log-gamma, and the multivariate normal distribution, respectively. 
 
 These choices of the CM distribution are themselves general. For example, when $\bm{\alpha} = \textbf{J}_{n,1}$, we obtain an exponential/multivariate negative-inverse-gamma model. Similarly, the binomial/multivariate logit-beta model has a Bernoulli/multivariate logit-beta model as a special case, which occurs when the number of Bernoulli trials that define the binomial distribution is equal to one. Likewise, when the number of successful Bernoulli trials is equal to one, the negative binomial/multivariate logit-beta model reduces to a geometric/multivariate logit-beta specification. This creates opportunity for analyzing many different types of dependent data.

\subsection{The LCM Model} The LCM model is proportional to the product of the following conditional and marginal distributions:
\begin{align}\label{summary}
&\mathrm{Data\hspace{5pt}Model:}\hspace{5pt} Z_{i}\vert \bfbeta,\bm{\eta}, \xi_{i} \ind \mathrm{EF}\left(\textbf{x}_{i}^{\prime}\bm{\beta} + \bm{\phi}_{i}^{\prime}\bm{\eta} + \xi_{i};\hspace{2pt} \psi_{j} \right);\hspace{15pt} i = 1,\ldots,n, j = 1,\ldots, 4\\
\nonumber
&\mathrm{Process\hspace{5pt}Model\hspace{5pt}1:}\hspace{5pt} \bm{\eta}\vert \textbf{V}, \alpha_{\eta},\kappa_{\eta}\sim \mathrm{CM}\left(\bm{0}_{r,1}, \textbf{V},\bm{\alpha}_{\eta},\bm{\kappa}_{\eta};\hspace{2pt} \psi_{k}\right); \hspace{15pt}\\
\nonumber
&\mathrm{Process\hspace{5pt}Model\hspace{5pt}2:}\hspace{5pt} \bm{\xi}\vert \bm{\alpha}_{\xi}, \bm{\kappa}_{\xi} \sim \mathrm{CM}\left(\bm{0}_{n,1}, \textbf{V}_{\xi},\bm{\alpha}_{\xi},\bm{\kappa}_{\xi};\hspace{2pt} \psi_{k}\right);\\
 	\nonumber
 	&{\mathrm{Parameter\hspace{5pt}Model\hspace{5pt}1:}\hspace{5pt} b\vert \alpha_{b}, \kappa_{b}\sim \mathrm{{CM}}\left(0,1,{\alpha}_{b},{\kappa}_{b};\hspace{2pt} \psi_{k}\right)I(b>0)}\\
 	\nonumber
&\mathrm{Parameter\hspace{5pt}Model\hspace{5pt}2:}\hspace{5pt} \bm{\beta}\vert \alpha_{\beta}, \kappa_{\beta}\sim \mathrm{CM}\left(\bm{0}_{p,1}, \textbf{V}_{\beta},\bm{\alpha}_{\beta},\bm{\kappa}_{\beta};\hspace{2pt} \psi_{k}\right); \hspace{2pt}k = 1,\ldots, 4,
\end{align}
\noindent
where $\psi_{j}$ and $\psi_{k}$ (for $j,k = 1,\ldots, 4$) are defined in Table 1 and the elements of $n$-dimensional vector $\bz \equiv \left(Z_{1},\ldots, Z_{n}\right)^{\prime}$ represent data that can be reasonably modeled using a member from the natural exponential family. Additionally for each $i$, $\textbf{x}_{i}$ is a known $p$-dimensional vector of covariates, $\bm{\beta} = (\beta_{1},\ldots, \beta_{p})^{\prime}\in \mathbb{R}^{p}$ is an unknown vector interpreted as fixed effects, $\bm{\phi}_{i}$ is a known $r$-dimensional real-valued vector (see Section~3.5 for an example), and the $r$-dimensional vector $\bm{\eta} = (\eta_{1},\ldots, \eta_{r})^{\prime}$ and $n$-dimensional vector $\bm{\xi} \equiv \left(\xi_{1},\ldots, \xi_{n}\right)^{\prime}$ are interpreted as real-valued random effects. We have not yet provided specifications of the hyperparameters and variance parameters: $\bm{\alpha}_{\beta} = (\alpha_{\beta,1},\ldots, \alpha_{\beta,p})^{\prime}$, $\bm{\alpha}_{\eta} = (\alpha_{\eta,1},\ldots, \alpha_{\eta,r})^{\prime}$, $\bm{\alpha}_{\xi}= (\alpha_{\xi,1},\ldots, \alpha_{\xi,n})^{\prime}$, $\bm{\kappa}_{\beta}= (\kappa_{\beta,1},\ldots, \kappa_{\beta,p})^{\prime}$, $\bm{\kappa}_{\eta}= (\kappa_{\eta,1},\ldots, \kappa_{\eta,r})^{\prime}$, $\bm{\kappa}_{\xi}=(\kappa_{\xi,1},\ldots, \kappa_{\xi,n})^{\prime}$, $\textbf{V}_{\beta}\in \mathbb{R}^{p}\times \mathbb{R}^{p}$, $\textbf{V}\in \mathbb{R}^{r}\times \mathbb{R}^{r}$, and $\textbf{V}_{\xi}\in \mathbb{R}^{n}\times \mathbb{R}^{n}$, where $\alpha_{\beta,i}/\kappa_{\beta,i}\in \mathcal{Y}$, $\alpha_{\eta,j}/\kappa_{\eta,j}\in \mathcal{Y}$, $\alpha_{\xi,k}/\kappa_{\xi,k} \in \mathcal{Y}$, $\kappa_{\beta,i}>0$, $\kappa_{\eta,j}>0$, and $\kappa_{\xi,k}>0$; $i = 1,\ldots, p$, $j = 1,\ldots, r$, $k = 1,\ldots, n$. These details are presented in Sections~2.7 and 2.8.
 
 {Parameter Model 1 in (\ref{summary}) is only included when $b$ is unknown (i.e., when the data model is specified to be either the negative binomial or gamma distributions). The truncated CM distribution is chosen because it is conjugate; see details in the Supplemental Appendix. In our experience (see Section 3.4), $b$ is difficult to learn, and the results are extremely sensitive to the choice of $\alpha_{b}$ and $\kappa_{b}$. Several priors have been suggested for the overdispersion parameter when the data are distributed according to a negative binomial distribution \citep[e.g.,][among others]{gelmanprior}, some of which have been developed based on the gamma-Poisson interpretation of the negative binomial distribution\citep[see][and the references therein]{zhou2015negative}. In this article, we focus on using CM priors, and hence, other choices of priors on $b$ (for LCM models) may lead to better results. We have found that the results are more favorable when specifying a different data model for the settings where $b$ is unknown. Specifically, for the negative binomial setting we suggest using a Poisson distribution, and when the data is distributed as gamma we suggest taking the log transform and using a normal distribution.}
 
 
 {Another important quantity that needs to be specified are the basis functions $\left\lbrace\bm{\phi}_{i}\right\rbrace$. This choice is very important and requires careful consideration. To illustrate the generality of our proposed model we consider three classes of basis function, each of which are demonstrated in Sections 3.3, 3.4, and 3.5, respectively. Many analyses let $\left\lbrace\bm{\phi}_{i}\right\rbrace$ consist of known covariates \citep[e.g., see][for a recent example]{reich2}. Another choice is to specify latent classes to model within-subject variability; in this setting, $\{\bm{\phi}_{i}\}$ is sometimes referred to as a ``random effect design matrix'' \citep[e.g., see][Chp. 1 for a discussion]{RPLM}. Consider the example where $g_{k}\subset\{1,\ldots, n\}$ represents the $k$-th group. In Section 3.3, $g_{k}$ represents the $k$-th herd of cows, and each element in $g_{k}$ represents a specific cow in the sample. Here, we shall specify $\bm{\phi}_{i} = \left(I(i \in g_{1}),\ldots, I(i \in g_{r})\right)^{\prime}$. For spatial and time-series datasets, it is often assumed that $\{\bm{\phi}_{i}\}$ consists of spatial/temporally varying functions, referred to as ``basis functions.'' For example, Fourier basis functions/wavelets are often used in the image analysis literature \citep[e.g., see][for a classic reference]{donahoe}. Similarly, radial basis functions, empirical orthogonal functions, and splines have been used to great effect in the spatial statistics, time-series, and spatio-temporal statistics literature \citep[e.g., see][for a different choices of basis functions]{wahba, bradley2014_comp,wikleHandbook, bradleyCAGE}. }

 {The value of $r$ is a feature of the observed dataset when specifying $\left\lbrace\bm{\phi}_{i}\right\rbrace$ to be either covariates or a random effects design matrix (see Sections 3.3 and 3.4 for examples). However, when using a known class of basis functions, $r$ must be specified. In this setting, selection criteria are often used to investigate both the sensitivity to the choice of $r$ and how many are necessary to give reasonable predictions \citep[e.g., see][among others]{wahba,selectr,bradley2011}. Spike and slab, horseshoe priors, and SSVS (among other similar techniques) are extensions of the LGP, which one might adapt to the LCM to select covariates and basis functions \citep{selectionReview}; however, we do not consider these extensions of the LCM in this article. When spatial basis functions depend on knot locations (thin-plate splines), a common rule-of-thumb is to specify equally spaced knots over the spatial domain \citep[e.g., see][among others]{nychka}. In Section 3.5, we demonstrate the use of a known kernel using a big Bernoulli dataset consisting of cloud fractions. Here, we use same the basis functions specified in \citep{aritrajsm}, where the knots were chosen to be equally spaced.} 
 
 \subsection{Methodological Properties of the LCM}
 An important point argued in Section~1 is that the LGP model is a \textit{special case} of an LCM. This can now easily be seen by letting $j = 1,\ldots, 4$, $k = 4$, $\bm{\alpha}_{\beta} = \bm{0}_{p,1}$, $\bm{\alpha}_{\eta} = \bm{0}_{r,1}$, and $\bm{\alpha}_{n} = \bm{0}_{n,1}$. This specification yields an LGP model. A difficulty with this specification is that we lose conjugacy by specifying $j \ne k$. \citet{bradleyPMSTM} showed that the multivariate log-gamma distribution they proposed can be made arbitrarily close to a multivariate normal distribution by specifying the shape and scale parameters to be large. This essentially allows one to use a LGP specification with a Poisson data model, and \textit{also use} the conjugacy that arises from the MLG distribution when $j = k = 3$ in (\ref{summary}). This important property of the MLG distribution can be extended to the more general CM distribution.\\
 
 \noindent
 \textit{Theorem 3: Suppose that $\psi \ne \psi_{4}$, and denote the first and second derivatives with $\psi^{\prime}$ and $\psi^{\prime\prime}$, $0<\psi^{\prime}<\infty$, and $0<\psi^{\prime\prime}<\infty$. Let the $n$-dimensional random vector $\textbf{Y}$  distributed according $\mathrm{CM}(\bm{\mu},\left(\psi^{''}(0)/\psi^{'}(0)\right)^{1/2}\alpha^{1/2}\textbf{V},$ $\alpha\textbf{J}_{n,1},\frac{\alpha}{\psi^{\prime}(0)}\textbf{J}_{n,1};\hspace{2pt}\psi)$ ignoring proportionality constants. Then $\textbf{Y}$ converges in distribution to a multivariate normal random vector with mean $\bm{\mu}$ and covariance matrix $\textbf{V}\textbf{V}^{\prime}$ as $\alpha$ approaches infinity.\\
 }
 
 \noindent
 The restriction of $\psi \ne \psi_{4}$ is sensible, since $\psi = \psi_{4}$ yields a CM exactly equal to a multivariate normal distribution. Also, Theorem 3 does not hold for the multivariate negative-inverse-gamma distribution, since $\psi_{1}^{\prime}(0) = -\infty$. 
 
 The ``best'' DY parameters, for the multivariate logit-beta distribution and the MLG distribution, might not lead to something that looks Gaussian. That is, we should be able to learn whether or not the multivariate normal distribution is appropriate for latent processes of binomial and Poisson data by observing whether or not posterior replicates of the DY parameters (i.e., $\alpha$ and $\kappa$) are large (which would invoke Theorem 3). Hence, from this point-of-view, it is very important that we place prior distributions on the DY parameters, as we describe in Section 2.8.
 
 These connections to the Gaussian distribution are important because it shows potential for the LCM to outperform a latent Gaussian process model. However, for the LCM to be as widely applicable as an LGP, we also require an important theoretical property referred to as Kolmogorov consistency \citep{Daniell,Kolmogorov}. That is, if the index on ${Z_{i}}$ is defined over space or time, for example, then we need the CM distribution to be well defined for every possible subset of locations \citep{gs}.\\
 
 \noindent
 \textit{Theorem 4: The CM distribution, as defined in Theorem~1, is Kolmogorov consistent.}\\
 
 \noindent
 Theorems 3 and 4 are important methodological properties; however, if it is more difficult to implement LCM over the LGP, then these results may have less of an impact in practice. In Section 2.5, we show that it is rather straightforward to implement the LCM using a collapsed Gibbs sampler.

\subsection{An Example Gibbs Sampler for the LCM} To simulate from a posterior distribution that is proportional to (\ref{summary}) we consider the following likelihood:
\begin{align}\label{augmented}
&\mathrm{Data\hspace{5pt}Model:}\hspace{5pt} Z_{i}\vert \bfbeta,\bm{\eta}, \xi_{i} \ind \mathrm{EF}\left(\textbf{x}_{i}^{\prime}\bm{\beta} + \bm{\phi}_{i}^{\prime}\bm{\eta} + \xi_{i} + \textbf{b}_{\beta,i}^{\prime}\textbf{q}_{\beta}+ \textbf{b}_{\eta,i}^{\prime}\textbf{q}_{\eta}+ \textbf{b}_{\xi,i}^{\prime}\textbf{q}_{\xi};\hspace{2pt} \psi_{j} \right)\\
\nonumber
&\mathrm{Process\hspace{5pt}Model\hspace{5pt}1:}\hspace{5pt} \bm{\eta}\vert \textbf{V}, \alpha_{\eta},\kappa_{\eta},\textbf{q}_{\eta}\sim \mathrm{CM}\left(-\textbf{V}\textbf{B}_{\eta}\textbf{q}_{\eta}, \textbf{V},\bm{\alpha}_{\eta},\bm{\kappa}_{\eta};\hspace{2pt} \psi_{k}\right); \hspace{15pt}\\
\nonumber
&\mathrm{Process\hspace{5pt}Model\hspace{5pt}2:}\hspace{5pt} \bm{\xi}\vert \bm{\alpha}_{\xi}, \bm{\kappa}_{\xi},\textbf{q}_{\xi} \sim \mathrm{CM}\left(-\textbf{V}_{\xi}\textbf{B}_{\xi}\textbf{q}_{\xi}, \textbf{V}_{\xi},\bm{\alpha}_{\xi},\bm{\kappa}_{\xi};\hspace{2pt} \psi_{k}\right);\\
\nonumber
&{\mathrm{Parameter\hspace{5pt}Model\hspace{5pt}1:}\hspace{5pt} b\vert \alpha_{b}, \kappa_{b}\sim \mathrm{{CM}}\left(0,1,{\alpha}_{b},{\kappa}_{b};\hspace{2pt} \psi_{k}\right)I(b>0)}\\
\nonumber
&\mathrm{Parameter\hspace{5pt}Model\hspace{5pt}2:}\hspace{5pt} \bm{\beta}\vert \alpha_{\beta}, \kappa_{\beta},\textbf{q}_{\beta}\sim \mathrm{CM}\left(-\textbf{V}_{\beta}\textbf{B}_{\beta}\textbf{q}_{\beta}, \textbf{V}_{\beta},\bm{\alpha}_{\beta},\bm{\kappa}_{\beta};\hspace{2pt} \psi_{k}\right)\\
\nonumber
&\mathrm{Parameter\hspace{5pt}Model\hspace{5pt}3:}\hspace{5pt} f(\textbf{q}_{\beta}) = 1\\
\nonumber
&\mathrm{Parameter\hspace{5pt}Model\hspace{5pt}3:}\hspace{5pt} f(\textbf{q}_{\eta}) = 1\\
\nonumber
&\mathrm{Parameter\hspace{5pt}Model\hspace{5pt}3:}\hspace{5pt} f(\textbf{q}_{\xi}) = 1;\hspace{15pt} i = 1,\ldots,n, j = 1,\ldots, 4, k = 1,\ldots, 4,
\end{align}
\noindent
where $\textbf{b}_{\beta,i}$, $\textbf{b}_{\eta,i}$, and $\textbf{b}_{\xi,i}$ are prespecified $n$-dimensional vectors and the $p\times n$ matrix $\textbf{B}_{\beta}$, the $r\times n$ matrix $\textbf{B}_{\eta}$, and the $n\times n$ matrix $\textbf{B}_{\xi}$ are also prespecified. There is an immediate connection between (\ref{summary}) and (\ref{augmented}), which introduces the improper $n$-dimensional random vector $\textbf{q}_{\beta}$, $n$-dimensional random vector $\textbf{q}_{\eta}$, and  $n$-dimensional random vector $\textbf{q}_{\xi}$. Specifically, when conditioning (\ref{augmented}) on the events $\textbf{q}_{\beta} = \bm{0}_{n,1}$, $\textbf{q}_{\eta}= \bm{0}_{n,1}$, and $\textbf{q}_{\xi} = \bm{0}_{n,1}$, we obtain a likelihood that is proportional to (\ref{summary}). Consequently, we suggest implementing the collapsed Gibbs sampler \citep{liu1994collapsed} outlined in the Pseudo-Code. In general, one can interpret $\textbf{q}_{\beta}$, $\textbf{q}_{\eta}$, and $\textbf{q}_{\xi}$ as location parameters for $\bm{\beta}$, $\bm{\eta}$, and $\bm{\xi}$, and are given non-informative priors.

As an example, consider deriving the full-conditional distribution in Step 2. Write the data model in (\ref{augmented}) as
\begin{align}\label{databeta}
f(\bz,\textbf{q}_{\beta}\vert \cdot) &\underset{\bz}{\propto} \mathrm{exp}\left\lbrace \bz^{\prime}\textbf{X}\bm{\beta} +\bz^{\prime}\bm{\Phi}\bm{\eta}   + \bz^{\prime}\bm{\xi}+\bz^{\prime}\textbf{B}_{\beta,1}\textbf{q}_{\beta}-b\textbf{J}_{n,1}^{\prime}\psi(\textbf{X}\bm{\beta}+\bm{\Phi}\bm{\eta}+\bm{\xi}+\textbf{B}_{\beta,1}\textbf{q}_{\beta})\right\rbrace h,
\end{align}
\noindent
where the $n\times n$ matrix $\textbf{B}_{\beta,1} = \left(\textbf{b}_{\beta,1},\ldots, \textbf{b}_{\beta,n}\right)^{\prime}$, $n\times p$ matrix $\textbf{X} \equiv \left(\textbf{x}_{1},\ldots, \textbf{x}_{n}\right)^{\prime}$, the $n\times r$ matrix $\bm{\Phi}\equiv \left(\bm{\phi}_{1},\ldots, \bm{\phi}_{n}\right)^{\prime}$, $h = \left\lbrace \prod_{i = 1}^{n}I(\textbf{x}_{i}^{\prime}\bm{\beta} + \bm{\phi}_{i}^{\prime}\bm{\eta}+\xi_{i} \in \mathcal{Y})\right\rbrace$, and $\underset{\bz}{\propto}$ denotes the ``proportional to as a function of $\bz$'' symbol. Using (\ref{augmented}) and Parameter Model 2 in (\ref{databeta}) we have that
\begin{align}\label{conjugate}
f(\bm{\beta},\textbf{q}_{\beta}\vert \cdot) &\underset{\bm{\beta}}{\propto}  \mathrm{CM}\left( \bm{\mu}_{\beta}^{*}, \textbf{V}_{\beta}^{*}, {\bm{\alpha}_{\beta}^{*}, \bm{\kappa}_{\beta}^{*}};\hspace{2pt}\psi \right) h,
\end{align}
where $\bm{\mu}_{\beta}^{*} = \textbf{V}_{\beta}^{*}\left(-\bm{\eta}^{\prime}\bm{\Phi}^{\prime} - \bm{\xi}^{\prime},\bm{0}_{p}^{\prime}\right)^{\prime}$, $\bm{\alpha}_{\beta}^{*} = (\textbf{Z}^{\prime},\bm{\alpha}_{\beta}^{\prime})^{\prime}$, $\bm{\kappa}_{\beta}^{*} = (b\textbf{J}_{n,1},\bm{\kappa}_{\beta}^{\prime})^{\prime}$, $\textbf{V}_{\beta}^{*-1} = (\textbf{H}_{\beta},\textbf{Q}_{\beta})$, the $(n+p)\times p$ matrix $\textbf{H}_{\beta} = (\textbf{X}^{\prime},\textbf{V}_{\beta}^{\prime})^{\prime}$, and the the $(n+p)\times n$ matrix $\textbf{Q}_{\beta} = (\textbf{B}_{\beta,1}^{\prime},\textbf{B}_{\beta}^{\prime})^{\prime}$. See the Supplemental Appendix for the algebra leading to (\ref{conjugate}). The full-conditional distribution in (\ref{conjugate}) is not well defined when $Z_{i} = 0$ for some $i$, because this produces a zero shape parameter. In this setting one can add an ``$\epsilon$'' to the elements of $\bz$ to force non-zero shape parameters. However, this choice changes the prior from a CM distribution to a $\mathrm{CM}_{c}$ distribution, and a considerable amount of book-keeping is required to derive the full-conditional distributions. For ease of exposition, we put these details in the Supplemental Appendix C.

If we prespecify $\textbf{Q}_{\beta}$ so that it is equal to the basis for the null space of $\textbf{H}_{\beta}$ (i.e., $\textbf{Q}_{\beta}^{\prime}\textbf{Q}_{\beta} = \textbf{I}_{n}$, $\textbf{Q}_{\beta}^{\prime}\textbf{H}_{\beta} = \bm{0}_{n,p}$, and $\textbf{H}_{\beta}(\textbf{H}_{\beta}^{\prime}\textbf{H}_{\beta})^{-1}\textbf{H}_{\beta}^{\prime} + \textbf{Q}_{\beta}\textbf{Q}_{\beta}^{\prime} = \textbf{I}_{n+p}$). Then,
\begin{equation}
\textbf{V}_{\beta}^{*} = (\textbf{H}_{\beta},\textbf{Q}_{\beta})^{-1} = \left(\begin{array}{c}
(\textbf{H}_{\beta}^{\prime}\textbf{H}_{\beta})^{-1}\textbf{H}_{\beta}^{\prime}\\
\textbf{Q}_{\beta}^{\prime}
\end{array}\right).
\end{equation}
From (\ref{linear_comb}) we see that to sample a value from $f(\bm{\beta},\textbf{q}_{\beta}\vert \bz, \bm{\eta},\bm{\xi}, b,\textbf{q}_{\eta} = \bm{0}_{n},\textbf{q}_{\xi} =\bm{0}_{n})$ we can compute
\begin{equation}
\left(\begin{array}{c}
\bm{\beta}\\
\bm{q}_{\beta}
\end{array}\right) = -\left(\begin{array}{c}
(\textbf{H}_{\beta}^{\prime}\textbf{H}_{\beta})^{-1}\textbf{H}_{\beta}^{\prime}(\bm{\Phi}\bm{\eta}+\bm{\xi})\\
\bm{0}_{n}
\end{array}\right)+ \left(\begin{array}{c}
(\textbf{H}_{\beta}^{\prime}\textbf{H}_{\beta})^{-1}\textbf{H}_{\beta}^{\prime}\textbf{w}\\
\textbf{Q}_{\beta}^{\prime}\textbf{w}
\end{array}\right),
\end{equation}
\noindent
where $\textbf{w}\sim\mathrm{CM}\left(\bm{0}_{n+p}, \textbf{I}_{n+p}, {\bm{\alpha}_{\beta}^{*}, \bm{\kappa}_{\beta}^{*}};\hspace{2pt}\psi \right)$, which can easily be generated using (\ref{linear_comb}). Thus, to simulate according to Step 2 of the collapsed Gibbs sampler we can compute,
\begin{equation}
\bm{\beta} = -(\textbf{H}_{\beta}^{\prime}\textbf{H}_{\beta})^{-1}\textbf{H}_{\beta}^{\prime}(\bm{\Phi}\bm{\eta}+\bm{\xi})+ (\textbf{H}_{\beta}^{\prime}\textbf{H}_{\beta})^{-1}\textbf{H}_{\beta}^{\prime}\textbf{w}.
\end{equation}
\begin{algorithm}[t]\caption{Pseudo-Code: Collapsed Gibbs sampler for the model in (\ref{augmented})}\label{euclid2}
	\begin{algorithmic}[1]
		\item Set $b = 1$ and initialize $\bm{\beta}^{[0]}$, $\bm{\eta}^{[0]}$, and $\bm{\xi}^{[0]}$.
		\item Sample $\bm{\beta}^{[g]}$ from $f(\bm{\beta}\vert \bz, \bm{\eta}^{[g-1]},\bm{\xi}^{[g-1]}, b^{[g-1]},\textbf{q}_{\eta} = \bm{0}_{n},\textbf{q}_{\xi} =\bm{0}_{n})$.
		\item Sample $\bm{\eta}^{[g]}$ from $f(\bm{\eta}\vert \bz, \bm{\beta}^{[g]},\bm{\xi}^{[g-1]}, b^{[g-1]},\textbf{q}_{\beta} = \bm{0}_{n},\textbf{q}_{\xi} = \bm{0}_{n})$.
		\item Sample $\bm{\xi}^{[g]}$ from $f(\bm{\xi}\vert \bz, \bm{\beta}^{[g]},\bm{\eta}^{[g]}, b^{[g-1]},\textbf{q}_{\beta} = \bm{0}_{n},\textbf{q}_{\eta} = \bm{0}_{n})$.
		\item Sample $b^{[g]}$ from $f(b\vert \bz, \bm{\beta}^{[g]},\bm{\eta}^{[g]}, 
		\bm{\xi}^{[g]},\textbf{q}_{\beta} = \bm{0}_{n},\textbf{q}_{\eta} = \bm{0}_{n},\textbf{q}_{\xi} = \bm{0}_{n})$.
		\item Repeat Steps 2, 3, and 4 until $g = G$ for a prespecified value of $G$.
	\end{algorithmic}
\end{algorithm}
\noindent
It is (computationally) easy to simulate in this manner provided that $p \ll n$. Recall that $\textbf{H}_{\beta}$ is $n\times p$, which implies that computing the $p\times p$ matrix $(\textbf{H}_{\beta}^{\prime}\textbf{H}_{\beta})^{-1}$ is computationally feasible when $p$ is ``small.'' {By small we mean a value such that the Gauss-Jordan elimination method for the inverse of a $p\times p$ matrix can be computed in real-time.} Furthermore, the $p$-dimensional random vector $\bm{\beta}$ is an orthogonal projection of the $n$-dimensional random vector $\textbf{w}$ onto the column space spanned by the columns of $\textbf{H}_{\beta}$. This provides a geometric interpretation of random vectors generated according to (\ref{augmented}).

\subsection{Properties of the Augmented LCM Model}As discussed in Section 2.2 and Supplementary Appendix A, it is difficult to simulate directly from a $\mathrm{CM_{c}}$ distribution since $\textbf{H}$ in (\ref{cond_pdf2}) is not square, and hence, one can not use Equation (\ref{linear_comb}). In Section 2.5, we instead consider simulating from $\mathrm{CM_{c}}$ after marginalizing across a location parameter with improper prior. This leads to the following result.\\


\noindent
\textit{Theorem 5: Let $\textbf{q}_{1}\vert \textbf{c}, \textbf{H}, \bm{\alpha},\bm{\kappa} \sim \mathrm{CM}_{c}(\textbf{c},\textbf{H},\bm{\alpha},\bm{\kappa})$, where $\textbf{H} \in \mathbb{R}^{M}\times \mathbb{R}^{r}$ is full column rank, $\bm{\alpha} = (\alpha_{1},\ldots, \alpha_{M})^{\prime}$, $\bm{\kappa} = (\kappa_{1},\ldots, \kappa_{M})^{\prime}$, $\alpha_{i}/\kappa_{i}\in \mathcal{Y}$, and $\kappa_{i}>0$ for $i = 1,\ldots, M$. Assume a re-parameterized value of $\textbf{c} = -\textbf{B}\textbf{q}_{2}+\bm{\mu}$, and the improper prior $f(\textbf{q}_{2}\vert \textbf{c}, \textbf{H},\textbf{B}, \bm{\alpha},\bm{\kappa})\propto 1$, where $\textbf{q}_{2}$ is $(M-r)$-dimensional. Also let $\textbf{B}, \in \mathbb{R}^{M}\times \mathbb{R}^{M-r}$ be the orthonormal basis for the null space of $\textbf{H}$, $\textbf{q} = (\textbf{q}_{1}^{\prime},\textbf{q}_{2}^{\prime})^{\prime}$, $\bm{\mu}\in \mathbb{R}^{M}$, $\textbf{I}_{n}$ be an $n\times n$ identity matrix, and let $\textbf{w}\sim \mathrm{CM}_{c}(\bm{\mu},\textbf{H},\bm{\alpha},\bm{\kappa})$. Define $\textbf{V}^{-1} = (\textbf{H},\textbf{B})$. }
	\begin{enumerate}[label=(\roman*)]
		\item \textit{Then,}
		 \begin{align}
		\label{cond_pdf4}
		\int f(\textbf{q}_{1}\vert \textbf{c} = -\textbf{B}\textbf{q}_{2}+\bm{\mu}, \textbf{H},\textbf{B}, &\bm{\alpha},\bm{\kappa})d\textbf{q}_{2}\propto \int \mathrm{exp}\left\lbrace\bm{\alpha}^{\prime}\textbf{V}^{-1}\textbf{q}-\bm{\kappa}^{\prime}\psi\left(\textbf{V}^{-1}\textbf{q}-\bm{\mu}\right)\right\rbrace d\textbf{q}_{2},
		\end{align}
		\textit{where $\psi$ is a unit log-partition function and the integrand on the right hand side of (\ref{cond_pdf4}) is proportional to $\mathrm{CM}(\textbf{V}\bm{\mu},\textbf{V}=(\textbf{H},\textbf{B})^{-1},\bm{\alpha},\bm{\kappa})$. Furthermore, the affine transformation $(\textbf{H}^{\prime}\textbf{H})^{-1}\textbf{H}^{\prime}\textbf{w}$ is a draw from the density in (\ref{cond_pdf4}).}
		\item \textit{The conditional mean and covariance can be computed as}
		\begin{align*}
		E(\textbf{Y}_{1}\vert \textbf{Y}_{2} = \textbf{d},\textbf{V},\bm{\alpha},\bm{\kappa})&=(\textbf{H}^{\prime}\textbf{H})^{-1}\textbf{H}^{\prime}\textbf{k}(\bm{\alpha},\bm{\kappa})\\ 
		\mathrm{cov}(\textbf{Y}_{1}\vert \textbf{Y}_{2} = \textbf{d},\textbf{V},\bm{\alpha},\bm{\kappa})&= (\textbf{H}^{\prime}\textbf{H})^{-1}\textbf{H}^{\prime}\textbf{K}(\bm{\alpha},\bm{\kappa})\textbf{H}(\textbf{H}^{\prime}\textbf{H})^{-1},
		\end{align*}
		\textit{where we have integrated across $g(\bm{\mu})$}.\\
	\end{enumerate} 
	\noindent
	The proof of Theorem 5$(i)$ is given in the Supplemental Appendix. The proof of Theorem $5(ii)$ follows immediately from Theorem $1(ii)$ and Theorem $5(i)$. Thus, we state Theorem $5(ii)$ without proof. 
	
	Theorem $5(i)$ offers a more formal statement of a heuristic described in the Rejoinder of \citet{bradleyPMSTM} for the MLG distribution. Thus, this result is an important contribution as it provides the necessary conditions required to argue the use of the sampler described in Section 2.5. As discussed at the end of Section 2.4, computational considerations are extremely important when proposing a new complex model. A collapsed Gibbs sampler will allow one to avoid Metropolis-Hastings updates, which in turn, increases the effective sample size and, consequently, the computational performance of the method.
	
	{The integrand on the left-hand-side of (\ref{cond_pdf4}) is proportional to a $\mathrm{CM_{c}}$, and is of the same form as the full-conditional distributions that arise in the LCM in Section 2.5.} Theorem~5 shows that it is (computationally) easy to simulate {from a pdf proportional the left-hand-side of (\ref{cond_pdf4})} provided that $r \ll n$ {and that $\bm{\mu}$ is marginalized}. Recall that $\textbf{H}$ is $n\times r$, which implies that computing the $r\times r$ matrix $(\textbf{H}^{\prime}\textbf{H})^{-1}$ is computationally feasible when $r$ is ``small.'' {By small we mean a value such that the Gauss-Jordan elimination method for the inverse of a $r\times r$ matrix can be computed in real-time.} Furthermore, Theorem 5 shows that the $r$-dimensional random vector $\textbf{q}$ is an orthogonal projection of the $n$-dimensional random vector $\textbf{w}$ onto the column space spanned by the columns of $\textbf{H}$. This provides a geometric interpretation of random vectors generated from $\mathrm{CM_{c}}$ {after marginalizing $\bm{\mu}$}.

 \subsection{Prior Distributions on Covariance Parameters}

A critical feature of our proposed distribution theory is the incorporation of dependence in non-Gaussian data from the exponential family. From this point-of-view it is especially important to learn about these dependencies, which are quantified by the unknown $n\times n$ real-valued matrix $\textbf{V}$ (or equivalently $\textbf{V}^{-1}$). Thus, we place a prior distribution on $\textbf{V}^{-1}$. Specifically, let $\textbf{V}^{-1}$ be an unknown lower unit triangle matrix. That is, let $\textbf{V}^{-1}\equiv \left\lbrace {v}_{i,j}\right\rbrace$, where $v_{i,j} = 1$ for $j = i$, $v_{i,j} = 0$ for $j > i$, and $v_{i,j} \in \mathbb{R}$ for $j < i$. It will be useful to organize the elements below the lower main diagonal into the $(i-1)$-dimensional vectors $\textbf{v}_{i}\equiv \left(v_{i,j}: j = 1,\ldots,i-1\right)^{\prime}$  for $i = 2,\ldots, n$.

We place a CM prior distribution on $\textbf{v}_{i}$ for each $i$. Specifically, let 
\begin{align}\label{gammaprior}
\textbf{v}_{i}&\ind \mathrm{CM}(\bm{0}_{i-1},\textbf{C}_{i},\bm{\alpha}_{i},\bm{\kappa}_{i};\hspace{2pt}\psi);\hspace{2pt} i = 2,\ldots, n,
\end{align}
\noindent
where, in practice, the $(i-1)\times (i-1)$ matrix $\textbf{C}_{i}$ is set equal to $\sigma_{v}\textbf{I}_{i-1}$, and $\bm{\alpha}_{i}$, $\bm{\kappa}_{i}$, and $\sigma_{v}$ are specified such that (\ref{gammaprior}) is relatively ``flat.'' This specification leads to a conjugate full-conditional distribution within a Gibbs sampler (see Supplemental Appendix C for the derivation).

The CM prior distribution on the modified Cholesky decomposition of the precision matrix is similar to priors considered by \citet{danielscov}, \citet{dunson}, and \citet{Pourahmadi} in the Gaussian setting. In fact, when $\psi = \psi_{4}$ the prior distribution in (\ref{gammaprior}) reduces to the prior distributions used in \citet{danielscov}, \citet{dunson}, and \citet{Pourahmadi}. Thus, (\ref{gammaprior}) constitutes a general non-Gaussian (natural exponential family) extension of such priors on modified Cholesky decompositions of precision and covariance matrices. 

There are certainly other prior distributions for $\textbf{V}^{-1}$ that may be more appropriate. For example, see \citet{bergerga} and \citet[][for the spatial setting]{bradleyPCOS} for a Givens angle prior on covariance parameters. The Wishart and inverse Wishart are also common alternatives \citep[e.g., see][for a standard reference]{gelmanbook}. However, conjugacy may not always be present depending on the choice of CM distribution. Thus, in this article, we investigate the fully conjugate form of the LCM and specify the prior for $\textbf{V}^{-1}$ as stated in (\ref{gammaprior}).

\subsection{Prior Distributions on DY Parameters}

Following the theme of the previous sections, we define conjugate priors for the DY parameters by defining a distribution with an exponential term and an exponential to the negative unit log partition function. That is, consider
\begin{equation}
\label{generalprior}
f\left(\alpha, \kappa\vert {\gamma}_{1}, {\gamma}_{2},\rho\right) \propto \mathrm{exp}\left[ \gamma_{1}\alpha +  \gamma_{2}\kappa - {\rho}\mathrm{log}\left\lbrace \frac{1}{K\left(\alpha, \kappa\right)}\right\rbrace\right],
\end{equation}
where $\gamma_{1}$, $\gamma_{2}$, and $\rho$ are hyperparameters. The parameter space for $\gamma_{1}$, $\gamma_{2}$, and $\rho$ that ensures that (\ref{generalprior}) is proper (i.e., can be normalized to define a probability measure) is an immediate consequence of a result from \citet{diaconis}. In particular, from Theorem 1 of \citet{diaconis}, the distribution in (\ref{generalprior}) is proper provided that $\mathcal{Y}$ is a nonempty real-valued open set, the range of $\psi$ is a nonempty real-valued open set, $\gamma_{1}/\rho \in \mathcal{Y}$, $\gamma_{2}/\rho \in \mathcal{Y}_{\psi}$, and $\rho > 0$, where $\mathcal{Y}_{\psi} \equiv \{M: M = -\psi(Y), Y \in \mathcal{Y}\}$. For the CM distribution associated with $\psi_{1}$ we see that $\mathcal{Y} =  \{Y: Y<0\}$ and the range of $\psi$ is $\mathbb{R}$; thus, for this setting $\gamma_{1} < 0$, $\gamma_{2} \in \mathbb{R}$, and $\rho > 0$ results in a proper prior in (\ref{generalprior}). For the CM distributions associated with $\psi_{2},\psi_{3},$ and $\psi_{4}$ we have that $\mathcal{Y} = \mathbb{R}$ and $\psi$ is a strictly positive; thus, for this setting $\gamma_{1} \in \mathbb{R}$, $\gamma_{2} <0$, and $\rho > 0$ ensures propriety of (\ref{generalprior}).

There are many interesting special cases of the prior distribution in (\ref{generalprior}). For example, when $\alpha$ is integer-valued and $\psi = \psi_{3}$ then the prior in (\ref{generalprior}) has a relationship with the Conway-Maxwell-Poisson distribution \citep{conwaymaxwell} and the gamma distribution. These special cases (listed in Table 3 of the Supplemental Appendix C) are particularly useful because they give rise to interpretations of the hyperparameters. In particular, for $\psi = \psi_{1}$, we have that $\rho$ can be interpreted as a dispersion parameter (in relation to the dispersion parameter of a Conway-Maxwell-Poisson distribution), $\gamma_{1}$ can be interpreted as a location parameter, and $\gamma_{2}$ can be interpreted as a scale parameter. When $\psi = \psi_{2}$ we have that $\gamma_{1}$ and $\gamma_{2}$ can be interpreted as functions of a proportion (i.e., the inverse logit or log of a proportion).  For $\psi = \psi_{3}$, we have that $\rho$ can be interpreted as a dispersion parameter, $\gamma_{2}$ can be interpreted as a location parameter, and $\gamma_{1}$ can be interpreted as a scale parameter. Finally, when $\psi = \psi_{4}$ we have that $\gamma_{1}$ is interpreted as a location parameter, $\rho$ represents a shape parameter, and $\gamma_{2}$ represents a scale parameter. 

The most familiar special case occurs when $\psi = \psi_{4}$ (i.e., a normal data model) and $\alpha = 0$. Namely, (\ref{generalprior}) reduces to independent gamma prior distributions on $\kappa$ with shape parameter $\rho/2 + 1$, and scale parameter $-\gamma_{2}$. When recognizing that $\kappa$ is equal to one-half the unknown variance of a normal random variable (See Table 1), we see that the conjugate prior distribution implies an inverse gamma distribution for the variance parameter, which is a common choice of a prior distribution on the variance parameter for normally distributed data \citep{gelmanprior}.


\section{Empirical Results}     
{ In Sections 3.1 and 3.2, we use simulations to demonstrate the performance of the LCM when analyzing binomial and Poisson data. To demonstrate the wide-applicability of the CM distribution, we also give several illustrations from a variety of disciplines; namely, we analyze an epidemiology dataset (Section 3.3), a federal statistics dataset (Section 3.4), and an environmental dataset (Section 3.5). Our computations were performed on a dual 10 core 2.8 GHz Intel Xeon E5-2680 v2 processor with 256 GB of RAM. All R code and Matlab code used in these examples are provided in the Supplemental Appendix. User-friendly R code is provided at: \url{https://github.com/JonathanBradley28/CM}.}

\subsection{Simulation Example}

{We compare predictions using the LGP versus predictions based on a LCM.} As discussed in Section~1, the LGP is the standard approach for Bayesian analysis of dependent data, and thus, the results in this section are meant to provide one comparison of the LCM to the current state-of-the-art. It is important to emphasize that if the LGP is more appropriate than the LCM, our model will be able to identify this {for some settings} because of Theorem 3; that is, if the posterior replicates of the DY parameters are large then Theorem 3 suggests that the latent processes are approximately Gaussian.

The $n\times p$ matrix $\textbf{X} \equiv \left(\textbf{x}_{1},\ldots, \textbf{x}_{p}\right)^{\prime}$, the $r\times r$ matrix $\bm{\Phi}\equiv \left(\bm{\phi}_{1},\ldots, \bm{\phi}_{r}\right)^{\prime}$, and the $r \times r$ lower unit triangle matrix $\textbf{V}^{-1}$ are randomly generated with {$p = 500$} and {$r = 10$}. {The choices for $p$ and $r$ were made to represent realistic values that one might see in practice. For example, see \citet{handbookbigdata}, Chp. 2, where they consider a dataset taken from a ``data exposition'' provided by the American Statistical Association's Sections on Statistical Computing and Statistical Graphics. This example had $n = 500,000$ and $p = 29$ and was considered to be a moderate $p$ {and} large $n$ setting. Also, see \citet{huangArxiv} for a recent example where $n = 2,153,888$ and $r = 60$ is considered to be a moderate $r$ and large $n$ setting. E}ach element of the $n\times p$ matrix $\textbf{X}$, the $n\times r$ matrix $\bm{\Phi}$, and the $r\times r$ matrix $\textbf{V}^{-1}$ are selected from a standard normal distribution. The elements of the fixed and random effects $\bm{\beta}$, $\bm{\eta}$, and $\bm{\xi}$ are randomly selected from a standard normal distribution as well. Then, we define $p_{i} = \frac{\mathrm{exp}(\textbf{x}_{i}^{\prime}\bm{\beta}+ \bm{\phi}_{i}^{\prime}\bm{\eta} + \xi_{i})}{1 + \mathrm{exp}(\textbf{x}_{i}^{\prime}\bm{\beta}+ \bm{\phi}_{i}^{\prime}\bm{\eta} + \xi_{i})}$ for $i = 1,\ldots,n$. We consider two different data models in this section. In particular, we consider observations $Z_{i}$ generated from a binomial distribution with sample size $t_{i}$ {(generated from a Poisson with mean 40)} and probability of success $p_{i}$ for $i = 1, \ldots, n$. For example, consider $n$ {households, where for each household $i$ there are $t_{i}$} individuals, and let $p_{i}$ represent the probability that an individual is female. Then $Z_{i}$ would represent the number of women living in {household} $i$. Here, one might choose $\bm{\phi}_{i} = (1,0)^{\prime}$ if the total income of the household is below the poverty line, and  $\bm{\phi}_{i} = (0,1)^{\prime}$ otherwise. Similarly, we consider observations $Z_{i}$ generated from a {Poisson distribution with mean $\mathrm{exp}(\textbf{x}_{i}^{\prime}\bm{\beta}+ \bm{\phi}_{i}^{\prime}\bm{\eta} + \xi_{i})$} for $i = 1, \ldots, n$. 

Using the Gibbs sampler outlined in Supplemental Appendix C we implement the LCM in Model 1 with $j = k$ and use the appropriate data model (i.e., binomial or {Poisson}). {We use the R-package \texttt{lme4} to implement a LGP. Default choices were used when possible in when using \texttt{lme4}.} For each $i$, denote the posterior mean of $p_{i}$ with $\widehat{p}_{i}$, the posterior mean of {$\mu$} with {$\widehat{\mu}_{i}$}, and define the total squared prediction errors to be
\begin{align}
\nonumber
& \sum_{i}(tp_{i} - t\widehat{p}_{i})^{2}\\
\label{prederror}
& {\sum_{i}\left(\mu_{i}-\widehat{\mu}_{i}\right)^{2}},
\end{align}
\noindent
used for the binomial and {Poisson} settings, respectively. We used a burn-in of 10,000 and generate $B = 20,000$ posterior replications for both data models that are considered.


        \begin{figure}[t]
        \begin{center}
        \begin{tabular}{c}
				\includegraphics[width=16.5cm,height=5cm]{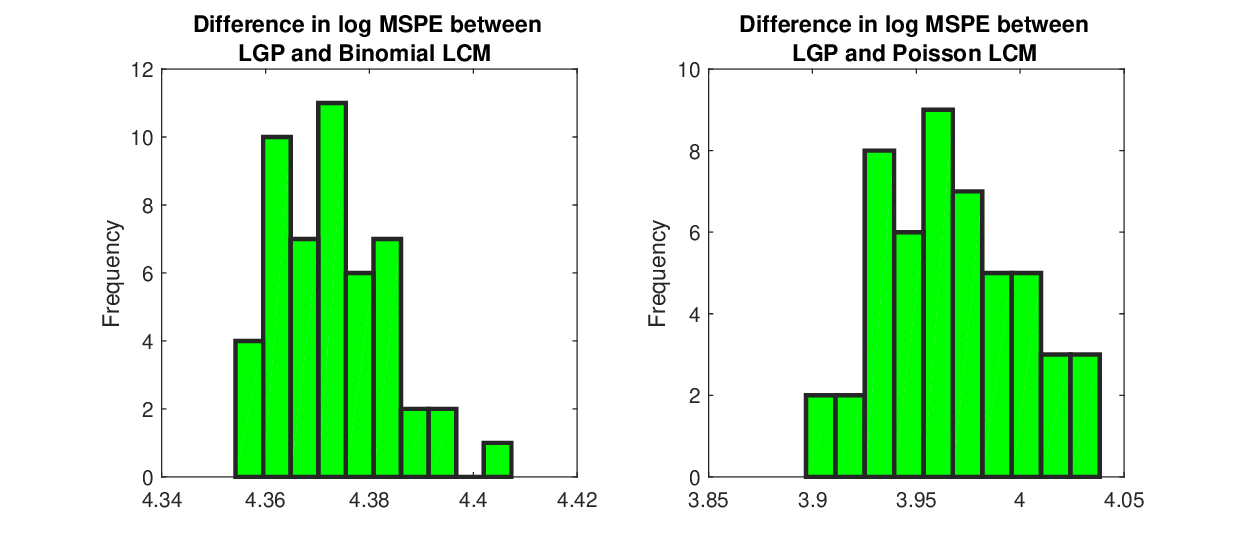}
        \end{tabular}
        \caption{\baselineskip=10pt{Histogram over the difference in log MSPE: For each of the fifty realizations of $\{Z_{1},\ldots, Z_{n}\}$ from a {(Poisson} distribution) binomial distribution, we produce ({$\{\widehat{\mu}_{i}\}$}) $\{\widehat{p}_{i}\}$ using the appropriate LCM model, produce ({$\{\widehat{\mu}_{i}\}$}) $\{\widehat{p}_{i}\}$ using the LGP model, and compute the difference in log MSPE. The difference in log MSPE for the {(Poisson} distribution) binomial setting is the log of the total squared prediction error of ({$\{\widehat{\mu}_{i}\}$}) $\{t\widehat{p}_{i}\}$ from the LGP model minus the log total squared prediction error of ({$\{\widehat{\mu}_{i}\}$}) $\{t\widehat{p}_{i}\}$ computed using the LCM model. The histogram in the (right) left panel is over the 50 independent replicates from the {(Poisson} distribution) binomial distribution.}}
        \end{center}
        \end{figure}

{We consider a large sample size of $n = 100,000$ and} simulate $\{Z_{1},\ldots, Z_{n}\}$ fifty times from the binomial distribution, and simulate another fifty independent replications of of $\{Z_{1},\ldots, Z_{n}\}$ from the {Poisson} distribution. In Figure 1(a,b) we plot the difference in log mean squared prediction error (MSPE) error of the LCM model and the total squared prediction error of the LGP model over the fifty independent replicates. A difference greater than zero indicates that the LCM has smaller total square prediction error. Here, we see that the differences in log MSPE are consistently larger than zero, and hence, the LCM clearly outperforms the LGP for this simulation design for both the binomial and Poisson settings. Thus, not only does the LCM lead to {practical} advantages {(no tuning is involved)} over the LGP for this example, there are also clear gains in predictive performance. {Thus, this simulation suggests that the LCM model yields precise predictions, and is computationally feasible for a large dataset {with moderate values for $p$ and $r$}. Note that the high predictive performance of the LCM model occurs in a setting where we do not generate the truth from a multivariate logit-beta distribution.}

\subsection{A Simulation Study} Real datasets often do not perfectly reflect the statistical model used for implementation. As such, it is necessary to provide evidence of the robustness of the LCM to model misspecification through simulation studies. We do this by considering several specifications of the simulation model in Section 3.1 and of the fitted model used to analyze the simulated data with $n=100$. Specifically, we consider the following factors in an analysis of variance (ANOVA) experiment:
\begin{itemize}
	\item \textbf{Factor 1 (Random Effects in the Simulation Model):} The simulated data are generated from the Poisson distribution in the same way as in Section 3.1 with: (\textit{Level 1}) Gaussian random effects and (\textit{Level 2}) multivariate log-gamma random effects.
	\item \textbf{Factor 2 (Number of Covariates in the Simulation Model):} The simulated data are generated from the Poisson distribution in the same way as in Section 3.1 with: (\textit{Level 1}) $p = 10$ and (\textit{Level 2}) $p = 50$.
	\item \textbf{Factor 3 (Number of Basis Functions in the Simulation Model):} The simulated data are generated from the Poisson distribution in the same way as in Section 3.1 with: (\textit{Level 1}) $r = 10$ and (\textit{Level 2}) $r = 50$.
	\item \textbf{Factor 4 (Distributional Assumptions of the Fitted model):} We make the following distributional assumptions when fitting a model to the simulated data:  (\textit{Level 1}) a Poisson LGP model, (\textit{Level 2}) a Poisson LCM model with $j = k = 3$, and (\textit{Level 3}) a negative binomial LCM with $j = k = 2$.
	\item \textbf{Factor 5 (Number of Covariates in the Fitted model):} We make the following assumptions when fitting a model to the simulated data: (\textit{Level 1}) $p = 10$ and (\textit{Level 2}) $p = 50$.
	\item \textbf{Factor 6 (Number of Basis Functions in the Fitted model):} We make the following distributional assumptions when fitting a model to the simulated data: (\textit{Level 1}) $r = 10$ and (\textit{Level 2}) $r = 50$.
\end{itemize}
\noindent
There are a total of $2^{5}\times 3 = 96$ factor level combinations (Factor 4 has three levels). The response in this experiment is the log total prediction error in (\ref{prederror}). The log transformation is done to aid in producing normality in an analysis of variance (ANOVA) experiment. Within each factor-level-combination we simulate 10 independent replicates of $\{Z_{1},\ldots, Z_{100}\}$, and compute the log total prediction error in (\ref{prederror}). This leads to a total of $10\times 96 = 960$ observations used in our ANOVA. Notice that we consider cases were we both correctly and incorrectly specify the covariates, basis functions, and distributional assumptions. This is done in an effort to assess robustness to model misspecification.

	\begin{figure}[t]
		\begin{center}
			\includegraphics[width=10cm,height=5cm]{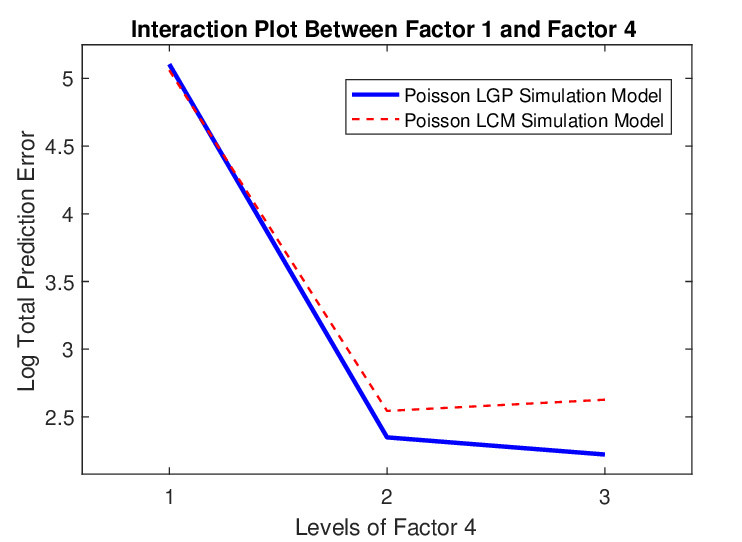}
			\caption{A two-way interaction plot for Factors 1 and 4, using the log total prediction error in (\ref{prederror}) as the response. The levels of Factor 1 are given in the legend, and the levels of factor one are listed on the $x$-axis. Notice that the models that are implemented (labeled on the $x$-axis), may be different from the models that the data are simulated from (indicated by the solid blue, and dashed red lines). Regardless of how the data are generated, fitting the LCM (levels 2 and 3 of Factor 4) appears leads to smaller log total prediction error than when fitting the Poisson LGP on average.}
		\end{center}
	\end{figure}
We implement an ANOVA with up to two-way interactions between the factors defined in the bulleted list above. The ANOVA table is provided in Supplemental Appendix D. The first and fourth main effect, and their interaction, have large F statistics. The remaining F statistics are not ``significant.'' This suggests that, for our simulation setup, the proposed model is fairly robust to misspecification of covariates and basis functions. However, the specification of the fitted model (i.e., an LGP or LCM) appears to explain most of the variability in the log total prediction error. In Figure 2, we plot the interaction plot associated with Factors 1 and 4. Here we see that even when the data are simulated with Gaussian random effects, we appear to outperform the LGP with either LCM. Both LCMs perform similarly in this setting. When the data are simulated with multivariate log-gamma random effects the ANOVA suggests that the Poisson LCM performs considerably better than the LGP, and slightly outperforms the negative binomial LCM.

	\begin{figure}[t]
	\begin{center}
		\begin{tabular}{c}
			\includegraphics[width=9cm,height=5.5cm]{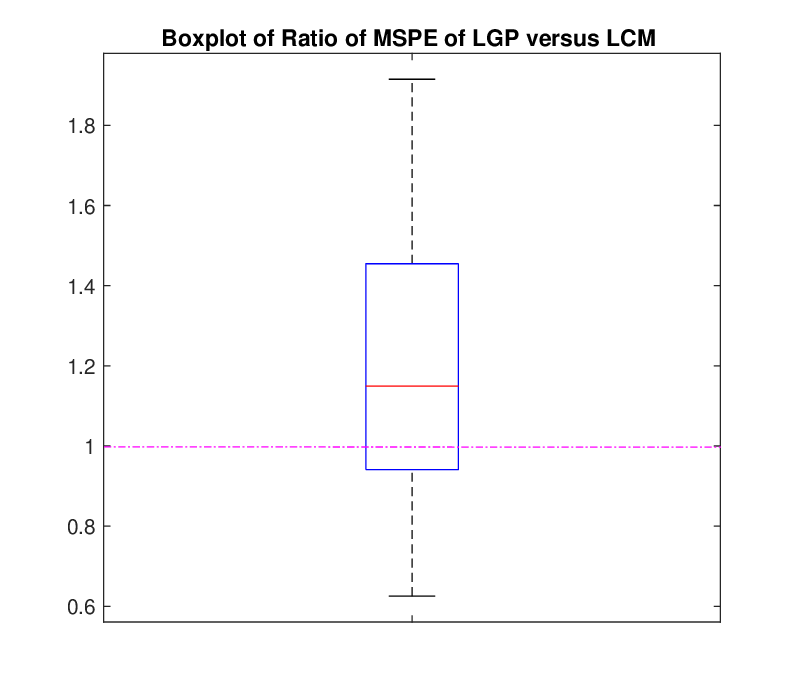}
		\end{tabular}
		\caption{\baselineskip=10pt{{The response is the ratio between mean squared prediction error using the LGP model and the mean squared prediction error of the LCM. We hold out roughly 5$\%$ of the observations. A boxplot is displayed over 50 different hold-out observations. The mean squared prediction error (MSPE) is between the predicted mean (e.g., posterior mean of $t_{i}\mathrm{exp}\left\lbrace \textbf{x}_{i}^{\prime}\bm{\beta} + \bm{\phi}_{i}^{\prime}\bm{\eta} + \xi_{i}\right\rbrace /\left[1+\mathrm{exp}\left\lbrace \textbf{x}_{i}^{\prime}\bm{\beta} + \bm{\phi}_{i}^{\prime}\bm{\eta} + \xi_{i}\right\rbrace\right]$) and the hold out dataset. Values greater than one (indicated by the dashed-dotted magenta line) suggest that the binomial LCM outperforms the binomial LGP.}}}
	\end{center}
\end{figure}

{\subsection{An Application to Contagious Bovine Pleuropneumonia in Ethiopian Highlands}Contagious bovine pleuropneumonia (CBPP) has been classified as a list-A disease by the World Organization for Animal Health. For this reason, \citet{Lesnoff} conducted an extensive study on herds of cows located within the Boji district of West Wellega, Ethiopia. They collected the incidence of CBPP among 15 herds over four time periods that span 16 months. They were interested in tracking the probability of contracting the disease as a function of time, and considered a generalized linear mixed model to assess this. Time was found to be an important fixed effect, and the herds were found to be an important random effect \citep{Lesnoff}. This is a small dataset consisting of 54 observations.
	
	We fit a binomial LCM to these data, where the response is the number of cows infected and the total number cows in the heard is known (i.e., $t_{i}$). We define $\textbf{x}_{i}$ to consist of indicators of the different time-periods. Let $g_{k}$ represent the $k$-th herd of cows, and let each element in $g_{k}$ represents a specific cow in the sample. Specify $\bm{\phi}_{i} = \left(I(i \in g_{1}),\ldots, I(i \in g_{r})\right)^{\prime}$ for cows $i = 1,\ldots, 54$. We compare our results to a LGP fitted using a standard R-package for generalized linear models; namely, the R-package \texttt{lme4}, and using the function ``glmer'' \citep{lme4}. In Figure 3, we plot the ratio of the mean squared prediction errors (i.e., MSPE associated with LGP and the MSPE associated with the LCM). Here, the paired t-test resulted in a p-value of 1.27$\times 10^{-5}$, which suggests that the LCM is outperforming the GLM in this setting. However, visually Figure 3 suggest that the LCM and LGP give similar results for this example.}

{
\subsection{An Application to Count-Valued ACS Public-Use Micro-Data}} The US Census Bureau has replaced the decennial census long-form with the American Community Survey (ACS), which is an ongoing survey that collects an enormous amount of information on US demographics. (To date there are over 64,000 variables published through the ACS.) The estimates published from the ACS have a unique multi-year structure. Specifically, the ACS produces 1-year and 5-year period estimates of US demographics, where 1-year period estimates are summaries (e.g., median income of a particular county) made available over populations over 65,000 and 5-year period estimates are made available for all published geographies \citep[e.g., see][for more information]{torrieri}. 

A difficulty with using ACS period estimates published over pre-defined geographies is that it is difficult to infer fine-level (i.e., household) information. As a result, the ACS provides a public$\--$use micro-sample (PUMS) over public$\--$use micro-areas (PUMAs). PUMS consists of individual and household information within each PUMA, where the location of the household within the PUMA is not released to the public. In this section, we focus on household level PUMS found within one particular PUMA; namely the PUMA that covers the metropolitan area of Tallahassee Florida (labeled as PUMA number 00701). 
        \begin{table}[t]
        \begin{center}
 \begin{tabular}{c c||*{8}{c|}}
 	\hfill & \multicolumn{9}{c}{\small Hold-Out Data Value} \\
 	\cline{3-10}
 	\multirow{11}*{\rotatebox{90}{\small Rounded Poisson LCM Predictions}} &\hfill
 	\hfill & {0} & {1} & {2} & {3} & {4} & {5} & {6} & {7}\\
 	\cmidrule{3-10}\morecmidrules\cmidrule{3-10}
 	\hfill& {0} & {\color{red}18} & 0 & 0 & 0 & 0 & 0 & 0 & 0\\
 	\cline{2-10}
 	\hfill& {1}& 0 & {\color{red}75} & 14 & 0 & 0 & 0 & 0 & 0\\
 	\cline{2-10}
 	\hfill& {2}& 0 & 14 & {\color{red}43} & 15 & 1 & 0 & 0 & 0\\
 	\cline{2-10}
 	\hfill& {3}& 0 & 0 & 3 & {\color{red}13} & 8 & 1 & 0 & 0\\
 	\cline{2-10}
 	\hfill& {4}& 0 & 0 & 0 & 0 & {\color{red}9} & 0 & 0 & 0\\
 	\cline{2-10}
 	\hfill& {5}& 0 & 0 & 0 & 0 & 4 & {\color{red}3} & 0 & 0\\
 	\cline{2-10}
 	\hfill& {6}& 0 & 0 & 0 & 0 & 0 & 1 & {\color{red}2} & 0\\
 	\cline{2-10}
 	\hfill& {7}& 0 & 0 & 0 & 0 & 0 & 0 & 1 & {\color{red}0}\\
 	\cline{2-10}
 	\hfill& {8}& 0 & 0 & 0 & 0 & 0 & 0 & 1 & 0\\
 	\cline{2-10}
 	\hfill& {9}& 0 & 0 & 0 & 0 & 0 & 0 & 0 & 1\\
 	\cline{2-10}
 \end{tabular}
        \caption{\baselineskip=10pt{{A cross-tabulation of a hold-out dataset with 216 observations and the corresponding rounded predicted values (i.e., the posterior mean estimated from the Gibbs sampler). These predictions are rounded to the nearest integer, since the hold-out dataset is known to be integer-valued. The red-values indicate that the rounded predictions and the hold-out data exactly agree.}}}
        \end{center}
        \end{table}
Consider 2005-2009 PUMS estimates of the number of individuals living in a household contained within PUMA 00701. This is a fairly large dataset (for multivariate statistics) consisting of 4,537 observations. An important inferential goal, besides giving an illustration of the LCM, is to accurately predict vacant households (i.e., predict zero people living in a household). Vacant households exhaust resources for those conducting surveys, and is of practical interest to the US Census Bureau (see, \url{http://www.census.gov/en.html}).

We would expect the number of individuals living in a household to be spatially correlated, since certain neighborhoods within Tallahassee are known to be more attractive for those with a family, and hence, have more people living within a household in these neighborhoods. However, the spatial correlation can not be leveraged, since the location \textit{within} PUMAs are not publicly available. Consequently, we model the dependencies within the PUMS using a generic multivariate distribution; namely, the CM distribution. In particular, we assume that the data follows a LCM. {We consider three types of LCMs: the first is a Poisson LCM (i.e., $j = k = 3$), the second is a negative binomial LCM (i.e., $j = k = 2$), and the third is a Poisson LGP (i.e., $j = 3$ and $k = 4$).} There are a large number of potential covariates (there are 358 in total) including fuel cost of the household, number of bedrooms in the household, and lot size, among others. For illustration, we picked a small subset of covariates using least angle regression \citep{lars}, which lead to ${41}$ covariates. {We consider defining each covariate as the coefficient of the random effects (i.e., a column of $\bm{\Psi}$) so that $r = 41$. Additionally, we include an intercept as a fixed effect (i.e., $\textbf{X} = \bm{1}_{p,1}$).} Convergence of the MCMC algorithm was assessed visually using trace plots, and no lack of convergence was detected.

To assess the quality of the predictions we randomly selected $216$ observations (roughly $5\%$ of the data). Using the remaining data we produce estimates of the mean number of individuals living in a household for the $216$ observations. {As an example, see Table 2 where we display} the hold-out dataset and the corresponding {Poisson LCM} predictions that were based on the remaining $4,321$ observations. Here, we see that a majority of the rounded (to the nearest integer) predictions are \textit{exactly equal} to the hold-out data. In fact, $163/226 \approx 72\%$ of the 226 hold-out dataset are \textit{exactly} equal to the corresponding rounded predicted value, and the remaining $30\%$ are within two counts of the corresponding hold-out data value. Furthermore, we are able to very accurately predict an empty household, which may have implications for sampling done by the US Census Bureau.

{This hold-out study was repeated 50 times, and the results of the mean absolute difference (MAD) between the hold-out data and the rounded predictions are presented in Figure 4. Here, we fit the LGP (or a Bayesian GLM) using the R-package \texttt{MCMCglmm} and the function ``MCMCglmm'' \citep{MCMCglmm}, and the remaining models were fitted using the Matlab (Version 9) code in the Supplemental Materials. Here, we see that both the Poisson LCM and the Negative binomial LCM outperforms the Poisson LGP. However, the negative binomial LCM performs worse than the Poisson LCM. The pairwise p-values for a paired $t$-tests (using 50 MAD values as the response) are as follows: A one-sided test between the Poisson LCM and the Poisson LGP resulted in a p-value of 0.0016; a one-sided test between the negative binomial LCM and the Poisson LGP resulted in a p-value of 0.0021; and a one-sided test between the negative binomial LCM and the Poisson LCM resulted in a p-value of $9.14\times 10^{-44}$. We found that the results for the negative binomial LCM to be sensitive to the prior on $b$ (i.e., the coefficient of the unit log-partition function); hence, we suggest using the Poisson LCM instead of the negative binomial LCM.}


		\begin{figure}[t]
			\begin{center}
				\begin{tabular}{c}
					\includegraphics[width=9cm,height=4cm]{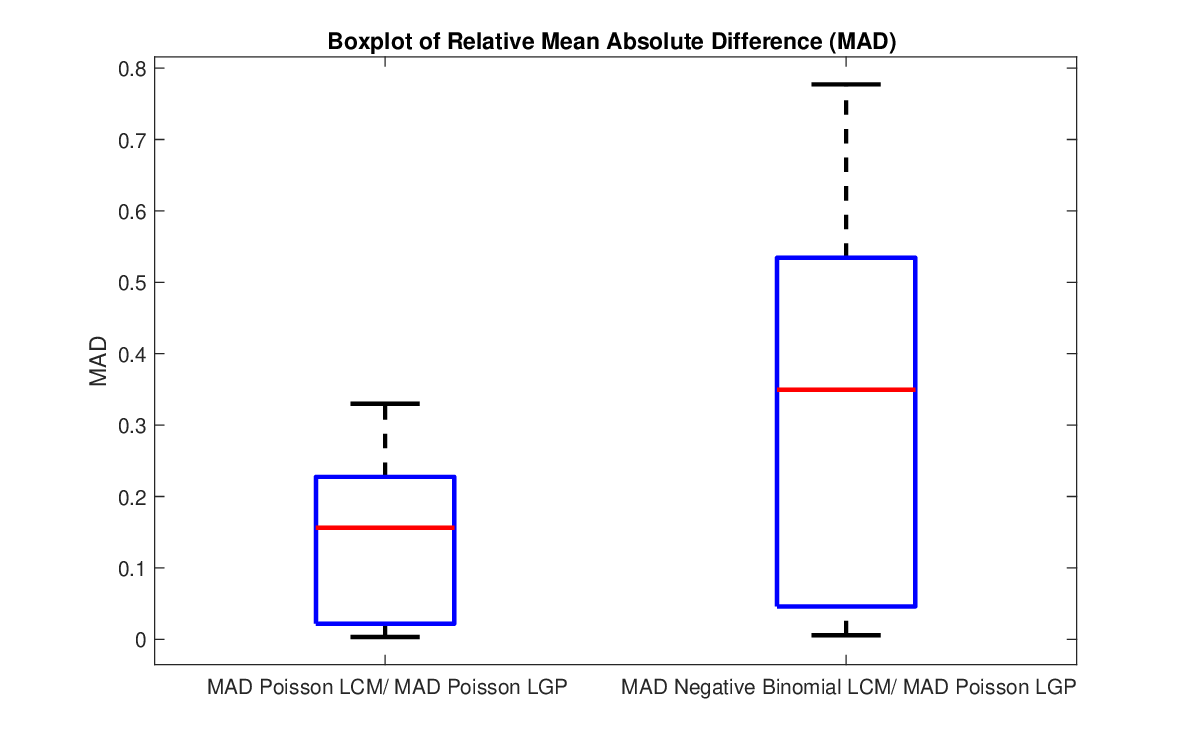}
				\end{tabular}
				\caption{\baselineskip=10pt{{The response is the ratio between mean absolute difference using the LCM model and the mean absolute difference of the LGP. We hold out roughly 5$\%$ of the observations. A boxplot is displayed over 50 different hold-out observations. The mean absolute difference (MAD) is between the predicted mean (e.g., posterior mean of $t_{i}\mathrm{exp}\left\lbrace \textbf{x}_{i}^{\prime}\bm{\beta} + \bm{\phi}_{i}^{\prime}\bm{\eta} + \xi_{i}\right\rbrace /\left[1+\mathrm{exp}\left\lbrace \textbf{x}_{i}^{\prime}\bm{\beta} + \bm{\phi}_{i}^{\prime}\bm{\eta} + \xi_{i}\right\rbrace\right]$) and the hold-out dataset. Values less than one suggest that the LCM outperforms the LGP. The left boxplot represents the 50 ratios of the MAD using the Poisson LCM and the MAD using the Poisson LGP. The right boxplot represents the 50 ratios of the MAD using the Negative Binomial LCM and the MAD using the Poisson LGP.}}}
			\end{center}
		\end{figure}	
\vspace{5pt}
{
\subsection{An Application to Moderate Resolution Imaging Spectroradiometer Cloud Data}} {On December 18, 1999 the National Aeronautics and Space Administration (NASA) launched the Terra satellite, which is part of the Earth Observing System (EOS). The Moderate Resolution Imaging Spectroradiometer (MODIS) is a remote sensing instrument attached to the Terra satellite and collects information on many environmental processes. In particular, the MODIS instrument converts spectral radiances into a level-2 (i.e., 1 km $\times$ 1 km spatial resolution) cloud mask using cloud detection algorithms. These cloud detection algorithms can not perfectly identify the presence of a cloud at each 1 km $\times$ 1 km region. \citet{aritrajsm} cast this as a big spatial data problem as, visually speaking, spatial correlations appear to be present (i.e., nearby observations tend to be more similar) and $n = 2, 748, 620$ is large.

		 \begin{figure}[t]
		 	\begin{center}
		 		\begin{tabular}{c}
		 			\includegraphics[width=15.5cm,height=5cm]{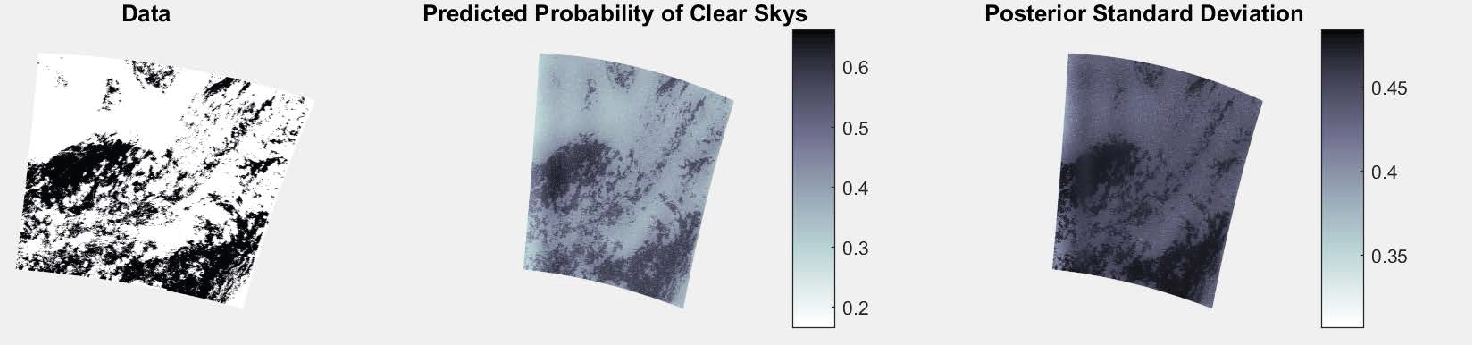}
		 		\end{tabular}
		 		\caption{\baselineskip=10pt{{In the left-most panel we have a plot of the data. White locations are observed clouds and black locations are observed clear skies. The middle panel are the posterior expected value of the probability of clear skies using the Bernoulli LCM, and the right-most panel is the corresponding posterior variance of the probability of clear skies. Posterior expected values and variances were computed using a training dataset consisting of 95$\%$ of the points in the left-most plot (these points were randomly selected).}}}
		 	\end{center}
		 \end{figure}
		 
In this article we consider fitting a Bernoulli LCM (i.e., $j = k = 2$ and $b = 1$) to the MODIS level-2 cloud mask data from \citet{aritrajsm}. This model takes approximately one day to run using the code in the Supplemental Materials. We use the same covariates and radial basis functions and covariates in \citet{aritrajsm}. Specifically, let $\bs_{1},\ldots \bs_{n} \in \mathrm{R}^{2}$ represent the observed data locations (latitude/longitude) seen in the left-panel in Figure 5. Set $\bm{\phi}_{i} = \left(\phi_{1}(\bs_{i}),\ldots, \phi_{j}(\bs_{i})\right)^{\prime}$, where
\begin{equation*}
\phi_{j}(\bs) = \left\lbrace 1 - \frac{||s - \textbf{g}_{j}||}{w_{j}}\right\rbrace^{2}I(||\bs - \textbf{g}_{j}||<w_{j});\hspace{2pt}j = 1,\ldots, r,
\end{equation*}
\noindent
where $\textbf{g}_{j}$, $j = 1,\ldots, r$, is the aforementioned knot points. This radial basis function is referred to as a bisquare function \citep{johan}. The knot locations are divided into three groups called ``resolutions.'' Then $w_{j}$ is set equal to 1.5 times the shortest great arc distance between the points that are in the same resolution as $\textbf{g}_{j}$. \citet{aritrajsm} chose $r = 137$ knots to have a ``quad-tree'' structure (or equally-spaced structure), where the knots of the different resolution all differ from one another \citep[e.g., see][among others]{kang-cressie-shi-2010}. 

The posterior predicted value of the probability of a clear sky is given in Figure 5 along with the posterior variance. The posterior predicted probabilities reflect the general pattern of the data. Also, the posterior variances are larger at smaller posterior predicted probabilities, which is to be expected as smaller probabilities tend to be more difficult to estimate. Thus, Figure 5 shows that it is possible to fit an LCM to a high-dimensional spatial dataset (with small $p$) and obtain reasonable in-sample results.

We consider only a single hold-out sample in this section, since the computation times for this dataset are so demanding. Here, we hold out $5\%$ of the observations from the left-most panel of Figure 5 and produced the posterior expected value of the probability of clear skies. We threshold the values of these posterior probabilities around the midpoint of the range to classify either clear sky or cloudy. The false positive rate is 0.22 and the false negative rate is moderately large at 0.28. We chose to compare these values to the misclassification rates using a standard binary classifier, support vector machines \citep[SVM,][]{htf} fitted using Matlab's ``fitcsvm'' function. SVM took approximately three days to run, the false positive rate is smaller at 0.11, and the false negative rate is much larger at 0.53. Thus, this single hold-out study suggests that the Bernoulli LCM leads to a classifier that is comparable to the current industry standard, SVM. Moreover, we are able to provide prediction uncertainty.}

		 \vspace{5pt}
\section{Discussion}

We have introduced methodology for jointly modeling dependent non-Gaussian data within the Bayesian framework. This methodology is rooted in the development of new distribution theory for dependent data that makes Bayesian inference possible to implement using a Gibbs sampler; hence, computationally intensive and ad hoc approaches needed for tuning and specifying proposal distributions are not needed. Specifically, we propose a multivariate version of the prior distributions introduced by \citet{diaconis}. Furthermore, the prior distributions similar to those used by \citet{danielscov}, \citet{dunson}, and \citet{Pourahmadi} are adapted to the non-Gaussian setting.

Several theoretical results were required to derive this \textit{conjugate multivariate distribution} (CM), and to develop its use for Bayesian inference of dependent data from the natural exponential family. The later is facilitated through the introduction of the latent CM (LCM) model. In particular, we show that full conditional distributions are of the same form of a conditional distribution of a CM random vector, and provide a way to simulate from this conditional distribution. Relationships between the LCM and the LGP also provide motivation for the use of the LCM. In particular, the latent Gaussian process (LGP) is a special case of the LCM. Furthermore, many types of LCMs can be well approximated by a LGP, by specifying certain parameters of a LCM to be ``large.'' This result shows that the LCM is not only computationally easier to implement, but is also more flexible than a LGPs.

Empirical exploration of the Poisson, binomial, Bernoulli, and negative binomial special cases were performed through simulations studies and through {analyses of several datasets from a variety of disciplines.} These examples indicate very small out-of-sample error when using LCM for prediction, and show gains in predictive performance over the LGP. Additionally, the LCM model is applicable for large datasets {(in the application we implemented the LCM on a MODIS level-2 cloud-mask data of size $2, 748, 620$)}. { In the first example we considered a small dataset of binomial counts of CBPP among herds of cows. We obtained obtained precise predictions and outperformed the LGP computed using a standard R-package.} In the {second} real data analysis section we predict the number of individuals within a household over the US city of Tallahassee Florida, and obtain very precise estimates (in terms of hold-out error). The predictions were very accurate even though $18/226 \approx 8\%$ of the hold-out dataset consisted of zero counts, which is known to cause difficulties in an LGP \citep{zeroinflated}. {In the third application, we obtain posterior predicted probabilities that reflected the pattern of data at observed locations, and a binary classifier that has misclassification rates that are comparable to support vector machines.}

Although there are many settings where the LCM improves both precision and computation, there are settings where it would not be feasible to implement the LCM. In particular, we consider one choice of $\psi$ that results in a case where $\mathcal{M}_{p}^{n+p}$, $\mathcal{M}_{r}^{n+r}$, and $\mathcal{M}_{n}^{2n}$ does not guarantee that $\textbf{x}_{i}^{\prime}\bm{\beta} + \bm{\phi}_{i}^{\prime}\bm{\eta}+\xi_{i} \in \mathcal{Y}$ for each $i$; namely $\psi_{1}$, which is the unit log partition function of a gamma data model. In this case, the full conditional distributions are \textit{truncated} $\mathrm{CM_{c}}$ distributions. Thus, in this setting the LCM is most easily implemented by doing a Gibbs sampler with component-wise updates due to the truncated support of the natural parameter. This is computationally less efficient than simply transforming the gamma data to the log scale and fitting an LGP, which can give precise predictions. {Additionally, we found that the negative binomial LCM to give poorer predictive results than the Poisson LCM. Thus, we suggest using the Poisson LCM when analyzing unbounded count values instead of the negative binomial LCM.}

As discussed in the Introduction, a general modeling framework for dependent data that can model non-Gaussian (natural exponential family) data as easily as Gaussian data, has important implications for applied statistics. Nevertheless, there are also many opportunities for new methodological results that are exciting, since a special case of our framework (i.e., the LGP) has been the central methodological tool used in the dependent data literature. In particular, we are interested in developing the LCM model within ``more specific'' dependent data settings such as time-series, spatial, spatio-temporal, and multivariate spatio-temporal arenas.

\section*{Acknowledgments} We would like to express our sincere gratitude to the editors, the associate editor, and the referees for their very helpful comments that improved this manuscript. We would also like to thank Drs. Matthew Simpson of SAS Inc. and Erin Schliep at the University of Missouri for helpful discussions. This research was partially supported by the U.S. National Science Foundation (NSF) and the U.S. Census Bureau under NSF grant SES-1132031, funded through the NSF-Census Research Network (NCRN) program. This article is released to inform interested parties of ongoing research and to encourage discussion of work in progress. The views expressed are those of the authors and not those of the NSF or the U.S. Census Bureau.

  \singlespacing
\bibliographystyle{jasa} 
\bibliography{myref}

\newpage
\chapter{\hfill}
\thispagestyle{empty} \baselineskip=28pt

\thispagestyle{empty} \baselineskip=28pt

\begin{center}
	{\LARGE{\bf Supplemental Appendix:  {Bayesian Hierarchical Models with Conjugate Full-Conditional Distributions for Dependent Data from the Natural Exponential Family}}}
	\end{center}

	\baselineskip=12pt

	\vskip 2mm
	\begin{center}
		Jonathan R. Bradley\footnote{(\baselineskip=10pt to whom correspondence should be addressed) Department of Statistics, Florida State University, 117 N. Woodward Ave, Tallahassee, Fl 32306, bradley@stat.fsu.edu},
		Scott H. Holan\footnote{\baselineskip=10pt  Department of Statistics, University of Missouri, 146 Middlebush Hall, Columbia, MO 65211-6100}\footnote{\baselineskip=10pt  U.S. Census Bureau, 4600 Silver Hill Road, Washington, D.C., 20233-9100},
		Christopher K. Wikle$^2$
		\end{center}
		
		\pagenumbering{arabic}
		
		\baselineskip=24pt
		
		\section*{Introduction} In Supplemental Appendix A, we provide additional discussion surrounding the DY and CM distributions introduced in the main text. In Supplemental Appendix B, we provide the {proofs of the results, propositions, and theorems} stated in the main text. Details surrounding the collapsed Gibbs sampler are provided in Supplemental Appendix C. Additional simulation results are presented in Supplemental Appendix D.
		
		\newpage
		\section*{Appendix A: Additional Discussion on the DY and CM Distribution}
		\renewcommand{\theequation}{A.\arabic{equation}}
		\setcounter{equation}{0}
		
		\subsection*{Appendix A.i: Example Univariate Distributions} In Table 1, we give examples of $\psi$, $\mathrm{EF}(Y;\psi)$, and $K(\alpha, \kappa)$.
		
		\begin{table}[h!]
			\centering
			{\renewcommand{\arraystretch}{6}%
				\noindent\adjustbox{max width = \textwidth}{%
					\begin{tabular}{ |p{0.4\textwidth}|c|c|c|p{0.4\textwidth}|  }
						\hline
						\centering\textbf{Data Model} & \textbf{Natural Parameter} & \textbf{Log Partition Function} (i.e., ${\psi}$ and $b$) & \textbf{Normalizing Constant}&\textbf{How to Simulate From the DY Distribution}\\ \hline
						\shortstack{Gamma($a,k$)\\\shortstack{$
								f(Z\vert \alpha,\kappa) = \frac{1}{\Gamma(a)k^{a}}\mathrm{exp}(-Z/k)$\\ $a>0, k>0, Z>0$}}
								& Negative Reciprocal: $Y = -\frac{1}{k}$. & \shortstack{$\psi_{1}(Y) = \mathrm{log}\left(-\frac{1}{Y}\right)$\\$b = a$} & $K(\alpha,\kappa) = \frac{\alpha^{\kappa+1}}{\Gamma(\kappa+1)}$& Let $W\sim \mathrm{Gamma}(\kappa+1,1/\alpha)$, where $\alpha > 0$,  and $\kappa > 0$. Then, $-W\sim \mathrm{DY}\left(\alpha,\kappa;\hspace{2pt}\psi_{1}\right)$.\\ \hline
								\shortstack{Bin($t,p$)\\ \shortstack{$
										f(Z\vert t,p) = {{t}\choose{Z}} p^{Z}(1-p)^{t-Z}$\\ $0<p<1, t = 1,2,\ldots, Z = 0,\ldots, t
										$}}& Logit: $Y = \mathrm{log}\left(\frac{p}{1-p}\right)$&  \shortstack{$\psi_{2}(Y) = \mathrm{log}\left(1 + \mathrm{exp}(Y)\right)$\\$b = t$} & $K(\alpha,\kappa) = \frac{\Gamma(\kappa)}{\Gamma(\alpha)\Gamma(\kappa - \alpha)}$ & Let $W\sim \mathrm{Beta}(\alpha,\kappa - \alpha)$, where  $\kappa > \alpha>0$ and ``Beta$(\alpha,\kappa - \alpha)$'' is a shorthand for the beta distribution with shape parameter $\alpha$ and scale parameter $\kappa - \alpha$. Then, $\mathrm{log}\left(\frac{W}{1-W}\right)\sim \mathrm{DY}\left(\alpha,\kappa;\hspace{2pt}\psi_{2}\right)$.\\ \hline
										\shortstack{NegBin($t,p$)\\ \shortstack{$
												f(Z\vert t,p) = {{Z+t-1}\choose{Z}} p^{Z}(1-p)^{t}$\\ $0\le p \le 1, t=1,2,\ldots, Z=0,1,\ldots,
												$}} & Logit: $Y = \mathrm{log}\left(\frac{p}{1-p}\right)$& \shortstack{$\psi_{2}(Y) = \mathrm{log}\left(1 + \mathrm{exp}(Y)\right)$\\ $b = t+Z$}  & $K(\alpha,\kappa) = \frac{\Gamma(\kappa)}{\Gamma(\alpha)\Gamma(\kappa - \alpha)}$ & Let $W\sim \mathrm{Beta}(\alpha,\kappa - \alpha)$, where $\kappa > \alpha>0$. Then, $\mathrm{log}\left(\frac{W}{1-W}\right)\sim \mathrm{DY}\left(\alpha,\kappa;\hspace{2pt}\psi_{2}\right)$.\\ \hline
												\centering \shortstack{Pois($\mu$)\\ \shortstack{$f(Z\vert \mu) = \frac{\mu^{Z}\mathrm{exp}(-\mu)}{Z!}$\\$ \mu \in \mathbb{R}^{+}, Z = 0,1,2,\ldots$}} & Log $Y = \mathrm{log}(\mu)$& \shortstack{$\psi_{3}(Y) = \mathrm{exp}\left(Y\right)$\\ $b = 1$} & $K(\alpha,\kappa) = \frac{\kappa^{\alpha}}{\Gamma(\alpha)}$ & Let $W\sim \mathrm{Gamma}(\alpha,1/\kappa)$, where $\alpha > 0$ and $\kappa > 0$. Then, $\mathrm{log}\left(W\right)\sim \mathrm{DY}\left(\alpha,\kappa;\hspace{2pt}\psi_{3}\right)$.\\ \hline
												\shortstack{Norm($\mu,s$)\\ \shortstack{$f(Z\vert \mu, s) = \left(\frac{1}{2\pi s^{2}}\right)^{1/2}\mathrm{exp}\left(-\frac{-(Z-\mu)^{2}}{2 s^{2}}\right)$ \\ $\mu \in \mathbb{R}, s\in \mathbb{R}^{+}, Z \in \mathbb{R}$}} & Linear: $Y = \frac{\mu}{s^{2}}$&  \shortstack{$\psi_{4}(Y) = Y^{2}$\\ $b = \frac{s^{2}}{2}$}& $K(\alpha,\kappa) = \left(\frac{\kappa}{\pi}\right)^{1/2}\mathrm{exp}(-\frac{\alpha^{2}}{4\kappa})$ & Let $W$ be a normal random variable with mean $\frac{\alpha}{2\kappa}$ and variance $\frac{1}{2\kappa}$. Then, $W\sim \mathrm{DY}\left(\alpha,\kappa;\hspace{2pt}\psi_{4}\right)$.\\ \hline
												\end{tabular}}}
												\caption{Univariate Distributions: The first column has the data model, the second column has the natural parameter, the third column contains quantities that define the log partition function, the fourth column has the normalizing constant, and the fifth column has instructions on how to simulate from the DY random variable with the corresponding $\psi$. Let $\mathbb{R}^{+} = \{x: x>0\}$.}
												\label{univ}
												\end{table}
												
												\subsection*{Appendix A.ii: A Metropolis-Hastings Approach to the Conditional CM distribution} {To use the affine transformation (i.e., $\textbf{q} =  (\textbf{H}^{\prime}\textbf{H})^{-1}\textbf{H}^{\prime}\textbf{w}$) as a means to generate from a pdf proportional to $\mathrm{CM}_{c}$, one does not necessarily have to marginalize across $\bm{\mu}$. This is because the unnormalized CM distribution is proportional to the marginal distribution from an improper extension of $\textbf{q}$.  Specifically, let $\rho$ be an unnormalized CM distribution with mean $\textbf{V}\bm{\mu}$ and covariance parameter $\textbf{V}^{-1} = [\textbf{H}, \frac{1}{\sigma_{2}}\textbf{Q}_{2}]$, where $\textbf{Q}_{2}$ is the $n \times (n-r)$ orthonormal basis for the null space of $\textbf{H}$. Then we introduce a latent $(n-r)$-dimensional random vector $\textbf{q}_{2}$ and augment the distribution of $\textbf{q}$ with,
													\begin{align}
													\nonumber
													&\rho(\textbf{q}, \textbf{q}_{2}\vert \textbf{c} = \textbf{V}\bm{\mu},\textbf{V},\bm{\alpha},\bm{\kappa}) = \mathrm{exp}\left\lbrace\bm{\alpha}^{\prime}\textbf{H}\textbf{q} - \bm{\kappa}^{\prime}\psi(\textbf{H}\textbf{q} - \bm{\mu})\right\rbrace\\
													\nonumber
													&= g(\textbf{q}\vert  \bm{\mu},\textbf{V},\bm{\alpha},\bm{\kappa})g(\textbf{q}_{2}\vert  \bm{\mu},\textbf{V},\bm{\alpha},\bm{\kappa}),
													\end{align}
													\noindent
													where
													\begin{align}\label{marginals22}
													g(\textbf{q}_{1}\vert  \bm{\mu},\textbf{V},\bm{\alpha},\bm{\kappa}) &= \mathrm{exp}\left\lbrace\bm{\alpha}^{\prime}\textbf{H}\textbf{q} - \bm{\kappa}^{\prime}\psi(\textbf{H}\textbf{q}-\bm{\mu})\right\rbrace \propto f(\textbf{q}_{1}\vert \textbf{q}_{2} = \bm{0}_{n-r,1},\bm{\mu},\textbf{H},\bm{\alpha},\bm{\kappa})\\
													\label{marginals23}
													g(\textbf{q}_{2}\vert \bm{\mu},\textbf{V},\bm{\alpha},\bm{\kappa}) &= 1.
													\end{align}
													\noindent
													Thus, the Metropolis-Hastings ratio with update $\textbf{q} = (\textbf{H}^{\prime}\textbf{H})^{-1}\textbf{H}^{\prime}\textbf{w}$ is one in the limit. That is, the following Metropolis-Hastings ratio approaches one as $\sigma_{2}$ increases,
													\begin{equation*}
													\frac{\mathrm{exp}\left\lbrace\bm{\alpha}^{\prime}\textbf{H}\textbf{q}^{*} - \bm{\kappa}^{\prime}\psi(\textbf{H}\textbf{q}^{*}-\bm{\mu})\right\rbrace}{\mathrm{exp}\left\lbrace\bm{\alpha}^{\prime}\textbf{H}\textbf{q}^{[m]} - \bm{\kappa}^{\prime}\psi(\textbf{H}\textbf{q}^{[m]}-\bm{\mu})\right\rbrace}\frac{\mathrm{exp}\left\lbrace\bm{\alpha}^{\prime}\textbf{H}\textbf{q}^{[m]}+\frac{1}{\sigma_{2}}\bm{\alpha}^{\prime}\textbf{Q}_{2}\textbf{q}_{2}^{[m]} - \bm{\kappa}^{\prime}\psi(\textbf{H}\textbf{q}^{[m]} + \frac{1}{\sigma_{2}}\textbf{Q}_{2}\textbf{q}_{2}^{[m]}-\bm{\mu})\right\rbrace}{\mathrm{exp}\left\lbrace\bm{\alpha}^{\prime}\textbf{H}\textbf{q}^{*}+\frac{1}{\sigma_{2}}\bm{\alpha}^{\prime}\textbf{Q}_{2}\textbf{q}_{2}^{*} - \bm{\kappa}^{\prime}\psi(\textbf{H}\textbf{q}^{*}+\frac{1}{\sigma_{2}}\textbf{Q}_{2}\textbf{q}_{2}^{*}-\bm{\mu})\right\rbrace},
													\end{equation*}
													\noindent
													where $\textbf{q}^{*}$ and $\textbf{q}_{2}^{*}$ are a proposed values of $\textbf{q}$ and $\textbf{q}_{2}$, and $\textbf{q}^{[m]}$ and $\textbf{q}_{2}^{[m]}$ are the previous values in the Markov chain. The argument in (\ref{marginals22}) and (\ref{marginals23}) is very similar to a result in \citet[][cf. Theorem 2]{bradleyPMSTM}, which was clarified in the rejoinder of \citet{bradleyPMSTM}. Although the $\mathrm{CM}_{c}$ is proper, it is crucial that we recognize that $\textbf{q}$ follows an unnormalized $\mathrm{CM}_{c}$ and is extended by an improper $\textbf{q}_{2}$. This \textit{improper extension} results in a lack of Kolmogorov consistency \citep{Daniell,Kolmogorov,bradleyPMSTM}. However, proper extensions of the CM distribution are Kolmogorov consistent (see Theorem 4).}\\

													\section*{Appendix B: Proofs}
													\renewcommand{\theequation}{B.\arabic{equation}}
													\setcounter{equation}{0}
													In this appendix we provide {proofs for the} technical results stated in the paper. 
													
													\noindent
													\paragraph{\large{Proof of Theorem 1($i$):}}
													\normalsize
													From (2) of the main text we see that the distribution of the random vector \textbf{w} in (7) is given by,\\
													\begin{equation*}
													\left(\prod_{i = 1}^{n}K(\alpha_{i},\kappa_{i})\right)\mathrm{exp}\left\lbrace\bm{\alpha}^{\prime}\textbf{w} - \bm{\kappa}^{\prime}\psi(\textbf{w})\right\rbrace;\hspace{10pt} \textbf{w} \in \mathbb{R}^{n}.
													\end{equation*}
													The inverse of the transform of (7) is given by $\textbf{w} = \textbf{V}^{-1}(\textbf{Y} - \bm{\mu})$, and the Jacobian is given by $|\mathrm{det}(\textbf{V}^{-1})|$. Then, by a change-of-variables \citep[e.g., see][]{casellaBerger}, we have that the pdf of $\textbf{Y}$ is given by,
													\begin{align}
													\nonumber
													&\mathrm{det}(\textbf{V}^{-1})\left(\prod_{i = 1}^{n}K(\alpha_{i},\kappa_{i})\right)\mathrm{exp}\left[\bm{\alpha}^{\prime}\textbf{V}^{-1}(\textbf{Y} - \bm{\mu}) - \bm{\kappa}^{\prime}\psi\left\lbrace\textbf{V}^{-1}(\textbf{Y} - \bm{\mu})\right\rbrace\right];\hspace{5pt} \textbf{Y} \in \mathcal{M}^{n}.
													\end{align}
													\noindent
													This completes the proof of Theorem~1($i$).\\

													\noindent
													\paragraph{\large{Proof of Theorem 2:}}
													\normalsize
													It follows from Proposition~1($i$) that the conditional distribution is given by
													
													\begin{align*}
													f(\textbf{Y}_{1}\vert \textbf{Y}_{2},\bm{\mu},\textbf{V},\bm{\alpha},\bm{\kappa}) &= \frac{\left[f(\textbf{Y}\vert \bm{\mu},\textbf{V},\bm{\alpha},\bm{\kappa})\right]_{\textbf{Y}_{2} = \textbf{d}}}{\left[\int f(\textbf{Y}\vert \bm{\mu},\textbf{V},\bm{\alpha},\bm{\kappa})d\textbf{Y}_{1}\right]_{\textbf{Y}_{2} = \textbf{d}}},\\
													&\propto\hspace{5pt}\mathrm{exp}\left[\bm{\alpha}^{\prime}\left(\textbf{H}\hspace{6pt} \textbf{B}\right)  \left(\begin{matrix}
													\textbf{Y}_{1}\\\textbf{d}
													\end{matrix} \right) - \bm{\kappa}^{\prime}\psi\left\lbrace\left(\textbf{H}\hspace{6pt}\textbf{B}\right)\left(\begin{matrix}
													\textbf{Y}_{1}\\\textbf{d}
													\end{matrix} \right)  - \textbf{V}^{-1}\bm{\mu}\right\rbrace\right],\\
													&\propto\hspace{5pt}\mathrm{exp}\left\lbrace\bm{\alpha}^{\prime}\textbf{H}\textbf{Y}_{1} - \bm{\kappa}^{\prime}\psi\left(\textbf{H}\textbf{Y}_{1}+\textbf{B}\textbf{d}  - \textbf{V}^{-1}\bm{\mu}\right)\right\rbrace,\\
													&=\hspace{5pt}\mathrm{exp}\left\lbrace\bm{\alpha}^{\prime}\textbf{H}\textbf{Y}_{1} - \bm{\kappa}^{\prime}\psi\left(\textbf{H}\textbf{Y}_{1}-\bm{\mu}^{*}\right)\right\rbrace;\hspace{5pt}\textbf{Y}_{1} \in \mathbb{R}^{n},
													\end{align*}
													\noindent
													which proves the result. The normalizing constant can be found using a change of variables
													\begin{align}
													M &= \frac{\mathrm{det}(\textbf{V}^{-1})\left\lbrace\prod_{i = 1}^{n}K(\alpha_{i},\kappa_{i})\right\rbrace\mathrm{exp}\left(\bm{\alpha}^{\prime}\textbf{B}\textbf{d} - \bm{\alpha}^{\prime}\textbf{V}^{-1}\bm{\mu}\right)}{\left[\int f(\textbf{Y}\vert \bm{\mu},\textbf{V},\bm{\alpha},\bm{\kappa})d\textbf{Y}_{1}\right]_{\textbf{Y}_{2} = \textbf{d}}}.
													\end{align}\\
													\noindent
													Although we do not find the expression of the integral $\left[\int f(\textbf{Y}\vert \bm{\mu},\textbf{V},\bm{\alpha},\bm{\kappa})d\textbf{Y}_{1}\right]_{\textbf{Y}_{2} = \textbf{d}}$, and consequently $M$, we know that $M$ is non-zero and finite. To see this, let $\mathcal{N}_{1} = \{\textbf{Y}_{2}: \left[\int f(\textbf{Y}\vert \bm{\mu},\textbf{V},\bm{\alpha},\bm{\kappa})d\textbf{Y}_{1}\right]_{\textbf{Y}_{2}} = 0\}$; then, by the definition of the CM distribution for $\by \in \mathcal{M}^{n}$ and $\textbf{Y}_{2} \in \mathcal{N}_{1}$
													\begin{align*}
													f\left(  \left[
													\begin{array}{c}
													\textbf{Y}_{1} \\ 
													\textbf{Y}_{2}
													\end{array}\right]\vert \bm{\mu},\textbf{V},\bm{\alpha},\bm{\kappa}\right)>0.
													\end{align*}
													Taking the integral with respect to $\textbf{Y}_{1}$ on both sides of the inequality gives $0>0$, which is a false statement. Thus, we have that $\left[\int f(\textbf{Y}\vert \bm{\mu},\textbf{V},\bm{\alpha},\bm{\kappa})d\textbf{Y}_{1}\right]_{\textbf{Y}_{2} = \textbf{d}}$ is non-zero, and hence, $M$ is finite. Similarly, let $\mathcal{N}_{2} = \{\textbf{Y}_{2}: \left[\int f(\textbf{Y}\vert \bm{\mu},\textbf{V},\bm{\alpha},\bm{\kappa})d\textbf{Y}_{1}\right]_{\textbf{Y}_{2}} = \infty\}$ be non-empty, and let $\mathcal{N}_{2}^{c}$ denote the set complement of $\mathcal{N}_{2}$. Then, if $\textbf{w} \sim \mathrm{CM}(\bm{0}_{n,1}, \textbf{I}_{n},\bm{\alpha},\bm{\kappa})$, a change of variables within the integral (see Proposition 1) gives,
													\begin{align*}
													1 &= \int f\left(\textbf{w}\vert \bm{\mu} = \bm{0}_{n,1},\textbf{V} = \textbf{I}_{n},\bm{\alpha},\bm{\kappa}\right)d\textbf{w} = \int f\left(\textbf{Y}\vert \bm{\mu},\textbf{V},\bm{\alpha},\bm{\kappa}\right)d\textbf{Y}= \int\int f(\textbf{Y}\vert \bm{\mu},\textbf{V},\bm{\alpha},\bm{\kappa})d\textbf{Y}_{1}d\textbf{Y}_{2}\\
													&= \int_{\mathcal{N}_{2} }\int f(\textbf{Y}\vert \bm{\mu},\textbf{V},\bm{\alpha},\bm{\kappa})d\textbf{Y}_{1}d\textbf{Y}_{2} + \int_{\mathcal{N}_{2} ^{c}}\int f(\textbf{Y}\vert \bm{\mu},\textbf{V},\bm{\alpha},\bm{\kappa})d\textbf{Y}_{1}d\textbf{Y}_{2}\\
													&=\infty,
													\end{align*}
													\noindent
													which is a contradiction. Thus, we have that the conditional distribution of $\textbf{Y}_{1}\vert \textbf{Y}_{2},\bm{\mu},\textbf{V},\bm{\alpha},\bm{\kappa}$ is proper.
													
													\paragraph{\large{Proof of Theorem 3:}}
													\normalsize
													Consider the transformation $Q = \left(\frac{\psi^{''}(0)}{\psi^{'}(0)}\right)^{1/2}\alpha^{1/2}W$, where $W$ follows an unnormalized $\mathrm{DY}\left(\alpha,\frac{\alpha}{\psi^{\prime}(0)};\hspace{2pt}\psi\right)$. Then we have that
													\begin{equation*}
													f(Q\vert \alpha,\kappa)\propto \mathrm{exp}\left[\left(\frac{\psi^{'}(0)}{\psi^{''}(0)}\right)^{1/2}\alpha^{1/2}Q - \frac{\alpha}{\psi^{\prime}(0)}\hspace{2pt}\psi\left\lbrace\left(\frac{\psi^{'}(0)}{\psi^{''}(0)}\right)^{1/2}\alpha^{-1/2}Q\right\rbrace\right],
													\end{equation*}
													\noindent
													and using the Taylor Series expansion of $\psi(x)$ we have
													\begin{align*}
													& f(Q\vert \alpha,\kappa)\\
													&\propto \mathrm{exp}\left[\left(\frac{\psi^{'}(0)}{\psi^{''}(0)}\right)^{1/2}\alpha^{1/2}Q\right.\\
													& \left. - \frac{\alpha}{\psi^{\prime}(0)}\hspace{2pt}\left\lbrace\psi^{'}(0)\left(\frac{\psi^{'}(0)}{\psi^{''}(0)}\right)^{1/2}\alpha^{-1/2}Q + \psi^{''}(0)\left(\frac{\psi^{'}(0)}{\psi^{''}(0)}\right)\alpha^{-1} \frac{Q^{2}}{2} + {O}\left(\frac{\psi^{'}(0)^{3/2}}{\psi^{''}(0)^{3/2}}\alpha^{-3/2} Q^{3}\right)\right\rbrace\right],
													\end{align*}
													\noindent
													where ``$O(\cdot)$'' is the ``Big-O'' notation \citep[e.g., see][among others]{lehman}. Then, letting $\alpha$ go to infinity yields,
													\begin{align*}
													& \underset{\alpha \rightarrow \infty}{\mathrm{lim}}f(Q\vert \alpha,\kappa)\propto \mathrm{exp}\left(-\frac{Q^{2}}{2}\right) \propto \mathrm{Normal}(0,1).
													\end{align*}
													\noindent
													Thus, $Q$ converges in distribution to a standard normal distribution as $\alpha$ goes to infinity. Now suppose $\textbf{w} = (w_{1},\ldots.,w_{n})^{\prime}$ follows an unnormalized $\mathrm{CM}\left(\bm{0}_{n},\alpha^{1/2}\textbf{I}_{n}, \alpha \textbf{J}_{n,1}, \frac{\alpha}{\psi^{'}(0)}\textbf{J}_{n,1};\hspace{2pt}\psi\right)$. Then it follows from the result above that $\left(\frac{\alpha}{\psi^{'}(0)}\right)^{1/2}\textbf{w}$ converges to a standard multivariate Gaussian distribution. Now, define the transformation $\textbf{Y} = \bm{\mu}+\textbf{V}(\alpha^{1/2}\textbf{w})$. It follows from Theorem~5.1.8 of \citet{lehman}, and the fact that  $\frac{\alpha^{1/2}}{\psi^{'}(0)}\textbf{w}$ converges to a standard Gaussian distribution, that $\textbf{Y}$ converges in distribution to a multivariate normal distribution with mean $\bm{\mu}$ and covariance matrix $\textbf{V}\textbf{V}^{\prime}$.
													\vspace{-10pt}
													\noindent
													{\paragraph{\large{Proof of Theorem 4:}} In the main-text we stated that the $\mathrm{CM}$ distribution is Kolmogorov consistent. We now prove that result. To prove Kolmogorov consistency we need to show the following:
														\begin{enumerate}
															\item For any finite set $\{1,...,n\}$ and for a generic permutation $\{i_{1},...,i_{n}\}$, we have \\$f\left\lbrace \left(Y_{i_{1}},...,Y_{i_{n}}\right)^{\prime}\vert \textbf{c},\textbf{V},\bm{\alpha},\bm{\kappa}\right\rbrace=f\left\lbrace \left(Y_{1},...,Y_{n}\right)^{\prime}\vert \textbf{c}, \textbf{V},\bm{\alpha},\bm{\kappa}\right\rbrace$.
															\item Let $\{j_{1},\ldots, j_{n}\}$ be a generic permutation of $ \{1,...,n\}$ and let $m <n$. Then we have that the marginal density $f\left(Y_{j_{1}},...,Y_{j_{m}}\vert \textbf{c},\textbf{V}, \bm{\alpha},\bm{\kappa}\right)=\int_{\mathcal{M}} \ldots \int_{\mathcal{M}} f\left(Y_{1},...,Y_{n}\vert \textbf{c},\textbf{V},\bm{\alpha},\bm{\kappa}\right) dY_{j_{m+1}}\ldots d Y_{j_{n}}$ exists.
															\end{enumerate}
															
															\noindent    
															Note that the conditions of the Kolmogorov extension theorem do not require that probability density functions exist. However, from Proposition 1($i$), we have an expression of the pdf of $\textbf{Y}$, which will be useful in our proof; hence, we can simplify the conditions of the Kolmogorov extension theorem to the setting where the joint probability density function exists.
															
															For Item 1, define a $n\times n$ permutation matrix $\bm{\Pi}$ such that $\left(Y_{i_{1}},...,Y_{i_{n}}\right)^{\prime}\equiv \textbf{Y}_{\pi}=\bm{\Pi}\textbf{Y}$. Recall that permutation matrices have the following properties:  $\bm{\Pi}\bm{\Pi}^{\prime}=\bm{\Pi}^{\prime}\bm{\Pi}=\textbf{I}_{n}$ and $\bm{\Pi}^{-1} = \bm{\Pi}^{\prime}$. From Equation (7) of the main text we have that,
															\begin{equation}
															\textbf{Y}_{\pi}=\bm{\Pi}\bm{c}+\bm{\Pi}\textbf{V}\textbf{w},
															\label{mulgammaform}
															\end{equation} 
															where $\textbf{w}$ consist of mutually independent DY random variables with respective shape and scale parameters organized into the $n$-dimensional vectors $\bm{\alpha}$ and $\bm{\kappa}$. 
															
															From Proposition 1($i$),
															\begin{align*}
															&f(\textbf{Y}_{\pi}|\bm{c},\bm{V},\bm{\alpha},\bm{\kappa})\\
															&=\mathrm{det}(\textbf{V}^{-1})\left(\prod_{i=1}^n K(\alpha_{i},\kappa_{i})\right)\exp[\bm{\alpha}'\bm{V}^{-1}\bm{\Pi}^{\prime}(\textbf{Y}_{\pi}-\bm{\Pi}\textbf{c})-\bm{\kappa}^{\prime}\psi\{\bm{V}^{-1}\bm{\Pi}^{\prime}(\textbf{Y}_{\pi}-\bm{\Pi}\textbf{c})\}]\\
															&=f(\textbf{Y}|\bm{c},\bm{V},\bm{\alpha},\bm{\kappa}),
															\end{align*}
															\noindent
															where the last equality holds since $\bm{\Pi}^{\prime}\bm{\Pi} = \textbf{I}_{n}$ and $\bm{\Pi}^{\prime}\textbf{Y}_{\pi} = \bm{\Pi}^{\prime}\bm{\Pi}\textbf{Y} = \textbf{Y}$. Thus, permutation holds.
															
															
															We now need to show that the marginal distribution stays the same regardless of what the ``extended'' proper joint distribution is defined as. Without loss of generality (due to Item 1) set $\textbf{P}_{m}^{\prime} = [\textbf{I}_{m}, \bm{0}_{m,n-m}]$ where $\bm{0}_{m,n-m}$ is a $m\times (n-m)$ matrix of zeros. Then define  $\textbf{V} = [\textbf{M}, \textbf{C}]^{\prime}$, $\textbf{M}^{\prime}$ to be a $m\times n$ is a real-valued matrix, $\textbf{C}$ to be any $n\times (n-m)$ real-valued matrix such that $\textbf{V}$ is invertible, $\textbf{Y} \in \mathbb{R}^{n}$, $\textbf{Y} = \textbf{c}+\textbf{V}\textbf{w} = (\textbf{Y}_{1}^{\prime}, \textbf{Y}_{2}^{\prime})^{\prime}$, $\textbf{Y}_{1}^{\prime}$ is $m$-dimensional, and $\textbf{Y}_{2}$ is $(n-m)$-dimensional.
															
															The joint distribution is determined by $\textbf{V} = [\textbf{M},\textbf{C}]$, \textbf{c}, $\bm{\alpha}$, and $\bm{\kappa}$. Thus, we need to show that joint probability density functions with different values of $\textbf{C}$ and $\textbf{c}$ results in the \textit{same} marginal probability density function upon integrating the joint probability density function. Let $\textbf{C}_{1}$ denote a generic real-valued matrix such that  $\textbf{V}_{1} = [\textbf{M}, \textbf{C}_{1}]^{\prime}$ is invertible and $\textbf{C} \ne \textbf{C}_{1}$. Let $\textbf{c}_{1} \in \mathbb{R}^{n}$. Define $\textbf{Y}^{(1)} = \textbf{c}_{1}+ \textbf{V}_{1}\textbf{w} = (\textbf{Y}_{1}^{(1)\prime}, \textbf{Y}_{2}^{(1)\prime})^{\prime}$, where $\textbf{Y}_{1}^{(1)}$ is $m$-dimensional, and $\textbf{Y}_{2}^{(1)}$ is $(n-m)$-dimensional. Then we have that 
															\begin{equation}\label{sigint}
															f(\textbf{Y}_{1}^{(1)}\vert \textbf{c}_{1},\textbf{V}_{1}, \bm{\alpha}, \bm{\kappa}) = \int f(\textbf{Y}^{(1)}\vert \textbf{c}_{1},\textbf{V}_{1}, \bm{\alpha},\bm{\kappa}) d\textbf{q}_{2}^{(1)},
															\end{equation}
															\noindent
															and a change of variables $\textbf{Y}=\textbf{V}\textbf{V}_{1}^{-1}\textbf{Y}^{(1)}-\textbf{V}\textbf{V}_{1}^{-1}\textbf{c}_{1}+\textbf{c}$ within (\ref{sigint}) gives,
															\begin{align*}
															&f(\textbf{Y}_{1}^{(1)}\vert \textbf{c}_{1}, \textbf{V}_{1}, \bm{\alpha},\bm{\kappa}) = \int f(\textbf{Y}^{(1)}\vert \textbf{c}_{1}, \textbf{V}_{1}, \bm{\alpha},\bm{\kappa}) d\textbf{Y}_{2}^{(1)}= \int f(\textbf{Y}\vert\textbf{c}, \textbf{V}, \bm{\alpha},\bm{\kappa}) d\textbf{Y}_{2}\\
															& = f(\textbf{Y}_{1}\vert \textbf{c},\textbf{V}, \bm{\alpha}, \bm{\kappa}).
															\end{align*}
															\noindent
															This completes the proof.
															}
															\noindent
															{
																\paragraph{\large{Proof of Theorem 5:}}
																\normalsize
																The distribution of $\textbf{q}$ is equal to $\mathrm{CM}_{c}(\textbf{c} = -\textbf{B}\textbf{q}_{2}+\bm{\mu}, \textbf{V} = (\textbf{H},\textbf{B})^{-1}, \bm{\alpha}, \bm{\kappa})$ $h(\textbf{q}_{2}\vert \bm{\mu},\textbf{V} = (\textbf{H},\textbf{B})^{-1}, \bm{\alpha}, \bm{\kappa})$, where recall we have reparameterized $\textbf{c} = -\textbf{B}\textbf{q}_{2}+\bm{\mu}$ and $f(\textbf{q}_{2}\vert \bm{\mu},\textbf{V} = (\textbf{H},\textbf{B})^{-1}, \bm{\alpha}, \bm{\kappa})\propto 1$. Thus,
																\begin{align*}
																f(\textbf{q}_{1},\textbf{q}_{2}\vert\bm{\mu}, \textbf{V},\bm{\alpha},\bm{\kappa})&\propto \mathrm{exp}\left\lbrace \bm{\alpha}^{\prime}\textbf{H}\textbf{q}_{1} + \bm{\alpha}^{\prime}\textbf{B}\textbf{q}_{2} - \bm{\alpha}^{\prime}\bm{\mu} -\bm{\kappa}^{\prime}\psi\left(\textbf{H}\textbf{q}_{1} + \textbf{B}\textbf{q}_{2} -\bm{\mu}\right)\right\rbrace \\
																&=\mathrm{exp}\left\lbrace \bm{\alpha}^{\prime}\textbf{V}^{-1}(\textbf{q}- \textbf{V}\bm{\mu}) -\bm{\kappa}^{\prime}\psi\left(\textbf{V}^{-1}(\textbf{q} - \textbf{V}\bm{\mu}\right)\right\rbrace.
																\end{align*}
																Integrating out $\textbf{q}_{2}$ we obtain,
																\begin{align}\label{int}
																f(\textbf{q}_{1}\vert\bm{\mu}, \textbf{V},\bm{\alpha},\bm{\kappa})\propto
																&\int \mathrm{exp}\left\lbrace \bm{\alpha}^{\prime}\textbf{V}^{-1}(\textbf{q} - \textbf{V}\bm{\mu}) -\bm{\kappa}^{\prime}\psi\left(\textbf{V}^{-1}(\textbf{q} - \textbf{V}\bm{\mu}\right)\right\rbrace d\textbf{q}_{2}.
																\end{align}
																Thus, $\textbf{q}_{1}$ is the marginal random vector associated with $\mathrm{CM}( \textbf{V}\bm{\mu}, \textbf{V} = (\textbf{H},\textbf{B})^{-1}, \bm{\alpha}, \bm{\kappa})$. Thus, we are left to show that $\textbf{q}_{1}= (\textbf{H}^{\prime}\textbf{H})^{-1}\textbf{H}^{\prime}\textbf{w}$ is a sample from this marginal distribution.
																
																Denote the QR decomposition of $\textbf{H} = \textbf{Q}\textbf{R}$,
																where the $M\times r$ matrix $\textbf{Q}$ satisfies $\textbf{Q}^{\prime}\textbf{Q} = \textbf{I}_{r}$ and $\textbf{R}$ is a $r\times r$ upper triangular matrix. Now recall the definition of the $M\times (M-r)$ matrix $\textbf{B}$, which satisfies $\textbf{B}^{\prime}\textbf{B} = \textbf{I}_{M-r}$ and $\textbf{B}^{\prime}\textbf{Q} = \bm{0}_{M-r,r}$. Then $\textbf{V}^{-1}$ can be written as
																\begin{align}\label{special_precision}
																& \textbf{V}^{-1}=\left[
																\begin{array}{cc}
																\textbf{Q} & \textbf{B}
																\end{array}\right]
																\left[
																\begin{array}{cc}
																\textbf{R} & \bm{0}_{r,M-r} \\ 
																\bm{0}_{M-r,r} & \textbf{I}_{M-r},
																\end{array}\right].\\
																\nonumber
																\end{align}
																\noindent
																It follows that
																\begin{equation*}
																\textbf{V}=
																\left[
																\begin{array}{cc}
																\textbf{R}^{-1} & \bm{0}_{r,M-r} \\ 
																\bm{0}_{M-r,r} & \textbf{I}_{M-r},
																\end{array}\right]\left[
																\begin{array}{c}
																\textbf{Q}^{\prime} \\ 
																\textbf{B}^{\prime},
																\end{array}\right]=
																\left[\begin{array}{c}
																(\textbf{H}^{*\prime}\textbf{H}^{*})^{-1}\textbf{H}^{*\prime} \\ 
																\textbf{B}^{\prime}
																\end{array}\right],
																\end{equation*}
																\noindent
																where the last equality in the above can be verified by substituting $\textbf{H} = \textbf{Q}\textbf{R}$ into $(\textbf{H}^{\prime}\textbf{H})^{-1}\textbf{H}^{\prime}$.Then, $\textbf{q}$ is distributed according to $\mathrm{CM}(\textbf{V}\bm{\mu}, \textbf{V}= (\textbf{H},\textbf{B})^{-1}, \bm{\alpha}, \bm{\kappa})$ and can be written as 
																\begin{equation}\label{margstep1}
																\left[
																\begin{array}{c}
																\textbf{q}_{1} \\ 
																\textbf{q}_{2}
																\end{array}\right] =    \left[
																\begin{array}{c}
																(\textbf{H}^{\prime}\textbf{H})^{-1}\textbf{H}^{\prime}\textbf{w} \\ 
																\textbf{B}^{\prime}\textbf{w}
																\end{array}\right],
																\end{equation}
																\noindent
																where the $n$-dimensional random vector $\textbf{w}$ is distributed according to $\mathrm{CM}(\bm{\mu}, \textbf{V} =\textbf{I}_{M}, \bm{\alpha}, \bm{\kappa})$. Multiplying both sides of (\ref{margstep1}) by $[\textbf{I}_{r},\bm{0}_{r,M-r}]$ we have
																\begin{equation}\label{marg}
																\textbf{q}_{1} = (\textbf{H}^{\prime}\textbf{H})^{-1}\textbf{H}^{\prime}\textbf{w},
																\end{equation}
																\noindent
																and hence the distribution associated with $(\textbf{H}^{\prime}\textbf{H})^{-1}\textbf{H}^{\prime}\textbf{w}$ is the marginal distribution associated with $\mathrm{CM}( \textbf{V}\bm{\mu}, \textbf{V} = (\textbf{H},\textbf{B})^{-1}, \bm{\alpha}, \bm{\kappa})$ as desired.\\
																}
																
																\section*{Appendix C: The Collapsed Gibbs Sampler}
																\renewcommand{\theequation}{C.\arabic{equation}}
																\setcounter{equation}{0}
																Adding a small number to the data to avoid zero counts changes the priors in the LCM stated in Section 5, and results in a considerable amount of bookkeeping. In Appendices C.i and C.ii, we give these technical details. While the model structure is complicated, it's implementation is computationally straightforward. In Appendix C.iii, we outline the steps involved for the collapsed Gibbs sampler for the model in Appendix C.i.
																
																\subsection*{Appendix C.i: Adding a Small Number to Zero Counts}
																
																The version of the LCM model that allows for zero counts, can be written as the product of the following conditional and marginal distributions:
																\begin{align*}
																\nonumber
																&\mathrm{Data\hspace{5pt}Model:}\hspace{5pt} Z_{i}\vert \bfbeta,\bm{\eta}, \xi_{i},b \ind \mathrm{EF}\left(\textbf{x}_{i}^{\prime}\bm{\beta} + \bm{\phi}_{i}^{\prime}\bm{\eta} + \xi_{i}+\textbf{b}_{\beta,i}^{\prime}\textbf{q}_{\beta}+\textbf{b}_{\eta,i}^{\prime}\textbf{q}_{\eta}+\textbf{b}_{\xi,i}^{\prime}\textbf{q}_{\xi};\hspace{2pt} \psi_{j} \right)\zeta_{\beta}(\textbf{q}_{\beta})\zeta_{\eta}(\textbf{q}_{\eta})\zeta_{\xi}(\textbf{q}_{\xi});\\
																\nonumber
																&\mathrm{Process\hspace{5pt}Model\hspace{5pt}1:}\hspace{5pt} \bm{\eta}\vert \textbf{V}, \alpha_{\eta},\kappa_{\eta}\sim \mathrm{{CM_{c}}}\left(-\textbf{B}_{\eta}\textbf{q}_{\eta}, {\textbf{M}},\bm{\alpha}_{\eta},\bm{\kappa}_{\eta};\hspace{2pt} \psi_{k}\right); \hspace{15pt}\\
																\nonumber
																&\mathrm{Process\hspace{5pt}Model\hspace{5pt}2:}\hspace{5pt} \bm{\xi}\vert \bm{\alpha}_{\xi}, \bm{\kappa}_{\xi} \sim \mathrm{{CM_{c}}}\left(-\textbf{B}_{\xi}\textbf{q}_{\xi}, {\textbf{M}}_{\xi},\bm{\alpha}_{\xi},\bm{\kappa}_{\xi};\hspace{2pt} \psi_{k}\right);\\
																\nonumber
																&{\mathrm{Parameter\hspace{5pt}Model\hspace{5pt}1:}\hspace{5pt} b\vert \alpha_{b}, \kappa_{b}\sim \mathrm{{CM}}\left(0,1,{\alpha}_{b},{\kappa}_{b};\hspace{2pt} \psi_{k}\right)I(b>0)}\\
																\nonumber
																&\mathrm{Parameter\hspace{5pt}Model\hspace{5pt}2:}\hspace{5pt} \bm{\beta}\vert \alpha_{\beta}, \kappa_{\beta}\sim \mathrm{{CM_{c}}}\left(-\textbf{B}_{\beta}\textbf{q}_{\beta}, {\textbf{M}}_{\beta},\bm{\alpha}_{\beta},\bm{\kappa}_{\beta};\hspace{2pt} \psi_{k}\right)\\
																\nonumber
																&{\mathrm{Parameter\hspace{5pt}Model\hspace{5pt}3:}\hspace{5pt} c\vert \alpha_{c}, \kappa_{c}\sim \mathrm{CM}\left(0, 1,\alpha_{c},\kappa_{c};\hspace{2pt} \psi_{k}\right);} \\
																\nonumber
																&{\mathrm{Parameter\hspace{5pt}Model\hspace{5pt}4:}\hspace{5pt} c_{\xi}\vert \alpha_{c}, \kappa_{c}\sim \mathrm{CM}\left(0, 1,\alpha_{c},\kappa_{c};\hspace{2pt} \psi_{k}\right);}\\
																&{\mathrm{Parameter\hspace{5pt}Model\hspace{5pt}5:}\hspace{5pt} c_{\beta}\vert \alpha_{c}, \kappa_{c}\sim \mathrm{CM}\left(0, 1,\alpha_{c},\kappa_{c};\hspace{2pt} \psi_{k}\right);}\\
																\nonumber
																&\mathrm{Parameter\hspace{5pt}Model\hspace{5pt}6:}\hspace{5pt} \textbf{v}_{i}\ind \mathrm{CM}(\bm{0},\sigma_{v}\textbf{I}_{i-1},\alpha_{v}\textbf{J}_{i-1,1},\kappa_{v}\textbf{J}_{i-1,1};\hspace{2pt}\psi_{k});\hspace{2pt} i = 2,\ldots, r,k = 1,2,3,4;\\
																\nonumber
																&{\mathrm{Parameter\hspace{5pt}Model\hspace{5pt}7:}\hspace{5pt}f(\alpha_{\beta},\kappa_{\beta}\vert \gamma_{\beta,1}, \gamma_{\beta,2}, \rho_{\beta})\propto \mathrm{exp}\left[ {\gamma}_{\beta,1}\alpha_{\beta}+ \gamma_{\beta,2}\kappa_{\eta} - {\rho}_{\beta}\mathrm{log}\left\lbrace \frac{1}{K\left(\alpha_{\beta}, \kappa_{\eta}\right)}\right\rbrace\right];}\\
																\nonumber
																&\mathrm{Parameter\hspace{5pt}Model\hspace{5pt}8:}\hspace{5pt}f({\alpha_{\eta},\kappa_{\eta}}\vert \gamma_{\eta,1}, \gamma_{\eta,2}, \rho_{\eta}) \propto \mathrm{exp}\left[ {\gamma}_{\eta,1}\alpha_{\eta,m}+ \gamma_{\eta,2}\kappa_{\eta,m} - {\rho}_{\eta}\mathrm{log}\left\lbrace \frac{1}{K\left(\alpha_{\eta,m}, \kappa_{\eta,m}\right)}\right\rbrace\right];\\
																\nonumber
																&\mathrm{Parameter\hspace{5pt}Model\hspace{5pt}9:}\hspace{5pt}f(\alpha_{\xi},\kappa_{\xi}\vert \gamma_{\xi,1}, \gamma_{\xi,2}, \rho_{\xi}) \propto \mathrm{exp}\left[ {\gamma}_{\xi,1}\alpha_{\xi}+ \gamma_{\xi,2}\kappa_{\xi} - {\rho}_{\xi}\mathrm{log}\left\lbrace \frac{1}{K\left(\alpha_{\xi}, \kappa_{\xi}\right)}\right\rbrace\right];\\
																\nonumber
																&{\mathrm{Parameter\hspace{5pt}Model\hspace{5pt}10:}\hspace{5pt}f(\alpha_{v},\kappa_{v}\vert \gamma_{v,1}, \gamma_{v,2}, \rho_{v})\propto \mathrm{exp}\left[ {\gamma}_{\beta,1}\alpha_{\beta}+ \gamma_{\beta,2}\kappa_{\eta} - {\rho}_{\beta}\mathrm{log}\left\lbrace \frac{1}{K\left(\alpha_{\beta}, \kappa_{\beta}\right)}\right\rbrace\right];}\\
																&{\mathrm{Parameter\hspace{5pt}Model\hspace{5pt}11:}\hspace{5pt} f(\textbf{q}_{\beta})=1;}\\
																&{\mathrm{Parameter\hspace{5pt}Model\hspace{5pt}12:}\hspace{5pt} f(\textbf{q}_{\eta})=1;}\\
																&{\mathrm{Parameter\hspace{5pt}Model\hspace{5pt}13:}\hspace{5pt} f(\textbf{q}_{\xi})=1;}\\
																&{\mathrm{Parameter\hspace{5pt}Model\hspace{5pt}14:}\hspace{5pt} f(\textbf{q}_{v,i})=1;\hspace{15pt} i = 1,\ldots,n, j = 1,2,3,4,},
																\end{align*}
																\noindent
																where $\psi_{j}$ and $\psi_{k}$ (for $j,k = 1,\ldots, 4$) are defined in Table 1 and the elements of $n$-dimensional vector $\bz \equiv \left(Z_{1},\ldots, Z_{n}\right)^{\prime}$ represent data that can be reasonably modeled using a member from the natural exponential family. Additionally for each $i$, $\textbf{x}_{i}$ is a known $p$-dimensional vector of covariates, $\bm{\beta} = (\beta_{1},\ldots, \beta_{p})^{\prime}\in \mathbb{R}^{p}$ is an unknown vector interpreted as fixed effects, $\bm{\phi}_{i}$ is a known $r$-dimensional real-valued vector (see Section~3 for examples), and the $r$-dimensional vector $\bm{\eta} = (\eta_{1},\ldots, \eta_{r})^{\prime}$ and $n$-dimensional vector $\bm{\xi} \equiv \left(\xi_{1},\ldots, \xi_{n}\right)^{\prime}$ are interpreted as real-valued random effects. {The hyperparameters and variance parameters are as follows: define the $(n+p)$-dimensional vector $\bm{\alpha}_{\beta} = (\epsilon_{\alpha},\ldots, \epsilon_{\alpha}, \alpha_{\beta,1},\ldots, \alpha_{\beta,p})^{\prime}$, the $(n+r)$-dimensional vector $\bm{\alpha}_{\eta} = (\epsilon_{\alpha},\ldots, \epsilon_{\alpha}, \alpha_{\eta,1},\ldots, \alpha_{\eta,r})^{\prime}$, the $(2n)$-dimensional vector $\bm{\alpha}_{\xi}= (\epsilon_{\alpha}, \ldots, \epsilon_{\alpha}, \alpha_{\xi,1},\ldots, \alpha_{\xi,n})^{\prime}$, the $(n+p)$-dimensional vector $\bm{\kappa}_{\beta}= (\epsilon_{\kappa,1},\ldots, \epsilon_{\kappa,n}, \kappa_{\beta,1},\ldots, \kappa_{\beta,p})^{\prime}$, the $(n+r)$-dimensional vector $\bm{\kappa}_{\eta}= (\epsilon_{\kappa,1},\ldots, \epsilon_{\kappa,n}, \kappa_{\eta,1},\ldots, \kappa_{\eta,r})^{\prime}$, the $2n$-dimensional vector  $\bm{\kappa}_{\xi}=(\epsilon_{\kappa,1},\ldots, \epsilon_{\kappa,n}, \kappa_{\xi,1},\ldots, \kappa_{\xi,n})^{\prime}$, the $(n+p)\times p$ real-valued matrix $\textbf{M}_{\beta} = (\textbf{X}^{\prime}, \textbf{V}_{\beta}^{\prime})^{\prime}$, the $(n+r)\times r$ real-valued matrix $\textbf{M} = (\bm{\Phi}^{\prime}, \textbf{V}_{\eta}^{\prime})^{\prime}$, the $(2n)\times n$ real-valued matrix $\textbf{M}_{\xi} = (\textbf{I}_{n}, \textbf{V}_{\xi}^{\prime})^{\prime}$, $\textbf{V}_{\beta}\in \mathbb{R}^{p}\times \mathbb{R}^{p}$, $\textbf{V}_{\eta}\in \mathbb{R}^{r}\times \mathbb{R}^{r}$, and $\textbf{V}_{\xi}\in \mathbb{R}^{n}\times \mathbb{R}^{n}$, where to ensure propriety (see Section 2.5) $\alpha_{\beta,i}/\kappa_{\beta,i}\in \mathcal{Y}$, $\alpha_{\eta,j}/\kappa_{\eta,j}\in \mathcal{Y}$, $\alpha_{\xi,k}/\kappa_{\xi,k} \in \mathcal{Y}$, $\kappa_{\beta,i}>0$, $\kappa_{\eta,j}>0$, and $\kappa_{\xi,k}>0$; $i = 1,\ldots, p$, $j = 1,\ldots, r$, $k = 1,\ldots, n$. 
																	
																	We have additionally assumed that $\alpha_{\beta,i}\equiv \alpha_{\beta}$, $\alpha_{\eta,i}\equiv \alpha_{\eta}$, $\alpha_{\xi,i}\equiv \alpha_{\xi}$, $\kappa_{\beta,i}\equiv \kappa_{\beta}$, $\kappa_{\eta,i}\equiv \kappa_{\eta}$, and $\kappa_{\xi,i}\equiv \kappa_{\xi}$. Using Theorem 3 from the main text, we argue that large values of $\alpha_{c}$, $\alpha_{\beta,c}$, $\alpha_{\xi,c}$, $\kappa_{b}$, $\kappa_{c}$, $\kappa_{\beta,c}$, $\kappa_{\xi,c}$,  and $\kappa_{b}$ imply a roughly normal prior on $c$, $c_{\beta}$, and $c_{\xi}$. Also, in our implementation we have assumed that $\textbf{V}_{\beta} = \textbf{I}_{p}$ and $\textbf{V}_{\xi} = \textbf{I}_{n}$, and that $\textbf{V}_{\eta}$ is a lower unit triangular matrix with $i$-th row $\textbf{v}_{i}$. 
																	
																	There are two specifications of the vectors $\textbf{b}_{\beta,i}$, $\textbf{b}_{\eta,i}$, and $\textbf{b}_{\xi,i}$. The first specification involves defining a real-valued $n\times n$ matrix $\textbf{B}_{\beta,1} = (\textbf{b}_{\beta,1}^{\prime},\ldots, \textbf{b}_{\beta,n}^{\prime})^{\prime}$, $n\times n$ matrix $\textbf{B}_{\eta,1} = (\textbf{b}_{\eta,1}^{\prime},\ldots, \textbf{b}_{\eta,n}^{\prime})^{\prime}$, and $n\times n$ matrix $\textbf{B}_{\xi,1} = (\textbf{b}_{\xi,1}^{\prime},\ldots, \textbf{b}_{\xi,n}^{\prime})^{\prime}$. Thus, in this setting $\textbf{q}_{\beta}$ is $n$-dimensional, $\textbf{q}_{\eta}$ is $n$-dimensional, and $\textbf{q}_{\xi}$ is $n$-dimensional. The second specification, increases the row and column dimensions, and involves defining a real-valued $n\times (2n)$ matrix $\textbf{B}_{\beta,1} = (\textbf{b}_{\beta,1}^{\prime},\ldots, \textbf{b}_{\beta,n}^{\prime})^{\prime}$, $n\times (2n)$ matrix $\textbf{B}_{\eta,1} = (\textbf{b}_{\eta,1}^{\prime},\ldots, \textbf{b}_{\eta,n}^{\prime})^{\prime}$, and $n\times (2n)$ matrix $\textbf{B}_{\xi,1} = (\textbf{b}_{\xi,1}^{\prime},\ldots, \textbf{b}_{\xi,n}^{\prime})^{\prime}$. In this setting $\textbf{q}_{\beta}$ is $(2n)$-dimensional, $\textbf{q}_{\eta}$ is $(2n)$-dimensional, and $\textbf{q}_{\xi}$ is $2n$-dimensional. The exact specifications of $\textbf{B}_{\beta,1}$, $\textbf{B}_{\eta,1}$, and $\textbf{B}_{\xi,1}$, will be given in Appendix C.ii. The random vector $\textbf{q}_{v,i}$ is $i$-dimensional.
																	
																	In a similar manner there are two specifications of $\textbf{B}_{\beta}$, $\textbf{B}_{\eta}$, and $\textbf{B}_{\xi}$. In the first setting, $\textbf{B}_{\beta}$ has dimensions $(n+p)\times n$, $\textbf{B}_{\eta}$ has dimension $(n+r)\times n$, and $\textbf{B}_{\xi}$ has dimension $(2n)\times n$. Additionally, the first $n$ rows of $\textbf{B}_{\beta}$, $\textbf{B}_{\eta}$, and $\textbf{B}_{\xi}$ are defined to be $\textbf{B}_{\beta,1}$, $\textbf{B}_{\eta,1}$, and $\textbf{B}_{\xi,1}$, respectively. In the second setting, $\textbf{B}_{\beta}$ has dimensions $(n+p)\times (2n)$, $\textbf{B}_{\eta}$ has dimension $(n+r)\times (2n)$, and $\textbf{B}_{\xi}$ has dimension $(2n)\times (2n)$. The exact specifications of $\textbf{B}_{\beta}$, $\textbf{B}_{\eta}$, and $\textbf{B}_{\xi}$, will be given in Appendix C.ii.
																	
																	The functions $\zeta_{\beta}:\mathbb{R}^{n+p}\rightarrow \mathbb{R}$, $\zeta_{\eta}:\mathbb{R}^{n+r}\rightarrow \mathbb{R}$, and $\zeta_{\xi}:\mathbb{R}^{2n}\rightarrow \mathbb{R}$ are defined in Appendix C.ii, and have the property that $\zeta_{\beta}(\bm{0}_{a,1})=1$, $\zeta_{\eta}(\bm{0}_{a,1})=1$, and $\zeta_{\xi}(\bm{0}_{a,1})=1$, where $a = n$ or $2n$ depending on the specifications of $\textbf{B}_{\beta,1}$, $\textbf{B}_{\eta,1}$, $\textbf{B}_{\xi,1}$, $\textbf{B}_{\beta}$, $\textbf{B}_{\eta}$, and $\textbf{B}_{\xi}$. These functions are needed so that $\epsilon_{\alpha}$ and $\epsilon_{\kappa,i}$ can be introduced and a Collapsed Gibbs sampler, similar to the one outlined in the Pseudo-Code in the main text, can be used. Recall, the values of $\epsilon_{\alpha}>0$ and $\epsilon_{\kappa,i}>0$ are needed to account for the case where $Z_{i}$ is equal to a boundary value on it's support (e.g., a zero Poisson count). {Other solutions to this boundary value problem exist in the Poisson setting \citep{bradleyPMSTM}, however we have found more consistent results using the approach in this paper. We perform inference using samples from the distribution of $\bm{\beta}$, $\bm{\eta}$, and $\bm{\xi}$ given the data $\textbf{Z}$ and the events $\textbf{q}_{\beta} = \bm{0}_{a,1}$, $\textbf{q}_{\eta} = \bm{0}_{a,1}$, $\textbf{q}_{\xi} = \bm{0}_{a,1}$, and $\textbf{q}_{v,i} = \bm{0}_{i,1}$. To simulate from this conditional distribution we implement a collapsed Gibbs sampler similar to the one outlined in Section 2.5 of the main text. The derivation of this collapsed Gibbs sampler is given in Appendix C.ii.
																		
																		\subsection*{Appendix C.ii: Derivation of the Full-Conditional Distributions within a Collapsed Gibbs Sampler}
																		
																		We assume $j = k$ in Appendix C.i and drop the subscript on the log partition function $\psi$. Let the $n\times p$ matrix $\textbf{X} \equiv \left(\textbf{x}_{1},\ldots, \textbf{x}_{n}\right)^{\prime}$, the $n\times r$ matrix $\bm{\Phi}\equiv \left(\bm{\phi}_{1},\ldots, \bm{\phi}_{n}\right)^{\prime}$, and $\underset{\bz}{\propto}$ denotes the ``proportional to as a function of $\bz$'' symbol. It follows that
																		\begin{align}
																		\nonumber
																		&f(\bz\vert \cdot, \textbf{q}_{\beta},\textbf{q}_{\eta} = \bm{0}_{a,1}, \textbf{q}_{\xi} = \bm{0}_{a,1},\textbf{q}_{v,i} = \bm{0}_{i,1}) \\
																		\label{databeta}
																		&\underset{\bm{\beta}}{\propto} \mathrm{exp}\left(\textbf{Z}^{\prime}\textbf{X}\bm{\beta} + \textbf{Z}^{\prime}\textbf{B}_{\beta,1}\textbf{q}_{\beta}- b\textbf{J}_{n,1}^{\prime}\psi\left(\textbf{X}\bm{\beta} + \textbf{B}_{\beta,1}\textbf{q}_{\beta}+ \bm{\Phi}\bm{\eta} + \bm{\xi}\right)\right) \zeta_{\beta}(\textbf{q}_{\beta}) h\\
																		\nonumber
																		&f(\bz\vert \cdot, \textbf{q}_{\beta}= \bm{0}_{a,1},\textbf{q}_{\eta}, \textbf{q}_{\xi} = \bm{0}_{a,1},\textbf{q}_{v,i} = \bm{0}_{i,1}) \\
																		\label{dataeta}
																		&\underset{\bm{\eta}}{\propto} \mathrm{exp}\left(\textbf{Z}^{\prime}\bm{\Phi}\bm{\eta} + \textbf{Z}^{\prime}\textbf{B}_{\eta,1}\textbf{q}_{\eta}- b\textbf{J}_{n,1}^{\prime}\psi\left(\bm{\Phi}\bm{\eta} + \textbf{B}_{\eta,1}\textbf{q}_{\eta}+ \textbf{X}\bm{\beta} +\bm{\xi}\right)\right) \zeta_{\eta}(\textbf{q}_{\eta}) h\\
																		\nonumber
																		&f(\bz\vert \cdot, \textbf{q}_{\beta} = \bm{0}_{a,1},\textbf{q}_{\eta} = \bm{0}_{a,1}, \textbf{q}_{\xi},\textbf{q}_{v,i} = \bm{0}_{i,1}) \\
																		\label{dataxi}
																		&\underset{\bm{\xi}}{\propto} \mathrm{exp}\left(\textbf{Z}^{\prime}\bm{\xi} + \textbf{Z}^{\prime}\textbf{B}_{\xi,1}\textbf{q}_{\xi}- b\textbf{J}_{n,1}^{\prime}\psi\left(\bm{\xi}+ \textbf{B}_{\xi,1}\textbf{q}_{\xi}+ \bm{\Phi}\bm{\eta} + \textbf{X}\bm{\beta}\right)\right) \zeta_{\xi}(\textbf{q}_{\xi}) h,
																		\end{align}
																		where $h = \left\lbrace \prod_{i = 1}^{n}I(\textbf{x}_{i}^{\prime}\bm{\beta} + \bm{\psi}_{i}^{\prime}\bm{\eta}+\xi_{i}+\textbf{b}_{\beta,i}^{\prime}\textbf{q}_{\beta}+\textbf{b}_{\eta,i}^{\prime}\textbf{q}_{\eta}+\textbf{b}_{\xi,i}^{\prime}\textbf{q}_{\xi}\in \mathcal{Y})\right\rbrace$, and $a = n$ or $2n$ depending on the specifications of $\textbf{B}_{\beta,1}$, $\textbf{B}_{\eta,1}$, $\textbf{B}_{\xi,1}$, $\textbf{B}_{\beta}$, $\textbf{B}_{\eta}$, and $\textbf{B}_{\xi}$. We have that 
																		\begin{align}\label{marginalbeta}
																		& f(\bm{\beta},\textbf{q}_{\beta}\vert\textbf{V}_{\beta},\bm{\alpha}_{\beta}, \bm{\kappa}_{\beta},\textbf{q}_{\eta} = \bm{0}_{a,1}, \textbf{q}_{\xi} = \bm{0}_{a,1},\textbf{q}_{v,i} = \bm{0}_{i,1})\\
																		\nonumber
																		&\propto \mathrm{exp}\left\lbrace\bm{\alpha}_{\beta}^{\prime}{\textbf{M}_{\beta}}\bm{\beta} + \bm{\alpha}_{\beta}^{\prime}{\textbf{B}_{\beta}}\textbf{q}_{\beta}-\bm{\kappa}_{\beta}^{\prime}\psi\left({\textbf{M}_{\beta}}\bm{\beta} + {\textbf{B}_{\beta}}\textbf{q}_{\beta}- {c_{\beta}\textbf{J}_{n+p,1}}\right)\right\rbrace,\\
																		\label{marginaleta}
																		& f(\bm{\eta},\textbf{q}_{\eta}\vert\textbf{V}_{\eta},\bm{\alpha}_{\eta}, \bm{\kappa}_{\eta},\textbf{q}_{\beta} = \bm{0}_{a,1}, \textbf{q}_{\xi} = \bm{0}_{a,1},\textbf{q}_{v,i} = \bm{0}_{i,1})\\
																		\nonumber
																		&\propto \mathrm{exp}\left\lbrace\bm{\alpha}_{\eta}^{\prime}{\textbf{M}}\bm{\eta} + \bm{\alpha}_{\eta}^{\prime}{\textbf{B}_{\eta}}\textbf{q}_{\eta}-\bm{\kappa}_{\eta}^{\prime}\psi\left({\textbf{M}}\bm{\eta} + {\textbf{B}_{\eta}}\textbf{q}_{\eta}- {c_{\eta}\textbf{J}_{n+r,1}}\right)\right\rbrace,\\
																		\label{marginalxi}
																		& f(\bm{\xi},\textbf{q}_{\xi}\vert\textbf{V}_{\xi},\bm{\alpha}_{\xi}, \bm{\kappa}_{\xi},\textbf{q}_{\eta} = \bm{0}_{a,1}, \textbf{q}_{\beta} = \bm{0}_{a,1},\textbf{q}_{v,i} = \bm{0}_{i,1})\\
																		\nonumber
																		&\propto \mathrm{exp}\left\lbrace\bm{\alpha}_{\xi}^{\prime}{\textbf{M}_{\xi}}\bm{\xi} + \bm{\alpha}_{\xi}^{\prime}{\textbf{B}_{\xi}}\textbf{q}_{\xi}-\bm{\kappa}_{\xi}^{\prime}\psi\left({\textbf{M}_{\xi}}\bm{\xi} + {\textbf{B}_{\xi}}\textbf{q}_{\xi}- {c_{\xi}\textbf{J}_{2n,1}}\right)\right\rbrace.
																		\end{align}
																		\noindent
																		Using (\ref{databeta}) and (\ref{marginalbeta}) we have that
																		\begin{align*}
																		\nonumber
																		&f(\bm{\beta},\textbf{q}_{\beta}\vert \cdot,\textbf{q}_{\eta} = \bm{0}_{a,1}, \textbf{q}_{\xi} = \bm{0}_{a,1},\textbf{q}_{v,i} = \bm{0}_{i,1}) \underset{\bm{\beta}}{\propto} f(\bz\vert \cdot)f(\bm{\beta}\vert \textbf{V}_{\beta},\bm{\alpha}_{\beta}, \bm{\kappa}_{\beta},c_{\beta})f(\textbf{q}_{\beta})\\
																		\nonumber
																		&\underset{\bm{\beta}}{\propto}  \mathrm{exp}\left\lbrace\bz^{\prime}\textbf{X}\bm{\beta}+\bz^{\prime}\textbf{B}_{\beta,1}\textbf{q}_{\beta}+\bm{\alpha}_{\beta}^{\prime}{\textbf{M}_{\beta}}\bm{\beta}+\bm{\alpha}_{\beta}^{\prime}{\textbf{B}_{\beta}}\textbf{q}_{\beta}\right.\\
																		&\left.-\bm{\kappa}_{\beta}^{\prime}\psi\left({\textbf{M}_{\beta}\bm{\beta}+{\textbf{B}_{\beta}}\textbf{q}_{\beta}  - {c_{\beta}\textbf{J}_{n+p,1}}}\right)- b\textbf{J}_{n,1}^{\prime}\psi\left(\textbf{X}\bm{\beta}+\textbf{B}_{\beta,1}\textbf{q}_{\beta} + \bm{\Phi}\bm{\eta} + \bm{\xi}\right)\right\rbrace \zeta_{\beta}(\textbf{q}_{\beta}) h\\
																		\nonumber
																		&= \mathrm{exp}\left\lbrace\bz^{\prime}\textbf{X}\bm{\beta} + {\epsilon\textbf{J}_{n,1}^{\prime}}\textbf{X}\bm{\beta}+\bm{\alpha}_{\beta{,-\epsilon}}^{\prime}\textbf{V}_{\beta}^{-1}\bm{\beta}\right.\\
																		&\left. -\bm{\kappa}_{\beta}^{\prime}\psi\left(\textbf{M}_{\beta}\bm{\beta}+ \textbf{B}_{\beta}\textbf{q}_{\beta}- {c_{\beta}\textbf{J}_{n+p,1}}\right)- b\textbf{J}_{n,1}^{\prime}\psi\left(\textbf{X}\bm{\beta} + \textbf{B}_{\beta,1}\textbf{q}_{\beta}+ \bm{\Phi}\bm{\eta} + \bm{\xi}\right)\right\rbrace \zeta_{\beta}(\textbf{q}_{\beta}) \omega_{\beta}(\textbf{q}_{\beta}) h\\
																		&\propto \mathrm{CM_{c}}\left\lbrace \bm{\mu}_{\beta}, \textbf{V}_{\beta}^{*}, {\bm{\alpha}_{\beta}^{*}, \bm{\kappa}_{\beta}^{*}};\hspace{2pt}\psi\right\rbrace h,
																		\end{align*}
																		where 
																		\begin{align*}
																		\omega_{\beta}(\textbf{q}_{\beta}) &= \mathrm{exp}(\textbf{Z}^{\prime}\textbf{B}_{\beta,1}\textbf{q}_{\beta} + \bm{\alpha}_{\beta}^{\prime}\textbf{B}_{\beta}\textbf{q}_{\beta})\\
																		\zeta_{\beta}(\textbf{q}_{\beta}) &= \frac{1}{\omega_{\beta}(\textbf{q})}\mathrm{exp}(\bm{\alpha}_{\beta}^{*\prime}\textbf{Q}_{\beta}\textbf{q}_{\beta}),
																		\end{align*}
																		\noindent
																		$\textbf{V}_{\beta}^{*} = (\textbf{H}_{\beta},\textbf{Q}_{\beta})^{-1}$, $\textbf{Q}_{\beta}$ is the null basis for $\textbf{H}_{\beta}$, {$\bm{\alpha}_{\beta,-\epsilon} = \left(\alpha_{\beta,1},\ldots, \alpha_{\beta,p}\right)^{\prime}$, and $\bm{\mu}_{\beta}$, $\textbf{H}_{\beta}$, $\bm{\alpha}_{\beta}^{*}$, and $\bm{\kappa}_{\beta}^{*}$ are defined in Table 2.} 
																		
																		Recall from Appendix C.i there are two specifications of $\textbf{B}_{\beta}$ and $\textbf{B}_{\beta,1}$. When $\textbf{H}_{\beta}$ is $(n+p)\times p$ (as defined in the first and third columns of Table 2), we use the first specification, and let $\textbf{B}_{\beta,1}$ be the first $n$ rows of $\textbf{B}_{\beta}$, and $\textbf{B}_{\beta}$ is set equal to the $(n+p)\times n$ matrix $\textbf{Q}_{\beta}$. When $\textbf{H}_{\beta}$ is $(2n+p)\times p$ (as defined in the second column of Table 2), we use the second specification of $\textbf{B}_{\beta}$ and $\textbf{B}_{\beta,1}$, and let the matrix $(\textbf{B}_{\beta,1}^{\prime}, \textbf{B}_{\beta}^{\prime})^{\prime}$ be set equal to the $(2n+p)\times 2n$ matrix $\textbf{Q}_{\beta}$.
																		
																		In a similar manner to Equations (15) through (17) of the main text, a sample from $f(\bm{\beta}\vert \cdot,\textbf{q}_{\eta} = \bm{0}_{a,1}, \textbf{q}_{\xi} = \bm{0}_{a,1},\textbf{q}_{v,i} = \bm{0}_{i,1})$ can be easily obtained with,
																		\begin{equation}\label{simbeta}
																		\bm{\beta}=(\textbf{H}_{\beta}^{\prime}\textbf{H}_{\beta})^{-1}\textbf{H}_{\beta}^{\prime}\bm{\mu}_{\beta} + (\textbf{H}_{\beta}^{\prime}\textbf{H}_{\beta})^{-1}\textbf{H}_{\beta}^{\prime}\textbf{w},
																		\end{equation}
																		\noindent
																		where $\textbf{w}\sim \mathrm{CM}(\bm{0}_{g,1},\textbf{I}_{g},\bm{\alpha}_{\beta}^{*},\bm{\kappa}_{\beta}^{*})$, $g$ is the number of rows in $\textbf{H}_{\beta}$, and $a = n$ or $2n$ depending on the specifications of $\textbf{B}_{\beta,1}$, $\textbf{B}_{\eta,1}$, $\textbf{B}_{\xi,1}$, $\textbf{B}_{\beta}$, $\textbf{B}_{\eta}$, and $\textbf{B}_{\xi}$.
																		
																		We can find the full conditional distributions associated with $\bm{\eta}$ and $\textbf{q}_{\eta}$, and $\bm{\xi}$ and $\textbf{q}_{\xi}$ in a similar manner. Using (\ref{dataeta}) and (\ref{marginaleta}) we have that
																		\begin{align*}
																		\nonumber
																		&f(\bm{\eta},\textbf{q}_{\eta}\vert \cdot,\textbf{q}_{\beta} = \bm{0}_{a,1}, \textbf{q}_{\xi} = \bm{0}_{a,1},\textbf{q}_{v,i} = \bm{0}_{i,1}) \underset{\bm{\eta}}{\propto} f(\bz\vert \cdot)f(\bm{\eta}\vert \textbf{V}_{\eta},\bm{\alpha}_{\eta}, \bm{\kappa}_{\eta},c_{\eta})f(\textbf{q}_{\eta})\\
																		\nonumber
																		&\underset{\bm{\eta}}{\propto}  \mathrm{exp}\left\lbrace\bz^{\prime}\bm{\Phi}\bm{\eta}+\bz^{\prime}\textbf{B}_{\eta,1}\textbf{q}_{\eta}+\bm{\alpha}_{\eta}^{\prime}{\textbf{M}}\bm{\eta}+\bm{\alpha}_{\eta}^{\prime}{\textbf{B}_{\eta}}\textbf{q}_{\eta}\right.\\
																		&\left.-\bm{\kappa}_{\eta}^{\prime}\psi\left({\textbf{M}\bm{\eta}+{\textbf{B}_{\eta}}\textbf{q}_{\eta}  - {c_{\eta}\textbf{J}_{n+r,1}}}\right)- b\textbf{J}_{n,1}^{\prime}\psi\left(\bm{\Phi}\bm{\eta}+\textbf{B}_{\eta,1}\textbf{q}_{\eta} + \textbf{X}\bm{\beta} + \bm{\xi}\right)\right\rbrace \zeta_{\eta}(\textbf{q}_{\eta}) h\\
																		\nonumber
																		&= \mathrm{exp}\left\lbrace\bz^{\prime}\bm{\Phi}\bm{\eta} + {\epsilon\textbf{J}_{n,1}^{\prime}}\bm{\Phi}\bm{\eta}+\bm{\alpha}_{\eta{,-\epsilon}}^{\prime}\textbf{V}_{\eta}^{-1}\bm{\eta}\right.\\
																		&\left. -\bm{\kappa}_{\eta}^{\prime}\psi\left(\textbf{M}\bm{\eta}+ \textbf{B}_{\eta}\textbf{q}_{\eta}- {c_{\eta}\textbf{J}_{n+r,1}}\right)- b\textbf{J}_{n,1}^{\prime}\psi\left(\bm{\Phi}\bm{\eta} + \textbf{B}_{\eta,1}\textbf{q}_{\eta}+ \textbf{X}\bm{\beta} + \bm{\xi}\right)\right\rbrace \zeta_{\eta}(\textbf{q}_{\eta}) \omega_{\eta}(\textbf{q}_{\eta}) h\\
																		&\propto \mathrm{CM_{c}}\left\lbrace \bm{\mu}_{\beta}, \textbf{V}_{\beta}^{*}, {\bm{\alpha}_{\beta}^{*}, \bm{\kappa}_{\beta}^{*}};\hspace{2pt}\psi\right\rbrace h,
																		\end{align*}
																		where 
																		\begin{align*}
																		\omega_{\eta}(\textbf{q}_{\eta}) &= \mathrm{exp}(\textbf{Z}^{\prime}\textbf{B}_{\eta,1}\textbf{q}_{\eta} + \bm{\alpha}_{\eta}^{\prime}\textbf{B}_{\eta}\textbf{q}_{\eta})\\
																		\zeta_{\eta}(\textbf{q}_{\eta}) &= \frac{1}{\omega_{\eta}(\textbf{q}_{\eta})}\mathrm{exp}(\bm{\alpha}_{\eta}^{*\prime}\textbf{Q}_{\eta}\textbf{q}_{\eta}),
																		\end{align*}
																		\noindent
																		$\textbf{V}_{\eta}^{*} = (\textbf{H}_{\eta},\textbf{Q}_{\eta})^{-1}$, $\textbf{Q}_{\eta}$ is the null basis for $\textbf{H}_{\eta}$, {$\bm{\alpha}_{\eta,-\epsilon} = \left(\alpha_{\eta,1},\ldots, \alpha_{\eta,p}\right)^{\prime}$, and $\bm{\mu}_{\eta}$, $\textbf{H}_{\eta}$, $\bm{\alpha}_{\eta}^{*}$, and $\bm{\kappa}_{\eta}^{*}$ are defined in Table 2.} 
																		
																		Recall from Appendix C.i there are two specifications of $\textbf{B}_{\eta}$ and $\textbf{B}_{\eta,1}$. When $\textbf{H}_{\eta}$ is $(n+r)\times r$ (as defined in the first and third columns of Table 2), we use the first specification, and let $\textbf{B}_{\eta,1}$ be the first $n$ rows of $\textbf{B}_{\eta}$, and $\textbf{B}_{\eta}$ is set equal to the $(n+r)\times n$ matrix $\textbf{Q}_{\eta}$. When $\textbf{H}_{\eta}$ is $(2n+r)\times r$ (as defined in the second column of Table 2), we use the second specification of $\textbf{B}_{\eta}$ and $\textbf{B}_{\eta,1}$, and let the matrix $(\textbf{B}_{\eta,1}^{\prime}, \textbf{B}_{\eta}^{\prime})^{\prime}$ be set equal to the $(2n+r)\times 2n$ matrix $\textbf{Q}_{\eta}$.
																		
																		In a similar manner to Equations (15) through (17) of the main text, a sample from $f(\bm{\eta}\vert \cdot,\textbf{q}_{\beta} = \bm{0}_{a,1}, \textbf{q}_{\xi} = \bm{0}_{a,1})$ can be easily obtained with,
																		\begin{equation}\label{simeta}
																		\bm{\eta}=(\textbf{H}_{\eta}^{\prime}\textbf{H}_{\eta})^{-1}\textbf{H}_{\eta}^{\prime}\bm{\mu}_{\eta} + (\textbf{H}_{\eta}^{\prime}\textbf{H}_{\eta})^{-1}\textbf{H}_{\eta}^{\prime}\textbf{w},
																		\end{equation}
																		\noindent
																		where $\textbf{w}\sim \mathrm{CM}(\bm{0}_{g,1},\textbf{I}_{g},\bm{\alpha}_{\eta}^{*},\bm{\kappa}_{\eta}^{*})$, $g$ is the number of rows in $\textbf{H}_{\eta}$, and $a = n$ or $2n$ depending on the specifications of $\textbf{B}_{\beta,1}$, $\textbf{B}_{\eta,1}$, $\textbf{B}_{\xi,1}$, $\textbf{B}_{\beta}$, $\textbf{B}_{\eta}$, and $\textbf{B}_{\xi}$.
																		
																		Using (\ref{dataxi}) and (\ref{marginalxi}) we have that
																		\begin{align*}
																		\nonumber
																		&f(\bm{\xi},\textbf{q}_{\xi}\vert \cdot,\textbf{q}_{\beta} = \bm{0}_{a,1}, \textbf{q}_{\eta} = \bm{0}_{a,1},\textbf{q}_{v,i} = \bm{0}_{i,1}) \underset{\bm{\xi}}{\propto} f(\bz\vert \cdot)f(\bm{\xi}\vert \textbf{V}_{\xi},\bm{\alpha}_{\xi}, \bm{\kappa}_{\xi},c_{\xi})f(\textbf{q}_{\xi})\\
																		\nonumber
																		&\underset{\bm{\xi}}{\propto}  \mathrm{exp}\left\lbrace\bz^{\prime}\bm{\xi}+\bz^{\prime}\textbf{B}_{\xi,1}\textbf{q}_{\xi}+\bm{\alpha}_{\xi}^{\prime}{\textbf{M}_{\xi}}\bm{\xi}+\bm{\alpha}_{\xi}^{\prime}{\textbf{B}_{\xi}}\textbf{q}_{\xi}\right.\\
																		&\left.-\bm{\kappa}_{\xi}^{\prime}\psi\left({\textbf{M}_{\xi}\bm{\xi}+{\textbf{B}_{\xi}}\textbf{q}_{\xi}  - {c_{\xi}\textbf{J}_{2n,1}}}\right)- b\textbf{J}_{n,1}^{\prime}\psi\left(\bm{\xi}+\bm{\Phi}\bm{\eta}+\textbf{B}_{\xi,1}\textbf{q}_{\xi} + \textbf{X}\bm{\beta}\right)\right\rbrace \zeta_{\xi}(\textbf{q}_{\xi}) h\\
																		\nonumber
																		&= \mathrm{exp}\left\lbrace\bz^{\prime}\bm{\xi} + {\epsilon\textbf{J}_{n,1}^{\prime}}\bm{\xi}+\bm{\alpha}_{\xi{,-\epsilon}}^{\prime}\textbf{V}_{\xi}^{-1}\bm{\xi}\right.\\
																		&\left. -\bm{\kappa}_{\xi}^{\prime}\psi\left(\textbf{M}_{\xi}\bm{\xi}+ \textbf{B}_{\xi}\textbf{q}_{\xi}- {c_{\xi}\textbf{J}_{2n,1}}\right)- b\textbf{J}_{n,1}^{\prime}\psi\left(\bm{\xi}+\bm{\Phi}\bm{\eta} + \textbf{B}_{\xi,1}\textbf{q}_{\xi}+ \textbf{X}\bm{\beta}\right)\right\rbrace \zeta_{\xi}(\textbf{q}_{\xi}) \omega_{\xi}(\textbf{q}_{\xi}) h\\
																		&\propto \mathrm{CM_{c}}\left\lbrace \bm{\mu}_{\xi}, \textbf{V}_{\xi}^{*}, {\bm{\alpha}_{\xi}^{*}, \bm{\kappa}_{\xi}^{*}};\hspace{2pt}\psi\right\rbrace h,
																		\end{align*}
																		where 
																		\begin{align*}
																		\omega_{\xi}(\textbf{q}_{\xi}) &= \mathrm{exp}(\textbf{Z}^{\prime}\textbf{B}_{\xi,1}\textbf{q}_{\xi} + \bm{\alpha}_{\xi}^{\prime}\textbf{B}_{\xi}\textbf{q}_{\xi})\\
																		\zeta_{\xi}(\textbf{q}_{\xi}) &= \frac{1}{\omega_{\xi}(\textbf{q}_{\xi})}\mathrm{exp}(\bm{\alpha}_{\xi}^{*\prime}\textbf{Q}_{\xi}\textbf{q}_{\xi}),
																		\end{align*}
																		\noindent
																		$\textbf{V}_{\xi}^{*} = (\textbf{H}_{\xi},\textbf{Q}_{\xi})^{-1}$, $\textbf{Q}_{\xi}$ is the null basis for $\textbf{H}_{\xi}$, {$\bm{\alpha}_{\xi,-\epsilon} = \left(\alpha_{\xi,1},\ldots, \alpha_{\xi,p}\right)^{\prime}$, and $\bm{\mu}_{\xi}$, $\textbf{H}_{\xi}$, $\bm{\alpha}_{\xi}^{*}$, and $\bm{\kappa}_{\xi}^{*}$ are defined in Table 2.} 
																		
																		Recall from Appendix C.i there are two specifications of $\textbf{B}_{\xi}$ and $\textbf{B}_{\xi,1}$. When $\textbf{H}_{\xi}$ is $(2n)\times n$ (as defined in the first and third columns of Table 2), we use the first specification, and let $\textbf{B}_{\xi,1}$ be the first $n$ rows of $\textbf{B}_{\xi}$, and $\textbf{B}_{\xi}$ is set equal to the $(2n)\times n$ matrix $\textbf{Q}_{\xi}$. When $\textbf{H}_{\xi}$ is $(3n)\times n$ (as defined in the second column of Table 2), we use the second specification of $\textbf{B}_{\xi}$ and $\textbf{B}_{\xi,1}$, and let the matrix $(\textbf{B}_{\xi,1}^{\prime}, \textbf{B}_{\xi}^{\prime})^{\prime}$ be set equal to the $(3n)\times 2n$ matrix $\textbf{Q}_{\xi}$.
																		
																		In a similar manner to Equations (15) through (17) of the main text, a sample from $f(\bm{\xi}\vert \cdot,\textbf{q}_{\beta} = \bm{0}_{a,1}, \textbf{q}_{\xi} = \bm{0}_{a,1})$ can be easily obtained with,
																		\begin{equation}\label{simxi}
																		\bm{\xi}=(\textbf{H}_{\xi}^{\prime}\textbf{H}_{\xi})^{-1}\textbf{H}_{\xi}^{\prime}\bm{\mu}_{\xi} + (\textbf{H}_{\xi}^{\prime}\textbf{H}_{\xi})^{-1}\textbf{H}_{\xi}^{\prime}\textbf{w},
																		\end{equation}
																		\noindent
																		where $\textbf{w}\sim \mathrm{CM}(\bm{0}_{g,1},\textbf{I}_{g},\bm{\alpha}_{\xi}^{*},\bm{\kappa}_{\xi}^{*})$, $g$ is the number of rows in $\textbf{H}_{\xi}$, and $a = n$ or $2n$ depending on the specifications of $\textbf{B}_{\beta,1}$, $\textbf{B}_{\eta,1}$, $\textbf{B}_{\xi,1}$, $\textbf{B}_{\beta}$, $\textbf{B}_{\eta}$, and $\textbf{B}_{\xi}$.
																		
																		If $b$ is unknown (e.g., the negative-binomial distribution) a prior for $b$ is introduced in Appendix C.i. The full-conditional distribution is given by,
																		\begin{align*}
																		& f(b\vert \cdot, \textbf{q}_{\beta} = \bm{0}_{a,1}, \textbf{q}_{\eta} = \bm{0}_{a,1}, \textbf{q}_{\xi} = \bm{0}_{a,1},\textbf{q}_{v,i} = \bm{0}_{i,1}) \\
																		&\propto K_{EF}(b,\textbf{Z}) \mathrm{exp}\left\lbrace\alpha_{b}b - \kappa_{b}\psi(b) - b\textbf{J}_{n,1}^{\prime}\psi(\textbf{X}\bm{\beta}+\bm{\Phi}\bm{\eta}+\bm{\xi})\right\rbrace I\left\lbrace b>\mathrm{max}(\textbf{Z})+1\right\rbrace,
																		\end{align*}
																		\noindent
																		where $K_{EF}(b,\textbf{Z})$ is the normalizing constant associated with the distribution of $\textbf{Z}$, and $\mathrm{max}(\textbf{Z})$ returns the maximum element of the vector $\textbf{Z}$. A slice sampler can be used to generate from this full-conditional distribution.
																		
																		{The full-conditional distribution for $c$ is given by,
																			\begin{align}
																			\nonumber
																			&f(c\vert \cdot, \textbf{q}_{\beta} = \bm{0}_{a,1}, \textbf{q}_{\eta} = \bm{0}_{a,1}, \textbf{q}_{\xi} = \bm{0}_{a,1},\textbf{q}_{v,i} = \bm{0}_{i,1}) \\
																			\nonumber
																			&\underset{c}{\propto} f(c)f(\bm{\eta}\vert \textbf{V},\bm{\alpha}_{\beta}, \bm{\kappa}_{\beta},c)\\
																			\nonumber
																			&\underset{c}{\propto}  \mathrm{exp}\left\lbrace \alpha_{c}c + \bm{\alpha}_{\eta}^{\prime}{\textbf{M}}\bm{\eta}-\bm{\kappa}_{\eta}^{\prime}\psi\left({\textbf{M}_{\eta}}\bm{\eta}-c\textbf{J}_{n+r,1}\right) - \kappa_{c}\psi(c)\right\rbrace I(c \in \mathcal{Y}) h\\
																			\nonumber
																			&\propto \mathrm{CM_{c}}\left\lbrace \bm{\mu}_{c}, \textbf{H}_{c}^{*}, {\bm{\alpha}_{c}^{*}, \bm{\kappa}_{c}^{*}};\hspace{2pt}\psi\right\rbrace  I(c \in \mathcal{Y}),
																			\end{align}
																			\noindent
																			where $\bm{\mu}_{c}$, $\textbf{H}_{c}^{*}$, $\bm{\alpha}_{c}^{*}$, and $\bm{\kappa}_{c}^{*}$ are defined in Table 2. The full-conditional distributions for $c_{\xi}$ and $c_{\beta}$ are found in a similar way. That is, 
																			\begin{align*}
																			f(c_{\beta}\vert \cdot, \textbf{q}_{\beta} = \bm{0}_{a,1}, \textbf{q}_{\eta} = \bm{0}_{a,1}, \textbf{q}_{\xi} = \bm{0}_{a,1},\textbf{q}_{v,i} = \bm{0}_{i,1})&\propto \mathrm{CM_{c}}\left\lbrace \bm{\mu}_{c,\beta}, \textbf{H}_{c,\beta}^{*}, {\bm{\alpha}_{c,\beta}^{*}, \bm{\kappa}_{c,\beta}^{*}};\hspace{2pt}\psi\right\rbrace  I(c_{\beta} \in \mathcal{Y})\\
																			f(c_{\xi}\vert \cdot, \textbf{q}_{\beta} = \bm{0}_{a,1}, \textbf{q}_{\eta} = \bm{0}_{a,1}, \textbf{q}_{\xi} = \bm{0}_{a,1},\textbf{q}_{v,i} = \bm{0}_{i,1})&\propto \mathrm{CM_{c}}\left\lbrace \bm{\mu}_{c,\xi}, \textbf{H}_{c,\xi}^{*}, {\bm{\alpha}_{c,\xi}^{*}, \bm{\kappa}_{c,\xi}^{*}};\hspace{2pt}\psi\right\rbrace  I(c_{\xi} \in \mathcal{Y}),
																			\end{align*}
																			where $\bm{\mu}_{c,\beta}$, $\textbf{H}_{c,\beta}^{*}$, $\bm{\alpha}_{c,\beta}^{*}$, $\bm{\kappa}_{c,\beta}^{*}$, $\bm{\mu}_{c,\xi}$, $\textbf{H}_{c,\xi}^{*}$, $\bm{\alpha}_{c,\xi}^{*}$, and $\bm{\kappa}_{c,\xi}^{*}$ are defined in Table 2. One could use the argument in Appendix A.ii to update $c$, $c_{\beta}$, and $c_{\xi}$. However, it is rather straightforward to update these parameters using the slice sampler.}\\

																			\begin{table}[htp]{{\textbf{Quantities Needed for Gibbs Sampling}}}
																				\centering
																				\noindent\adjustbox{max width=\textwidth}{%
																					\begin{tabular}{ |l|l|l|  }
																						\hline
																						{ \textbf{No Boundary Adjustments}}& \hspace{100pt} {$\psi_{2}$} & \hspace{100pt}{$\psi_{3}$}\\ \hline
																						$\bm{\mu}_{\beta} = (-\bm{\eta}^{\prime}\bm{\Phi}^{\prime} - \bm{\xi}^{\prime},{c_{\beta}\textbf{J}_{1,p}})^{\prime}$&	    	$\bm{\mu}_{\beta} = ({c_{\beta}\textbf{J}_{1,n},}-\bm{\eta}^{\prime}\bm{\Phi}^{\prime} - \bm{\xi}^{\prime},{c_{\beta}\textbf{J}_{1,p}})^{\prime}$&	    	$\bm{\mu}_{\beta} = {\bm{0}_{n+p,1}}$\\ \hline
																						$\bm{\mu}_{\eta} = (-\bm{\beta}^{\prime}\textbf{X}^{\prime} - \bm{\xi}^{\prime},{c\textbf{J}_{1,r}})^{\prime}$ &	$\bm{\mu}_{\eta} = ({c\textbf{J}_{1,n},}-\bm{\beta}^{\prime}\textbf{X}^{\prime} - \bm{\xi}^{\prime},{c\textbf{J}_{1,r}})^{\prime}$ &$\bm{\mu}_{\eta} = {\bm{0}_{n+r,1}}$ \\ \hline
																						$\bm{\mu}_{\xi} = (-\bm{\beta}^{\prime}\textbf{X}^{\prime} -\bm{\eta}^{\prime}\bm{\Phi}^{\prime},{c_{\xi}\textbf{J}_{1,n}})^{\prime}$&	$\bm{\mu}_{\xi} = ({c_{\xi}\textbf{J}_{1,n},}-\bm{\beta}^{\prime}\textbf{X}^{\prime} -\bm{\eta}^{\prime}\bm{\Phi}^{\prime},{c_{\xi}\textbf{J}_{1,n}})^{\prime}$&$\bm{\mu}_{\xi} = {\bm{0}_{2n,1}}$ \\\hline
																						$\bm{\mu}_{\gamma,i} = (\eta_{i},\bm{0}_{1,i-1})^{\prime}; \hspace{10pt}i = 2,\ldots, r$ & $\bm{\mu}_{\gamma,i} = (\eta_{i},\bm{0}_{1,i-1})^{\prime}; \hspace{10pt}i = 2,\ldots, r$ &$\bm{\mu}_{\gamma,i} = (\eta_{i},\bm{0}_{1,i-1})^{\prime}; \hspace{10pt}i = 2,\ldots, r$ \\\hline
																						{$\bm{\mu}_{c} = (\bm{\eta}^{\prime}\bm{\Phi}^{\prime},0)^{\prime}$} & {$\bm{\mu}_{c} = (\bm{\eta}^{\prime}\bm{\Phi}^{\prime},0)^{\prime}$} &{$\bm{\mu}_{c} = (\bm{\eta}^{\prime}\bm{\Phi}^{\prime},0)^{\prime}$} \\\hline
																						{$\bm{\mu}_{c,\xi} = (\bm{\xi}^{\prime},0)^{\prime}$} & {$\bm{\mu}_{c,\xi} = (\bm{\xi}^{\prime},0)^{\prime}$} &{$\bm{\mu}_{c,\xi} = (\bm{\xi}^{\prime},0)^{\prime}$} \\\hline
																						{$\bm{\mu}_{c,\beta} = (\bm{\beta}^{\prime}\textbf{X}^{\prime},0)^{\prime}$} & {$\bm{\mu}_{c,\beta} = (\bm{\beta}^{\prime}\textbf{X}^{\prime},0)^{\prime}$} &{$\bm{\mu}_{c,\beta} = (\bm{\beta}^{\prime}\textbf{X}^{\prime},0)^{\prime}$} \\\hline
																						$\textbf{H}_{\beta} = (\textbf{X}^{\prime},\textbf{V}_{\beta}^{-1\prime})^{\prime}$&	 $\textbf{H}_{\beta} = ({\textbf{X}^{\prime},}\textbf{X}^{\prime},\textbf{V}_{\beta}^{-1\prime})^{\prime}$& $\textbf{H}_{\beta} = (\textbf{X}^{\prime},\textbf{V}_{\beta}^{-1\prime})^{\prime}$ \\ \hline
																						$\textbf{H}_{\eta} = (\bm{\Phi}^{\prime},\textbf{V}_{\eta}^{-1\prime})^{\prime}$		  & $\textbf{H}_{\eta} = ({\bm{\Phi}^{\prime},}\bm{\Phi}^{\prime},\textbf{V}_{\eta}^{-1\prime})^{\prime}$		  &$\textbf{H}_{\eta} = (\bm{\Phi}^{\prime},\textbf{V}_{\eta}^{-1\prime})^{\prime}$		  \\ \hline
																						$\textbf{H}_{\xi} = (\textbf{I}_{n},\textbf{V}_{\xi}^{-1\prime})^{\prime}$		&  $\textbf{H}_{\xi} = ({\textbf{I}_{n},}\textbf{I}_{n},\textbf{V}_{\xi}^{-1\prime})^{\prime}$		&$\textbf{H}_{\xi} = (\textbf{I}_{n},\textbf{V}_{\xi}^{-1\prime})^{\prime}$		\\ \hline
																						$\textbf{H}_{\gamma,i} = \left\lbrace \left(\eta_{1},\ldots, \eta_{i-1},  \right)^{\prime},\textbf{C}_{i}^{\prime}\right\rbrace^{\prime}; \hspace{10pt}i = 2,\ldots, r$  & $\textbf{H}_{\gamma,i} = \left\lbrace \left(\eta_{1},\ldots, \eta_{i-1},  \right)^{\prime},\textbf{C}_{i}^{\prime}\right\rbrace^{\prime}; \hspace{10pt}i = 2,\ldots, r$  & $\textbf{H}_{\gamma,i} = \left\lbrace \left(\eta_{1},\ldots, \eta_{i-1},  \right)^{\prime},\textbf{C}_{i}^{\prime}\right\rbrace^{\prime}; \hspace{10pt}i = 2,\ldots, r$  \\ \hline
																						{$\textbf{H}_{c}^{*} =- \textbf{J}_{n+r,1}$} & {$\textbf{H}_{c}^{*} =-\textbf{J}_{n+r,1}$} & {$\textbf{H}_{c}^{*} =-\textbf{J}_{n+r,1}$}\\        \hline
																						{$\textbf{H}_{c,\xi}^{*} =-\textbf{J}_{2n,1}$} & {$\textbf{H}_{c,\xi}^{*} =-\textbf{J}_{2n,1}$} & {$\textbf{H}_{c,\xi}^{*} =-\textbf{J}_{2n,1}$}\\        \hline
																						{$\textbf{H}_{c,\beta}^{*} =-\textbf{J}_{n+p,1}$} & {$\textbf{H}_{c,\beta}^{*} =-\textbf{J}_{n+p,1}$} & {$\textbf{H}_{c,\beta}^{*} =-\textbf{J}_{n+p,1}$}\\      
																						\hline
																						{$\bm{\alpha}_{\beta}^{*} = (\textbf{Z}^{\prime},{\alpha}_{\beta,1},\ldots, \alpha_{\beta,p})^{\prime}$}	&	$\bm{\alpha}_{\beta}^{*} = ({\frac{1}{2}\textbf{Z}^{\prime} + \frac{\epsilon_{\alpha}}{2}\textbf{J}_{1,n},\frac{1}{2}\textbf{Z}^{\prime} + \frac{\epsilon_{\alpha}}{2}\textbf{J}_{1,n},{\alpha}_{\beta,1},\ldots, \alpha_{\beta,p})^{\prime}}$	&$\bm{\alpha}_{\beta}^{*} = {(\textbf{Z}^{\prime} + \epsilon_{\alpha}\textbf{J}_{1,n},{\alpha}_{\beta,1},\ldots, \alpha_{\beta,p})^{\prime}}$	  \\ \hline
																						{$\bm{\alpha}_{\eta}^{*} = (\textbf{Z}^{\prime},{\alpha}_{\eta,1},\ldots, \alpha_{\eta,r})^{\prime}$}	& $\bm{\alpha}_{\eta}^{*} = ({\frac{1}{2}\textbf{Z}^{\prime} + \frac{\epsilon_{\alpha}}{2}\textbf{J}_{1,n},\frac{1}{2}\textbf{Z}^{\prime} + \frac{\epsilon_{\alpha}}{2}\textbf{J}_{1,n},{\alpha}_{\eta,1},\ldots, \alpha_{\eta,r})^{\prime}}$	&$\bm{\alpha}_{\eta}^{*} ={ (\textbf{Z}^{\prime} + \epsilon_{\alpha}\textbf{J}_{1,n},{\alpha}_{\eta,1},\ldots, \alpha_{\eta,r})^{\prime}}$	\\ \hline
																						{$\bm{\alpha}_{\xi}^{*} = (\textbf{Z}^{\prime},,{\alpha}_{\xi,1},\ldots, \alpha_{\xi,n})^{\prime}$}	&	$\bm{\alpha}_{\xi}^{*} = ({\frac{1}{2}\textbf{Z}^{\prime} + \frac{\epsilon_{\alpha}}{2}\textbf{J}_{1,n},\frac{1}{2}\textbf{Z}^{\prime} + \frac{\epsilon_{\alpha}}{2}\textbf{J}_{1,n},{\alpha}_{\xi,1},\ldots, \alpha_{\xi,n})^{\prime}}$	&${\bm{\alpha}_{\xi}^{*} = (\textbf{Z}^{\prime} + \epsilon_{\alpha}\textbf{J}_{1,n},{\alpha}_{\xi,1},\ldots, \alpha_{\xi,n})^{\prime}}$	  \\ 
																						\hline
																						$\bm{\alpha}_{\gamma,i} = (\alpha_{\eta,i},\bm{\alpha}_{i}^{\prime})^{\prime}$;\hspace{10pt} $i = 2,\ldots, r$ & $\bm{\alpha}_{\gamma,i} = (\alpha_{\eta,i},\bm{\alpha}_{i}^{\prime})^{\prime}$;\hspace{10pt} $i = 2,\ldots, r$ & $\bm{\alpha}_{\gamma,i} = (\alpha_{\eta,i},\bm{\alpha}_{i}^{\prime})^{\prime}$;\hspace{10pt} $i = 2,\ldots, r$ \\\hline
																						{$\bm{\alpha}_{c}^{*} ={\alpha}_{c} \textbf{J}_{n+r,1}$} & {$\bm{\alpha}_{c}^{*} ={\alpha}_{c} \textbf{J}_{n+r,1}$} & {$\bm{\alpha}_{c}^{*} ={\alpha}_{c} \textbf{J}_{n+r,1}$}\\        \hline
																						{$\bm{\alpha}_{c,\xi}^{*} ={\alpha}_{c,\xi} \textbf{J}_{2n,1}$} & {$\bm{\alpha}_{c,\xi}^{*} ={\alpha}_{c,\xi} \textbf{J}_{2n,1}$} & {$\bm{\alpha}_{c,\xi}^{*} ={\alpha}_{c,\xi} \textbf{J}_{2n,1}$}\\        \hline
																						{$\bm{\alpha}_{c,\beta}^{*} ={\alpha}_{c,\beta} \textbf{J}_{n+p,1}$} & {$\bm{\alpha}_{c,\beta}^{*} ={\alpha}_{c,\beta} \textbf{J}_{n+p,1}$} & {$\bm{\alpha}_{c,\beta}^{*} ={\alpha}_{c,\beta} \textbf{J}_{n+p,1}$}\\        \hline
																						$\bm{\kappa}_{\beta}^{*} = {(\textbf{b}^{\prime},\kappa_{\beta,1},\ldots, \kappa_{\beta,p})^{\prime}}$	  & $\bm{\kappa}_{\beta}^{*} = {(\textbf{b}^{\prime},\textbf{b}^{\prime},\kappa_{\beta,1},\ldots, \kappa_{\beta,p})^{\prime}}$	  & $\bm{\kappa}_{\beta}^{*} = {(\mathrm{exp}(\bm{\eta}^{\prime}\bm{\Phi}^{\prime} + \bm{\xi}^{\prime})+\bm{\epsilon}_{\kappa}^{\prime},\kappa_{\beta,1},\ldots, \kappa_{\beta,p})^{\prime}}$	  \\ \hline
																						$\bm{\kappa}_{\eta}^{*} = {(\textbf{b}^{\prime},{\kappa}_{\eta,1},\ldots, \kappa_{\eta,r})^{\prime}}$ &	$\bm{\kappa}_{\eta}^{*} = {(\textbf{b}^{\prime},\textbf{b}^{\prime},{\kappa}_{\eta,1},\ldots, \kappa_{\eta,r})^{\prime}}$ & $\bm{\kappa}_{\eta}^{*} = {(\mathrm{exp}(\bm{\beta}^{\prime}\textbf{X}^{\prime} + \bm{\xi}^{\prime})+\bm{\epsilon}_{\kappa}^{\prime},{\kappa}_{\eta,1},\ldots, \kappa_{\eta,r})^{\prime}}$   \\ \hline
																						$\bm{\kappa}_{\xi}^{*} = {(\textbf{b}^{\prime},\kappa_{\xi,1},\ldots, \kappa_{\xi,n})^{\prime}}$	&	$\bm{\kappa}_{\xi}^{*} ={(\textbf{b}^{\prime},\textbf{b}^{\prime},\kappa_{\xi,1},\ldots, \kappa_{\xi,n})^{\prime}}$	& $\bm{\kappa}_{\xi}^{*} ={(\mathrm{exp}(\bm{\beta}^{\prime}\textbf{X}^{\prime} +\bm{\eta}^{\prime}\bm{\Phi}^{\prime})+\bm{\epsilon}_{\kappa}^{\prime},\kappa_{\xi,1},\ldots, \kappa_{\xi,n})^{\prime}}$  \\ 
																						\hline
																						$\bm{\kappa}_{\gamma,i} = (\kappa_{\eta,i},\bm{\kappa}_{i}^{\prime})^{\prime}$;\hspace{10pt} $i = 2,\ldots, r$ & $\bm{\kappa}_{\gamma,i} = (\kappa_{\eta,i},\bm{\kappa}_{i}^{\prime})^{\prime}$;\hspace{10pt} $i = 2,\ldots, r$ & $\bm{\kappa}_{\gamma,i} = (\kappa_{\eta,i},\bm{\kappa}_{i}^{\prime})^{\prime}$;\hspace{10pt} $i = 2,\ldots, r$ \\        \hline
																						{$\bm{\kappa}_{c}^{*} = (\bm{\kappa}_{\eta}^{\prime},{\kappa}_{c})^{\prime}$} & {$\bm{\kappa}_{c}^{*} = (\bm{\kappa}_{\eta}^{\prime},{\kappa}_{c})^{\prime}$} & {$\bm{\kappa}_{c}^{*} = (\bm{\kappa}_{\eta}^{\prime},{\kappa}_{c})^{\prime}$}\\        \hline
																						{$\bm{\kappa}_{c,\xi}^{*} = (\bm{\kappa}_{\xi}^{\prime},{\kappa}_{c,\xi})^{\prime}$} & {$\bm{\kappa}_{c,\xi}^{*} = (\bm{\kappa}_{\xi}^{\prime},{\kappa}_{c,\xi})^{\prime}$} & {$\bm{\kappa}_{c,\xi}^{*} = (\bm{\kappa}_{\xi}^{\prime},{\kappa}_{c,\xi})^{\prime}$}\\        \hline
																						{$\bm{\kappa}_{c,\beta}^{*} = (\bm{\kappa}_{\beta}^{\prime},{\kappa}_{c,\beta})^{\prime}$} & {$\bm{\kappa}_{c,\beta}^{*} = (\bm{\kappa}_{\beta}^{\prime},{\kappa}_{c,\beta})^{\prime}$} & {$\bm{\kappa}_{c,\beta}^{*} = (\bm{\kappa}_{\beta}^{\prime},{\kappa}_{c,\beta})^{\prime}$}\\        \hline
																						\end{tabular}}
																						\caption{{A comprehensive list of matrices, vectors, and constants to define the full-conditional distributions in Theorem 3. If $Z_{i}$ does not lay on the boundary of it's support then use the left-hand column. The other columns should be used when $j = k = 2$ and $j = k = 3$ and when there exists $Z_{i}$ on the boundary of it's support (i.e., there exists an $i$ such that $Z_{i} = 0$ or $t_{i}$ for $j = k = 2$ and $Z_{i} = 0$ for $j = k = 3$). The $i$-th element of $\textbf{b}$ is the value of $b$ associated with $Z_{i}$, where we note that this value is assumed to be the same for all $i$. In the left-most column $\epsilon_{\alpha} = \epsilon_{\kappa,i} \equiv 0$. In the middle column $\epsilon_{\alpha}$ is chosen to be ``small'' and $(\epsilon_{\kappa,1},\ldots, \epsilon_{\kappa,n})^{\prime} = \textbf{b}$. In the third column the elements of $\bm{\epsilon}_{\kappa} \equiv (\epsilon_{\kappa,1},\ldots, \epsilon_{\kappa,n})^{\prime}$ and $\epsilon_{\alpha}$ are chosen to be ``small.'' {When $\psi = \psi_{3}$, set $c = c_{\eta} = c_{\xi} = 0$.}}}
																						\label{gprop}
																						\end{table}
																						
																						\begin{sidewaystable}[htp]
																							\centering
																							{\renewcommand{\arraystretch}{7}%
																								\noindent\adjustbox{max width=\textwidth}{%
																									\begin{tabular}{|c|c|c|p{0.5\textwidth}|  }
																										\hline
																										\textbf{Unit Log Partition Function}& \textbf{Form of the Prior Distribution on ${\alpha}$ and ${\kappa}$ (i.e., $f(\alpha,\kappa\vert \gamma_{1},\gamma_{2},\rho)$)}&\textbf{Suggested Hyperparameters} &\textbf{Special Case of the Prior Distribution}\\ \hline
																										$\psi_{1}(Y) = \mathrm{log}\left(-\frac{1}{Y}\right)$ & \shortstack{$\mathrm{exp}\left\lbrace\gamma_{1}\alpha + \gamma_{2}\kappa -\rho\mathrm{log}(\Gamma(\kappa+1)) - \rho(\kappa+1)\mathrm{log}(\alpha)\right\rbrace$ \\ $= \frac{1}{\Gamma(\kappa+1)^{\rho}}\left(\alpha^{-\rho}\mathrm{exp}(\gamma_{2})\right)^{\kappa+1}\mathrm{exp}\left(\gamma_{1}\alpha\right)$}  &\shortstack{\shortstack{$\gamma_{1} = -1000$\\ $\gamma_{2} = 1000$} \\ $\rho = 10^{-15}$}& If $\kappa$ is integer-valued then the conditional distribution of $\kappa\vert \alpha$ is Conway-Maxwell-Poisson with parameters $\alpha^{\rho}\mathrm{exp}(\gamma_{2})$ and $\rho$, and the conditional distribution of $\alpha\vert \kappa$ is $\mathrm{Gamma}((\kappa+1)\rho+1, -1/\gamma_{1})$ provided that $\gamma_{2}\in \mathbb{R}$, $\gamma_{1}$ is negative, and $\rho \ge 0$.\\ \hline
																										$\psi_{2}(Y) = \mathrm{log}\left(1+\mathrm{exp}(Y)\right)$ & 
																										\shortstack{$\mathrm{exp}\left[\gamma_{1}\alpha + \gamma_{2}\kappa + \rho \mathrm{log}\left\lbrace\Gamma(\kappa)\right\rbrace - \rho\mathrm{log}\left\lbrace\Gamma(\alpha)\right\rbrace - \rho \mathrm{log}\left\lbrace\Gamma(\kappa - \alpha)\right\rbrace\right]$\\ $=\left(\frac{\Gamma(\kappa)}{\Gamma(\alpha)\Gamma(\kappa - \alpha)}\right)^{\rho}\mathrm{exp}(\gamma_{1})^{\alpha}\mathrm{exp}(\gamma_{2})^{\kappa}$}
																										&\shortstack{\shortstack{$\gamma_{1} = 0$\\ $\gamma_{2} = -1000$} \\ $\rho = 1$}& Let $\rho = 1$, $\gamma_{1} \in \mathbb{R}$, and $\gamma_{2} < 0$. If $\alpha$ and $\kappa$ are integer-valued, then the conditional distribution of $(\alpha - 1)\vert \kappa$ is binomial with $\kappa$ number of Bernoulli trials, and probability of success $\mathrm{exp}(\gamma_{1})/(1+\mathrm{exp}(\gamma_{1}))$. Also, $(\kappa - \alpha - 1)\vert \alpha$ follows a negative binomial distribution with $\alpha + 1$ number of successful Bernoulli trials, and probability of success $\mathrm{exp}(\gamma_{2})$.\\ \hline
																										$\psi_{3}(Y) = \mathrm{exp}(Y)$ & 
																										$\mathrm{exp}\left\lbrace\gamma_{1}\alpha + \gamma_{2}\kappa -\rho\mathrm{log}(\Gamma(\alpha)) - \rho(\alpha)\mathrm{log}(\kappa)\right\rbrace = \frac{1}{\Gamma(\alpha)^{\rho}}\left(\kappa^{-\rho}\mathrm{exp}(\gamma_{1})\right)^{\alpha}\mathrm{exp}\left(\gamma_{2}\kappa\right)$ &\shortstack{\shortstack{$\gamma_{1} = 1$\\ $\gamma_{2} = -10^{-15}$} \\ $\rho =1$}& If $\alpha$ is integer-valued then the conditional distribution of $(\alpha - 1)\vert \kappa$ is Conway-Maxwell-Poisson with parameters $\kappa^{\rho}\mathrm{exp}(\gamma_{1})$ and $\rho$, and the conditional distribution of $\kappa\vert \alpha$ is $\mathrm{Gamma}(\alpha\rho+1, -1/\gamma_{2})$ provided that $\gamma_{1}\in \mathbb{R}$, $\gamma_{2}$ is negative, and $\rho \ge 0$.\\ \hline
																										$\psi_{4}(Y) = Y^{2}$ & $ \mathrm{exp}\left(\gamma_{1}\alpha + \gamma_{2}\kappa+\frac{\rho}{2}\mathrm{log}{\kappa} - \frac{\alpha^{2}}{4\kappa}\right)= \kappa^{\rho/2+1 - 1}\mathrm{exp}(\gamma_{2}\kappa)\mathrm{exp}(-\frac{(\alpha - 2\kappa\gamma_{1})^{2}}{4\kappa})$ &\shortstack{\shortstack{Set $\alpha = 0$\\ \shortstack{$\gamma_{1} = 0$\\ $\gamma_{2} = -\frac{1}{2}$}} \\ $\rho = 2$}& We have that $\kappa$ is distributed as Gamma$(\rho/2 + 1, -1/\gamma_{2})$ and is independent of $\alpha$, which is distributed as normal with mean $2\kappa\gamma_{1}$ and variance $2\kappa$. The suggested hyperparameters result in an inverse-gamma prior distribution on the variance of a normal random variable with shape 2 and scale 1, which yields mean 1 and variance infinity.\\ \hline
																										\end{tabular}}}
																										\caption{Special Cases: We list the form of the the prior distribution in Equation (2) of the main text by $\psi_{j}$ for $j = 1,\ldots, 4$. The first column has the unit log partition function, the second column has the form of the prior distributions (up to a proportionality constant), the third column gives suggested hyperparameters, and the fourth column gives special cases of the conditional distributions $\alpha\vert \kappa$ and $\kappa\vert \alpha$.}
																										\end{sidewaystable}
																										
																										Using induction we find that,
																										\begin{align*}
																										& f(\bm{\eta}\vert c\textbf{J}_{a,1},{\textbf{M}},\bm{\alpha}_{\eta},\bm{\kappa}_{\eta})\underset{\textbf{V}}{\propto}\mathrm{exp}\left[{\alpha}_{\eta}\textbf{J}_{r,1}^{\prime}\textbf{V}^{-1}\bm{\eta} - {\kappa}_{\eta}\textbf{J}_{r,1}^{\prime}\psi\left\lbrace\textbf{V}^{-1}\bm{\eta}-c\textbf{J}_{r,1}\right\rbrace\right]\\
																										&\propto\mathrm{exp}\left[\sum_{i = 2}^{r}\sum_{j = 1}^{i-1}\alpha_{\eta}v_{i,j}\eta_{j} - \sum_{i = 2}^{r}\kappa_{\eta}\psi\left(\sum_{j = 1}^{i-1}\eta_{j}v_{i,j} + \eta_{i} - c\right)\right]\\
																										&=\prod_{i = 2}^{r}\mathrm{exp}\left\lbrace \alpha_{\eta}\bm{\Sigma}_{i}\textbf{v}_{i} - \kappa_{\eta}\psi(\bm{\Sigma}_{i}\textbf{v}_{i} + Y_{i}-c)\right\rbrace
																										\end{align*}
																										\noindent
																										where $\bm{\Sigma}_{i}^{\prime} = \left( \eta_{j}: j = 1,\ldots, i-1\right)^{\prime}$. Thus, the full conditional distribution is given by
																										\begin{align*}
																										& f(\textbf{v}_{2},\ldots,\textbf{v}_{r}\vert \cdot)\underset{\textbf{V}}{\propto}f(\bm{\eta}\vert c\textbf{J}_{a},\textbf{V} ,\bm{\alpha}_{\eta},\bm{\kappa}_{\eta})\prod_{i = 2}^{r}f(\textbf{v}_{i})\\
																										&\underset{\textbf{V}}{\propto}\prod_{i = 2}^{r}\mathrm{exp}\left[ \alpha_{\eta}\textbf{J}_{i,1}^{\prime}\textbf{H}_{\gamma,i}\textbf{v}_{i} - \kappa_{\eta}\textbf{J}_{i,1}^{\prime}\psi\left\lbrace\textbf{H}_{\gamma,i}\textbf{v}_{i} - \bm{\mu}_{\gamma,i}\right\rbrace\right],\\
																										&\underset{\textbf{V}}{\propto} \prod_{i = 2}^{n}\mathrm{CM_{c}}\left(\bm{\mu}_{\gamma,i},\textbf{H}_{\gamma,i},\alpha_{\eta}\textbf{J}_{i,1},\kappa_{\eta}\textbf{J}_{i,1};\hspace{2pt}\psi\right),
																										\end{align*}
																										\noindent
																										where
																										\begin{align*}
																										&\textbf{H}_{\gamma,i}\equiv\left[\begin{array}{c}
																										\bm{\Sigma}_{i} \\ 
																										\sigma_{\nu}\textbf{I}_{i-1}
																										\end{array}\right],\\
																										&\bm{\mu}_{\gamma,i} = \left(\eta_{i},\hspace{5pt}\bm{0}_{1,i-1}\right)^{\prime}.
																										\end{align*}
																										\noindent
																										The Metropolis-Hasting algorithm in Appendix A.ii provides a way to sample $\textbf{v}_{i}$ from $f(\textbf{v}_{i}\vert \cdot,\textbf{q}_{\beta} = \bm{0}_{a,1},\textbf{q}_{\eta} = \bm{0}_{a,1}, \textbf{q}_{\xi} = \bm{0}_{a,1})f(\textbf{q}_{v,i})$, which leads to the following update for $\textbf{v}_{i}$,
																										\begin{equation}\label{simv}
																										\textbf{v}_{i}=(\textbf{H}_{\gamma,i}^{\prime}\textbf{H}_{\gamma,i})^{-1}\textbf{H}_{\gamma,i}^{\prime}\bm{\mu}_{\gamma,i} + (\textbf{H}_{\gamma,i}^{\prime}\textbf{H}_{\gamma,i})^{-1}\textbf{H}_{\gamma,i}^{\prime}\textbf{w},
																										\end{equation}
																										\noindent
																										where $\textbf{w}\sim \mathrm{CM}(\bm{0}_{i,1},\textbf{I}_{i},\alpha_{\eta}\textbf{J}_{i,1},\kappa_{\eta}\textbf{J}_{i,1}; \psi)$. \\

																										\subsection*{Appendix C.iii: Step-by-Step Implementation} 
																										The Gibbs sampler associated with (20) of the main text requires one to compute certain quantities. These values are listed in Table 2. To aid the reader, we provide step-by-step instructions for implementing the Gibbs sampler associated with (20) of the main text as follows.
																										
																										\begin{enumerate}
																											\item Initialize $\bm{\beta}$, $\bm{\eta}$, $\bm{\xi}$, $c$, $c_{\beta}$, $c_{\xi}$, $\{\textbf{v}_{i}\}$, ${\alpha}_{\eta}$, $\alpha_{\xi}$, ${\kappa}_{\eta}$, and $\kappa_{\xi}$. Denote these initializations with $\bm{\beta}^{[0]}$, $\bm{\eta}^{[0]}$, $\bm{\xi}^{[0]}$, $c^{[0]}$, $c_{\beta}^{[0]}$, $c_{\xi}^{[0]}$, $\{\textbf{v}_{i}^{[0]}\}$, ${\alpha}_{\eta}^{[0]}$, $\alpha_{\xi}^{[0]}$, ${\kappa}_{\eta}^{[0]}$, and $\kappa_{\xi}^{[0]}$. Set $m = 1$.
																											\item Set $\bm{\beta}^{[m]}$ equal to the right hand side of (\ref{simbeta}). The matrix $\textbf{H}_{\beta}$ and the vector $\mu_{\beta}$ are defined in Table 2. The $r$-dimensional vector $\bm{\eta}$ is set equal to $\bm{\eta}^{[m-1]}$, the $n$-dimensional vector $\bm{\xi}$ is set equal to $\bm{\xi}^{[m-1]}$, $\alpha_{\beta}$ is set equal to $\alpha_{\beta}^{[m-1]}$, and $\kappa_{\beta}$ is set equal to $\kappa_{\beta}^{[m-1]}$.
																											\item Set $\bm{\eta}^{[m]}$ equal to the right hand side of (\ref{simeta}). The matrix $\textbf{H}_{\eta}$ and the vector $\mu_{\eta}$ are defined in Table 2. The $p$-dimensional vector $\bm{\beta}$ is set equal to $\bm{\beta}^{[m]}$, the $n$-dimensional vector $\bm{\xi}$ is set equal to $\bm{\xi}^{[m-1]}$, $\alpha_{\eta}$ is set equal to $\alpha_{\eta}^{[m-1]}$, $\kappa_{\eta}$ is set equal to $\kappa_{\eta}^{[m-1]}$, and for each $i$ the $i$-dimensional vector $\textbf{v}_{i}$ is set equal to $\textbf{v}_{i}^{[m-1]}$.
																											\item Set $\bm{\xi}^{[m]}$ equal to the right hand side of (\ref{simxi}). The matrix $\textbf{H}_{\beta}$ and the vector $\mu_{\beta}$ are defined in Table 2. The $r$-dimensional vector $\bm{\eta}$ is set equal to $\bm{\eta}^{[m]}$, the $p$-dimensional vector $\bm{\beta}$ is set equal to $\bm{\beta}^{[m]}$, $\alpha_{\xi}$ is set equal to $\alpha_{\xi}^{[m-1]}$, and $\kappa_{\xi}$ is set equal to $\kappa_{\xi}^{[m-1]}$.
																											\item For $i = 2,\ldots, r$ set $\textbf{v}_{i}^{[m]}$ equal to a value generated to the right hand side of (\ref{simv}). The matrix $\textbf{H}_{\gamma,i}$ and the vector $\mu_{\gamma,i}$ are defined in Table 2. The $r$-dimensional vector $\bm{\eta}$ is set equal to $\bm{\eta}^{[m]}$.
																											\item {Set $c^{[m]}$ equal to a draw from $\mathrm{CM_{c}}(\bm{\mu}_{c},\textbf{H}_{c}^{*},\bm{\alpha}_{c}^{*},{\kappa}_{c}^{*})$ using a slice sampler, where $\bm{\mu}_{c}$, $\textbf{H}_{c}^{*}$, $\bm{\alpha}_{c}^{*}$, and $\bm{\kappa}_{c}^{*}$ are computed using Table 2 and the most current values of the remaining parameters. We have found that $c$ is weakly identifiable, and hence, truncating the support of the prior or using an informative prior often leads to better results.}
																											\item {Set $c_{\beta}^{[m]}$ equal to a draw from $\mathrm{CM_{c}}(\bm{\mu}_{c,\beta},\textbf{H}_{c,\beta}^{*},\bm{\alpha}_{c,\beta}^{*},{\kappa}_{c,\beta}^{*})$ using a slice sampler, where $\bm{\mu}_{c,\beta}$, $\textbf{H}_{c,\beta}^{*}$, $\bm{\alpha}_{c,\beta}^{*}$, and $\bm{\kappa}_{c,\beta}^{*}$ are computed using Table 2 and the most current values of the remaining parameters. We have found that $c_{\beta}$ is weakly identifiable, and hence, truncating the support of the prior or using an informative prior often leads to better results.}
																											\item {Set $c_{\xi}^{[m]}$ equal to a draw from $\mathrm{CM_{c}}(\bm{\mu}_{c,\xi},\textbf{H}_{c,\xi}^{*},\bm{\alpha}_{c,\xi}^{*},{\kappa}_{c,\xi}^{*})$ using a slice sampler, where $\bm{\mu}_{c,\xi}$, $\textbf{H}_{c,\xi}^{*}$, $\bm{\alpha}_{c,\xi}^{*}$, and $\bm{\kappa}_{c,\xi}^{*}$ are computed using Table 2 and the most current values of the remaining parameters. We have found that $c_{\xi}$ is weakly identifiable, and hence, truncating the support of the prior or using an informative prior often leads to better results.}
																											\item Use a slice sampler {(or Metropolis)} to set $\alpha_{\beta}^{[m]}$ and $\kappa_{\beta}^{[m]}$ to a value generated from the pdf:
																											\begin{align*}
																											& f\left(\alpha_{\beta}, \kappa_{\beta} \vert \cdot\right)&\\
																											\nonumber
																											&\propto \mathrm{exp}\left[ (\gamma_{\beta,1}+{\textbf{J}_{1,p}}\textbf{V}_{\beta}^{[m]-1}\bm{\beta}^{[m]})\alpha_{\beta} +  \left\lbrace\gamma_{\beta,2}-{\textbf{J}_{1,g}}\bm{\psi}(\textbf{M}_{\beta}^{[m]}\bm{\beta}^{[m]} - c_{\beta}^{[m]}\textbf{J}_{g,1})\right\rbrace \kappa_{\beta} \right.\\
																											&\hspace{100pt}\left.- ({\rho}_{\beta}+g)\mathrm{log}\left\lbrace \frac{1}{K\left(\alpha_{\beta}, \kappa_{\beta}\right)}\right\rbrace\right],
																											\end{align*}
																											\noindent
																											{where $g = p$ if no boundary value update is needed, $g = n+p$ if $\psi = \psi_{3}$, and $g = 2n+p$ if $\psi = \psi_{2}$.}
																											\item Use a slice sampler {(or Metropolis)} to set $\alpha_{\eta}^{[m]}$ and $\kappa_{\eta}^{[m]}$ to a value generated from the pdf:
																											\begin{align*}
																											& f\left(\alpha_{\eta}, \kappa_{\eta} \vert \cdot\right)&\\
																											\nonumber
																											&\propto \mathrm{exp}\left[ (\gamma_{\eta,1}+{\textbf{J}_{1,r}}\textbf{V}^{[m]-1}\bm{\eta}^{[m]})\alpha_{\eta} +  \left\lbrace\gamma_{\eta,2}-{\textbf{J}_{1,g}}\bm{\psi}(\textbf{M}^{[m]}\bm{\eta}^{[m]}- c^{[m]}\textbf{J}_{g,1})\right\rbrace \kappa_{\eta}\right.\\
																											&\hspace{100pt}\left. - ({\rho}_{\eta}+g)\mathrm{log}\left\lbrace \frac{1}{K\left(\alpha_{\eta}, \kappa_{\eta}\right)}\right\rbrace\right],
																											\end{align*}
																											\noindent
																											{where $g = r$ if no boundary value update is needed, $g = n+r$ if $\psi = \psi_{3}$, and $g = 2n+r$ if $\psi = \psi_{2}$.}
																											\item Using a slice sampler {(or Metropolis)} to set $\alpha_{\xi}^{[m]}$ and $\kappa_{\xi}^{[m]}$ equal to values generated from the pdf:
																											\begin{align*}
																											& f\left(\alpha_{\xi}, \kappa_{\xi}\vert \cdot\right)& \\
																											&\propto \mathrm{exp}\left[ (\gamma_{\xi,1}+\textbf{J}_{1,n}\bm{\xi})\alpha_{\xi} +  \left\lbrace\gamma_{\xi,2}-\textbf{J}_{1,2n}\psi(\textbf{M}_{\xi}\bm{\xi}- c_{\xi}^{[m]}\textbf{J}_{g,1}) \right\rbrace\kappa_{\xi}\right.\\
																											&\hspace{100pt}\left. - ({\rho}_{\xi}+g)\mathrm{log}\left\lbrace \frac{1}{K\left(\alpha_{\xi}, \kappa_{\xi}\right)}\right\rbrace\right],
																											\end{align*}
																											\noindent
																											{where $g = n$ if no boundary value update is needed, $g = 2n$ if $\psi = \psi_{3}$, and $g = 3n$ if $\psi = \psi_{2}$.}
																											\item Set $m = m+1$.
																											\item Repeat steps 2 through 12 until convergence of the Gibbs sampler.
																											\end{enumerate}
																											
																											\noindent    
																											It is straightforward to adjust this Gibbs sampler in variety of ways to be more appropriate for a particular problem. For example, one could consider different hyperparameters, different basis functions $\{\bm{\phi}_{j}\}${, update the shape and scale of the prior on $\textbf{V}^{-1}$,} and assume heterogeneous DY parameters associated with $\bm{\beta}$, $\bm{\eta}$, and $\bm{\xi}$.    
																											
																											{It is important to note that many software packages have built in functions to simulate from beta and gamma distributions, which are needed when $j = k = 2$ and $j = k = 3$, respectively. However, it is common for the Gibbs sampler to produce small values of shape and scale parameters, which may lead to computational errors when simulating from a beta or a gamma distribution. In this setting, we simulate beta and gamma random variables using strategies outlined in \citet[][pgs. 181, 182, and 419]{Devroye}. Additionally, if the shape and scale parameters are so small (i.e., close to zero) that it is not possible to simulate the beta and gamma random variables using the techniques in \citet{Devroye}, we reject the proposed sample. However, after a sufficient burn-in period of the Gibbs sampler, the acceptance rate is approximately equal to one.} 
																											
																											Finally, the updates for shape and rate parameters can be simplified in many settings. These simplifications often require additional assumptions, such as, the shape parameter is assumed to be integer-valued. We refer the reader to Table 3 to see a list of special cases by log-partition function.

																											\section*{Appendix D: The ANOVA Table for the Simulation Study in Section 3.2 of the Main Text}
																											\renewcommand{\theequation}{D.\arabic{equation}}
																											\setcounter{equation}{0}
																											
																											The ANOVA Table associated with the simulation study in Section 3.2 is given in Table 4. The assumptions for this ANOVA may not hold, and hence, we interpret large F statistics subjectively.
																											\begin{table}[htp]
																												\begin{center}
																													\begin{tabular}{ ccccc}
																														\hline
																														Source &	DF & SS & MS & F \\\hline
																														Factor 1 &	1 & 8.174 & 8.174 & \textbf{941.13} \\
																														Factor 2 &	1 & $\approx 0$  & $\approx 0$ & 0.05 \\
																														Factor 3 &	1 & 0.01 & 0.007 & 0.86\\
																														Factor 4 &	2 & 1487.85 & 743.926 & \textbf{85649.6}\\ 	
																														Factor 5 &	1 & $\approx 0$ & $\approx 0$ & $\approx 0$\\
																														Factor 6 &	1 & 0.07 & 0.066 & 7.63\\
																														Factor 1 $\times$ Factor 2 &	1 & 0.12 & 0.117 & 13.51\\
																														Factor 1 $\times$ Factor 3 &	1 & $\approx 0$ & $\approx 0$ & 0.01\\
																														Factor 1 $\times$ Factor 4 &	2 & 8.08 & 4.041 & \textbf{465.24}\\
																														Factor 1 $\times$ Factor 5 &	1 & 0.04 & 0.039 & 4.53\\
																														Factor 1 $\times$ Factor 6 &	1 & $\approx 0$ & 0.001 & 0.16\\
																														Factor 2 $\times$ Factor 3 &	1 & 0.01 & 0.011 & 1.25\\
																														Factor 2 $\times$ Factor 4 &	2 & 0.01 & 0.003 & 0.3\\
																														Factor 2 $\times$ Factor 5 &	1 & $\approx 0$ & $\approx 0$ & 0.01\\
																														Factor 2 $\times$ Factor 6 &	1 & $\approx 0$ & $\approx 0$ & 0.01\\
																														Factor 3 $\times$ Factor 4 &	2 & $\approx 0$ & 0.001 & 0.17\\
																														Factor 3 $\times$ Factor 5 &	1 & $\approx 0$ & $\approx 0$ & 0.04\\
																														Factor 3 $\times$ Factor 6 &	1 & 0.03 & 0.035 & 3.99\\
																														Factor 4 $\times$ Factor 5 &	2 & $\approx 0$ & $\approx 0$ & 0.04\\
																														Factor 4 $\times$ Factor 6 &	2 & $\approx 0$ & 0.002 & 0.28\\
																														Factor 5 $\times$ Factor 6 &	1 & 0.01 & 0.011 & 1.31\\
																														Residual &	925 & 8.03 & 0.009 & \hfill \\ \hline
																														\end{tabular}
																														\end{center}
																														\caption{Analysis of variance (ANOVA). The response in this experiment is the log total prediction error in (21) of the main text. The six factors are listed in Section 3.2 of the main text, for up to two-way interactions. In the table, the column ``Source'' contains the source of variability; ``DF'' stands for degrees of freedom; ``SS'' denotes the sum of squared error; ``MS'' stands for mean squared error; and ``F'' denotes the F-statistic. There are 96 factor-level combinations each containing 10 replicates. We denote ``approximately equal to zero'' with ``$\approx 0$.'' Large F-statistics are bold.}
																														\end{table}
																														
																														\section*{Acknowledgments} We would like to express our sincere gratitude to the editors, the associate editor, and the referees for their very helpful comments that improved this manuscript. We would also like to thank Drs. Matthew Simpson of SAS Inc. and Erin Schliep at the University of Missouri for helpful discussions. This research was partially supported by the U.S. National Science Foundation (NSF) and the U.S. Census Bureau under NSF grant SES-1132031, funded through the NSF-Census Research Network (NCRN) program. This article is released to inform interested parties of ongoing research and to encourage discussion of work in progress. The views expressed are those of the authors and not those of the NSF or the U.S. Census Bureau.
																														%
																														%
																														
																														\baselineskip=14pt \vskip 4mm\noindent
																														
																														\singlespacing
																														\bibliographystyle{jasa}  
																														\bibliography{myref}

\end{document}

%% file: notations4.tex
\def\bs{\textbf{s}}

\def\bz{\textbf{Z}}

\def\by{\textbf{Y}}

\def\bfbeta{\bm{\beta}}